\documentclass[twocolumn]{aastex6}

\usepackage{xspace}
\usepackage{graphicx}
\usepackage{natbib}
\usepackage{amssymb}

\newcommand{\nraoblurb}{The National Radio Astronomy Observatory is
a facility of the National Science Foundation operated under cooperative
agreement by Associated Universities, Inc.}
\newcommand{\hide}[1]{}

\newcommand{\gl}{\ensuremath{\ell}\xspace}
\newcommand{\gb}{\ensuremath{{\it b}}\xspace}

\newcommand{\lb}{\ensuremath{(\gl,\gb)}\xspace}
\newcommand{\lv}{\ensuremath{(\gl,v)}\xspace}

\newcommand{\ra}{\ensuremath{\alpha}\xspace}
\newcommand{\dec}{\ensuremath{\delta}\xspace}

\newcommand{\kms}{\ensuremath{\,{\rm km\,s^{-1}}}\xspace}

\newcommand{\microns}{\ensuremath{\,\mu{\rm m}}\xspace}

\newcommand{\kpc}{\ensuremath{\,{\rm kpc}}\xspace}

\newcommand{\ghz}{\ensuremath{\,{\rm GHz}}\xspace}

\newcommand{\degper}{\ensuremath{\rlap.{^{\circ}}}}

\newcommand{\rgal}{\ensuremath{\,R_{\rm Gal}}\xspace}   

\newcommand{\hi}{{\rm H\,{\footnotesize I}}\xspace}
\newcommand{\hii}{{\rm H\,{\footnotesize II}}\xspace}

\newcommand{\ammonia}{\ensuremath{\rm NH_3}\xspace}

\shorttitle{High-Mass Star Formation in the OSC}
\shortauthors{Armentrout et al.}
\begin{document}

\title{High-Mass Star Formation in the Outer Scutum-Centaurus Arm}

\author{W.~P.~Armentrout\altaffilmark{1, 2}}
\author{L.~D.~Anderson\altaffilmark{1, 2, 3}}
\author{Dana~S.~Balser\altaffilmark{4}}
\author{T.~M.~Bania\altaffilmark{5}}
\author{T.~M.~Dame\altaffilmark{6}}
\author{Trey~V.~Wenger\altaffilmark{4,7}}

\altaffiltext{1}{Department of Physics and Astronomy, West Virginia
  University, Morgantown, West Virginia 26505, USA}
\altaffiltext{2}{Center for Gravitational Waves and Cosmology, West Virginia University, Chestnut Ridge Research Building, Morgantown, WV 26505}
\altaffiltext{3}{Adjunct Astronomer at the Green Bank Observatory, PO Box 2, Green
  Bank, WV 24944, USA}
\altaffiltext{4}{National Radio Astronomy Observatory, 520 Edgemont Road, Charlottesville, VA 22903-2475, USA}
\altaffiltext{5}{Institute for Astrophysical Research, Department of Astronomy, Boston University, 725 Commonwealth Avenue, Boston, MA 02215, USA}
\altaffiltext{6}{Harvard-Smithsonian Center for Astrophysics, 60 Garden Street, Cambridge, MA 02138, USA}
\altaffiltext{7}{Astronomy Department, University of Virginia, P.O. Box 400325, Charlottesville, VA 22904-4325, USA}

\begin{abstract}
The Outer Scutum-Centaurus (OSC) spiral arm is the most distant molecular spiral arm in the Milky Way, but until recently little was known about this structure. Discovered by Dame and Thaddeus (2011), the OSC lies $\sim$15 kpc from the Galactic Center. Due to the Galactic warp, it rises to nearly 4$^{\circ}$ above the Galactic Plane in the first Galactic quadrant, leaving it unsampled by most Galactic plane surveys. Here we observe \hii\ region candidates spatially coincident with the OSC using the Very Large Array to image radio continuum emission from 65 targets and the Green Bank Telescope to search for ammonia and water maser emission from 75 targets. This sample, drawn from the WISE Catalog of Galactic \hii\ Regions, represents every \hii\ region candidate near the longitude-latitude \lv\ locus of the OSC. Coupled with their characteristic mid-infrared morphologies, detection of radio continuum emission strongly suggests that a target is a bona fide \hii\ region. Detections of associated ammonia or water maser emission allow us to derive a kinematic distance and determine if the velocity of the region is consistent with that of the OSC. Nearly 60\% of the observed sources were detected in radio continuum, and over 20\% have ammonia or water maser detections. The velocities of these sources mainly place them beyond the Solar orbit. These very distant high-mass stars have stellar spectral types as early as O4. We associate high-mass star formation at 2 new locations with the OSC, increasing the total number of detected \hii\ regions in the OSC to 12. 
\end{abstract}



\section{Introduction\label{sec:intro}}

\begin{figure*}[!t]
\includegraphics[width=\textwidth]{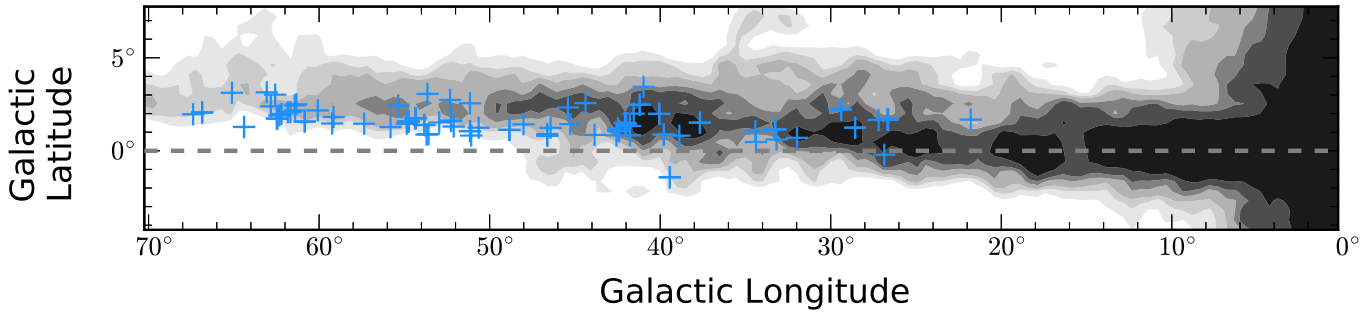}
\caption{The Outer Scutum-Centaurus Arm as traced by HI emission. The HI LAB survey was integrated over a 14 \kms\ wide window following the center velocity given by $V_{LSR}$ = $-$1.6 km~s$^{-1}$ deg$^{-1}\times\gl\/$ in \citet{dame11}. This clearly shows the Galactic warp bending the arm towards positive latitudes. Overplotted are our observed \hii\ region targets (blue), selected from the WISE Catalog of Galactic \hii\ regions near the \lb\ locus of the OSC arm. \label{fig:dameTargets}}
\end{figure*}

An \hii\ region is an area of ionized hydrogen surrounding massive stars. Stars of O and early B-type emit enough Lyman-continuum radiation ($\lambda<91.2$ nm, $E_{ph}>13.6$ eV) to completely ionize the interstellar hydrogen surrounding them for up to tens of parsecs \citep{stromgren39}. They are the archetypical tracers of star formation. Because of the short lifetimes of OB stars, \hii\ regions are zero-age objects compared to the age of the Milky Way; they therefore trace the state of the Galaxy at the current epoch. \hii\ regions form preferentially in spiral arms where interstellar gas is overdense, and because of their relatively short lifetimes, they do not stray far from their birth places. \hii\ regions can be detected across the entire Galactic disk and are the brightest objects in the Milky Way at mid-Infrared wavelengths. They can therefore be used to trace structure throughout the Galaxy.

The structure of the Milky Way is still an area of active study, though most Galactic models assume a symmetric barred spiral. The discovery of the symmetric near and far components of the 3-kpc arms adds support to this argument \citep{dame08}, but other searches for symmetric far-side features have not been as successful. This is largely a function of their extreme distance from us; signals from the molecular gas and stars tracing these far-side features would traverse over 10 kpc of in-plane gas and dust, leading to attenuation and source confusion. The Galaxy, however, warps above the plane in the first quadrant. This means especially distant structures on the outer edges of the Milky Way should rise out of the plane to positive latitudes, making them distinct from the bulk of the Galactic disk. \citet{dame11} were the first to discover a large-scale structure within this distant warp, possibly a symmetric counterpart to the third quadrant Perseus Arm. They called this structure the ``Outer Scutum-Centaurus Arm," or OSC.

The OSC appears to be a continuation of the Scutum-Centaurus spiral arm which apparently begins on the near end of the Galactic bar at Galactic longitude $\gl\,\sim30^{\circ}$ and crosses between the Galactic center and the Sun into the fourth quadrant. \citet{dame11} first identified the arm in the first quadrant using the LAB 21 cm survey \citep{kalberla05}. Guided by these \hi\ data, further measurements using the Harvard-Smithsonian Center for Astrophysics 1.2 meter telescope revealed molecular gas within the arm. Emission from CO was detected at 10 of 220 targeted positions, with velocities consistent with this gas being located in the extreme outer Galaxy, i.e. $\rgal\,\gtrsim13$ \kpc\ based on a Brand Galactic rotation curve \citep[R$_0$=8.5 kpc, $\Theta_0$=220 \kms\/;][]{brand93}. In the first quadrant, the OSC arm warps significantly above the midplane, arcing to a Galactic latitude of nearly 4$^{\circ}$ at a Galactic longitude of 70$^{\circ}$. \citet{sun15} extended the search for CO emission in the OSC into the second quadrant, showing evidence for the distant spiral arm continuing to nearly 150$^{\circ}$ Galactic longitude.  While \citet{izumi14} and \citet{kobayashi08} found evidence for stellar mass stars embedded within Digel Clouds 1 and 2 in the extreme outer Galaxy of the second quadrant, no high-mass star formation has been observed beyond the OSC.

This arm may represent the outermost limit of high-mass star formation within the Milky Way. \citet{anderson15b} identified 6 Galactic \hii\ regions with radio recombination line (RRL) velocities within 15 \kms\ of the longitude-velocity \lv\ locus of the OSC, defined by \citet{dame11} as $V_{LSR}$ = $-$1.6 km~s$^{-1}$ deg$^{-1}\times\gl\/$. An additional 4 \hii\ regions from the WISE Catalog of Galactic \hii\ Regions have RRL velocities consistent with the OSC \citep{anderson12c}, bringing the OSC \hii\ region count to 10 before the present analysis.

\section{Target Selection}
\label{sec:targets}
Targets for these observations were taken from the Wide-field Infrared Survey Explorer (WISE) Catalog of Galactic \hii\ Regions. To produce this catalog, \citet{anderson14} identified \hii\ region candidates by their mid-infrared emission, primarily from the 12 and 22-\microns\ bands of WISE. The characteristic morphology of \hii\ regions in the mid-infrared is a 22-\microns\ core of emission surrounded by a more diffuse 12-\microns\ envelope of small-grain emission \citep{watson08, anderson11}. Searching within $8^{\circ}$ of the Galactic midplane, we identified more than 6000 \hii\ region candidates with this morphology in the WISE Catalog. Of these 6000 targets, $\sim$2000 show coincident radio continuum emission. The sources lacking radio continuum data are denoted ``radio quiet." In earlier searches for \hii\ regions, primarily with the Green Bank Telescope \citep[GBT HRDS;][]{bania10} and the Arecibo Observatory \citep{bania12}, only sources with detected coincident radio continuum emission were observed. Since \hii\ regions emit thermal bremsstrahlung radiation in the radio regime, WISE candidates with coincident radio continuum detections were deemed likely \hii\ regions.

A source lacking detected radio continuum emission can generally be attributed to a lack of survey sensitivity, although some targets might not be \hii\ regions at all. Since the majority of Galactic plane surveys do not have wide latitude coverage, candidates that deviate from the mid-plane by over a degree are often not covered by sensitive surveys in the radio regime. Sources with faint thermal bremsstrahlung continuum emission could be either at a large distance from the Sun or have a low intrinsic luminosity. That is, some infrared-identified candidate \hii\ regions could be associated with lower mass stars than those that create \hii\ regions, i.e. later B-stars. Throughout this work, any infrared-identified source with the characteristic morphology of \hii\ regions described above will be referred to as a ``candidate," whether it has previously detected radio continuum emission or not.

To define our target search area, we integrated data from the LAB 21 cm survey over a 14 \kms\ wide window following the \lv\ locus of the OSC. We considered all WISE \hii\ region candidates between 20$^{\circ}$ and 70$^{\circ}$ Galactic longitude. Candidates spatially coincident with the integrated LAB emission were added to our target list. These sources roughly followed the longitude-latitude \lb\ locus of the OSC, defined by \citet{dame11} as $b = 0.375^{\circ} + 0.075\times$\gl\/. In addition, we added \hii\ regions from \citet{anderson15b} with velocities within 15 \kms\ of the \lv\ locus of the OSC to our target list. In total, we identified 75 OSC \hii\ region candidates. Of these, 65 comprised our candidate list for the National Radio Astronomy Observatory's Karl G. Jansky Very Large Array (VLA) in Socorro, NM. An additional 10 sources were added to the candidate list for observations with the Robert C. Byrd Green Bank Telescope (GBT) in Green Bank, WV the following year. Shown in Figure~\ref{fig:dameTargets}, these candidates are overlaid on \hi\ emission, integrated along the expected velocity of the OSC to highlight the structure of the arm.

One source in particular, G039.183$-$01.422, lies significantly below the plane. We measured the velocity of this source as part of the WISE HRDS, and although it lies $\sim3^{\circ}$ below the OSC locus, its velocity is consistent with that of the OSC at the same longitude.

\begin{deluxetable}{rcccc}
\tabletypesize{\scriptsize} 
\tablecaption{Observation Summary \label{tab:observeParameters}}
\tablecolumns{5}
\tablenum{1}
\tablewidth{0pt}
\tablehead{\colhead{Facility} &  \colhead{\# Sources} & \colhead{$t_{int} \mathrm{(min)}$} & \colhead{$\nu~(\ghz\/)$} & \colhead{Transition} }
\startdata
VLA & 65 & 4 & 8-10 & Continuum \\
 &  & &  & H87$\alpha$ $-$ H93$\alpha$ \\
GBT & 75 & 6-36 & 22.2351 & H$_2$O 6(1,6)$\rightarrow$5(2,3)\\
& & & 23.6945 & NH$_{3}$ (J,K) = (1,1)\\
& & & 23.7226 & NH$_{3}$ (J,K) = (2,2)\\
& & & 23.8701 & NH$_{3}$ (J,K) = (3,3)\\
\enddata
\end{deluxetable}

\vspace{-25pt}
\section{Observations\label{sec:obs}}

Radio recombination line emission unambiguously identifies a candidate as an \hii\ region, but this emission is intrinsically weak. Distant \hii\ regions require prohibitively long observations for RRL detections. Other options for detecting distant \hii\ regions include observing radio continuum emission or molecular line emission. \hii\ regions emit thermal bremsstrahlung continuum radiation in the radio regime. This locates an \hii\ region on the sky but does not allow us to determine a source velocity. Detection of molecular gas emission will provide this missing source velocity. Thus, we followed a two tier observing strategy and observed both radio continuum and molecular line emission of OSC \hii\ region targets as summarized in Table~\ref{tab:observeParameters}. Our total sample included 75 \hii\ region candidates. We observed 65 of these candidates using the VLA in the Summer of 2014 and all 75 with the GBT in early 2015 and early 2016. We mapped radio continuum emission around \hii\ region candidates with the VLA and searched for molecular gas emission with the GBT.

Ammonia has a relatively high critical density ($\sim2\times10^3$ cm$^{-3}$ at 10 K), and observations of ammonia inversion transitions can trace the dense molecular cores of \hii\ regions. We posit that the regions we detect in ammonia are sites of recent star formation, however, \citet{balser11} and \citet{balser15} showed that \hii\ regions further from the Galactic center have lower metallicities on average. Consequently, as we observe farther and farther from the Galactic center, there simply may not be enough metals to produce detectable quantities of molecular gas, especially at the great distance of the OSC.

\subsection{VLA Radio Continuum Snapshots \label{sec:VLAcont}}

We observed OSC \hii\ region candidates at X-band with the VLA in the 14A semester (VLA14A-194 in July-August 2014). With a minimum baseline of 35 m, D-configuration allowed us to be sensitive to large scale emission. The largest angular scale resolvable by continuum snapshots with the VLA in D-configuration at X-band was $\sim70^{\prime\prime}$. This was well-matched to our expected \hii\ region sizes. Two overlapping basebands, covering from 8.012 to 10.041 \ghz\/, gave us a total bandwidth of over 2 \ghz\/. We also had 8 high spectral resolution, 128 MHz bandwidth spectral windows tuned to radio recombination lines H87$\alpha$ through H93$\alpha$, but we did not expect to have the sensitivity to detect these RRLs with our short integrations. We integrated for 4 minutes per source. The VLA data have a typical spatial resolution of $10^{\prime\prime}$.

Our typical noise was on order 0.1 mJy/beam. An \hii\ region created by a single B1.5 star at a Solar distance of 22 kpc, the farthest detected distance for a Galactic \hii\ region to date \citep{anderson15b}, would result in a flux density of $\sim$0.3 mJy for unresolved sources. This would be a 3$\sigma$ detection at our sensitivity. We therefore could detect any \hii\ region within the Milky Way.

\subsection{GBT Molecular Line Observations \label{sec:GBTammonia}}

We observed the (J,K) = (1,1), (2,2), and (3,3) ammonia inversion transitions (at 23.6945, 23.7226, and 23.8701 \ghz\/, respectively) plus the water maser transition H$_{2}$O 6(1,6)$\rightarrow$5(2,3) (at 22.2351 $\ghz\/$) with the GBT. These observations spanned two years and included projects GBT13B-403 in January 2015, GBT15B-232 in January-March 2016, and GBT16B-414 in January 2017. Observations in 2015 made use of the now decommissioned Autocorrelation Spectrometer (ACS), while 2016 and 2017 observations used the Versatile GBT Astronomical Spectrometer (VEGAS). We had a bandwidth of 23.44 MHz in GBT mode 22 which gave a native spectral resolution of 1.4 kHz. For all observations, we nodded between two beams using the 7 beam K-Band Focal Plane Array (KFPA), with an integration time of at least 6 minutes per source. This beam nodding routine ensured that one beam was always on source.  We observed faint detections longer as needed to raise their significance to the 3$\sigma$ level.

\begin{deluxetable*}{hrhhrrhrrrr}
\tablecaption{VLA Radio Continuum Parameters \label{tab:contParams}}
\tabletypesize{\footnotesize} 
\tablecolumns{11}
\tablenum{2}
\tablehead{&\colhead{Name}&&&\colhead{$\ra_{J2000}$}&\colhead{$\dec_{J2000}$}&&\colhead{$S_{int}$}&\colhead{$S_{peak}$}&\colhead{$\sigma S_{peak}$}&\colhead{Region Area}\\
&&&& \colhead{(hh:mm:ss)} & \colhead{(dd:mm:ss)} & & \colhead{(mJy)} & \colhead{(mJy beam$^{-1}$)} & \colhead{(mJy beam$^{-1}$)} & \colhead{(arcsec$^2$)}}
\startdata
DB225&G021.541+01.675&21.541&1.676&18:24:25.27&$-$09:20:38.6&$\ldots$&$\ldots$&$\ldots$&0.09&$\ldots$\\
DB003&G026.380+01.677&36.380&1.677&18:33:32.75&$-$05:05:02.5&$\ldots$&$\ldots$&$\ldots$&0.53&$\ldots$\\
DB005&G026.417+01.683&26.418&1.683&18:33:30.61&$-$05:01:17.2&Y&6.76&2.37&0.17&1110\\
DB008&G026.942+01.657&26.942&1.658&18:34:34.07&$-$04:34:05.0&$\ldots$&$\ldots$&$\ldots$&0.07&$\ldots$\\
DB100&G028.320+01.243&28.320&1.243&18:38:34.88&$-$03:32:04.8&Y&20.09&16.92&0.32&310\\
DB009&G029.138+02.218&29.138&2.219&18:36:36.47&$-$02:21:38.3&$\ldots$&$\ldots$&$\ldots$&0.08&$\ldots$\\
DB015&G033.007+01.150&33.008&1.151&18:47:28.74&00:35:33.0&Y&28.93&2.45&0.22&3670\\
DB016&G034.172+01.054&34.172&1.054&18:49:56.80&01:35:03.0&$\ldots$&$\ldots$&$\ldots$&0.07&$\ldots$\\
DB018&G037.419+01.513&37.419&1.514&18:54:13.87&04:41:00.5&Y&7.80&0.39&0.08&5380\\
FQ431&G038.627+00.812&38.627&0.813&18:58:56.89&05:26:18.8&Y&7.10&0.49&0.07&2880\\
FQ009&G039.183$-$01.422&39.183&$-$1.422&19:07:56.99&04:54:29.9&Y&53.54&8.27&0.05&1850\\
FQ438&G039.536+00.872&39.536&0.872&19:00:24.43&06:16:26.4&Y&7.61&1.55&0.06&950\\
DB021&G039.801+01.984&39.801&1.985&18:56:54.33&07:01:03.3&Y&2.17&0.25&0.07&3880\\
DB026&G040.723+03.442&40.723&3.442&18:53:21.25&08:29:59.4&Y&1.00&0.30&0.07&790\\
DB201&G040.954+02.473&40.955&2.473&18:57:16.22&08:15:59.2&$\ldots$&$\ldots$&$\ldots$&0.06&$\ldots$\\
DB027&G041.515+01.333&41.515&1.333&19:02:24.16&08:14:40.0&Y&2.38&0.22&0.06&2560\\
DB202&G041.522+00.826&41.522&0.827&19:04:13.99&08:01:08.1&Y&2.14&0.34&0.06&1350\\
DB032&G042.154+01.045&42.154&1.046&19:04:37.06&08:40:51.2&Y&5.45&0.50&0.03&3360\\
DB102&G042.209+01.081&42.210&1.081&19:04:35.75&08:44:48.1&Y&6.77&6.21&0.03&340\\
DB033&G042.224+01.205&42.225&1.205&19:04:10.65&08:49:00.7&Y&2.19&0.45&0.11&120\\
DB203&G042.310+00.831&42.311&0.831&19:05:40.93&08:43:18.2&Y&0.61&0.25&0.02&760\\
DB204&G043.598+00.855&43.598&0.855&19:07:59.62&09:52:32.1&Y&12.88&3.25&0.03&1480\\
DB037&G044.128+02.564&44.129&2.565&19:02:48.04&11:07:52.1&Y&2.45&0.21&0.05&21830\\
DB040&G045.019+01.434&45.019&1.435&19:08:33.57&11:24:13.7&$\ldots$&$\ldots$&$\ldots$&0.04&$\ldots$\\
DB041&G045.161+02.330&45.161&2.331&19:05:34.46&11:56:28.4&$\ldots$&$\ldots$&$\ldots$&0.04&$\ldots$\\
DB045&G046.178+01.236&46.179&1.236&19:11:28.05&12:20:27.7&$\ldots$&$\ldots$&$\ldots$&0.12&$\ldots$\\
DB207&G046.367+00.801&46.368&0.802&19:13:24.04&12:18:27.4&$\ldots$&$\ldots$&$\ldots$&0.35&$\ldots$\\
DB205&G046.375+00.896&46.375&0.897&19:13:04.16&12:21:28.3&Y&11.04&0.69&0.18&9300\\
DB047&G047.765+01.424&47.765&1.424&19:13:47.83&13:50:00.7&Y&1.78&0.14&0.02&5900\\
DB048&G048.589+01.125&48.589&1.125&19:16:28.10&14:25:26.4&Y&4.22&3.57&0.03&390\\
FQ473&G050.394+01.241&50.394&1.242&19:19:32.07&16:04:27.5&Y&5.29&1.77&0.03&910\\
DB210&G050.830+00.820&50.830&0.820&19:21:56.32&16:15:39.5&$\ldots$&$\ldots$&$\ldots$&0.05&$\ldots$\\
LA802&G050.900+01.055&50.901&1.056&19:21:12.59&16:26:04.8&Y&22.06&1.60&0.04&4470\\
DB055&G050.901+02.554&50.902&2.554&19:15:40.65&17:08:06.3&Y&0.80&0.09&0.02&3300\\
DB059&G051.853+01.304&51.854&1.305&19:22:09.91&17:23:33.3&$\ldots$&$\ldots$&$\ldots$&0.04&$\ldots$\\
DB061&G052.021+01.628&52.021&1.628&19:21:17.92&17:41:30.5&Y&18.68&0.96&0.06&5180\\
DB208&G052.072+02.737&52.073&2.737&19:17:16.67&18:15:21.8&Y&7.61&0.43&0.02&5950\\
DB063&G052.705+01.526&52.706&1.526&19:23:01.94&18:14:52.7&$\ldots$&$\ldots$&$\ldots$&0.09&$\ldots$\\
DB212&G053.334+00.894&53.334&0.895&19:26:37.68&18:30:07.7&$\ldots$&$\ldots$&$\ldots$&0.04&$\ldots$\\
DB064&G053.393+03.057&53.393&3.057&19:18:40.43&19:34:18.7&Y&23.85&1.40&0.04&6640\\
DB213&G053.449+00.871&53.449&0.871&19:26:56.87&18:35:30.6&Y&2.35&0.24&0.07&4910\\
DB065&G053.580+01.387&53.581&1.388&19:25:17.44&18:57:12.8&$\ldots$&$\ldots$&$\ldots$&0.08&$\ldots$\\
DB067&G054.093+01.748&54.094&1.748&19:24:58.58&19:34:32.5&Y&28.62&7.40&0.03&760\\
DB070&G054.490+01.579&54.491&1.579&19:26:24.49&19:50:41.8&Y&37.93&15.31&0.09&980\\
DB071&G054.543+01.560&54.544&1.561&19:26:34.95&19:52:58.9&Y&3.91&1.06&0.03&2510\\
DB072&G054.616+01.452&54.616&1.452&19:27:08.16&19:53:40.2&$\ldots$&$\ldots$&$\ldots$&0.07&$\ldots$\\
S83&G055.114+02.422&55.114&2.422&19:24:29.93&20:47:33.2&Y&601.17&48.66&0.65&2920\\
DB214&G055.560+01.271&55.560&1.272&19:29:43.90&20:38:16.0&$\ldots$&$\ldots$&$\ldots$&0.06&$\ldots$\\
DB105&G057.107+01.457&57.107&1.457&19:32:13.58&22:04:58.9&Y&2.80&2.72&0.07&470\\
DB075&G058.902+01.819&58.903&1.819&19:34:37.13&23:49:46.8&Y&7.16&3.70&0.08&680\\
DB216&G058.988+01.469&58.988&1.469&19:36:08.19&23:44:01.9&$\ldots$&$\ldots$&$\ldots$&0.07&$\ldots$\\
DB076&G059.828+02.171&59.829&2.171&19:35:14.25&24:48:34.9&$\ldots$&$\ldots$&$\ldots$&0.05&$\ldots$\\
DB077&G060.592+01.572&60.592&1.572&19:39:11.27&25:10:58.2&Y&97.63&20.01&0.34&1240\\
DB217&G061.085+02.502&61.085&2.502&19:36:39.45&26:04:02.5&$\ldots$&$\ldots$&$\ldots$&0.03&$\ldots$\\
DB221&G061.153+02.169&61.154&2.170&19:38:05.83&25:57:56.0&$\ldots$&$\ldots$&$\ldots$&0.02&$\ldots$\\
DB218&G061.179+02.446&61.180&2.447&19:37:04.64&26:07:24.5&$\ldots$&$\ldots$&$\ldots$&0.02&$\ldots$\\
DB082&G061.587+02.073&61.587&2.074&19:39:24.89&26:17:44.5&Y&4.40&0.15&0.05&16900\\
DB223&G061.954+01.982&61.955&1.983&19:40:34.61&26:34:16.0&Y&1.11&0.91&0.06&280\\
DB224&G062.074+01.900&62.075&1.901&19:41:09.64&26:38:05.9&$\ldots$&$\ldots$&$\ldots$&0.03&$\ldots$\\
DB083&G062.193+01.714&62.194&1.715&19:42:08.84&26:38:46.9&Y&10.02&0.41&0.05&7360\\
DB219&G062.325+03.019&62.325&3.020&19:37:19.91&27:24:05.1&$\ldots$&$\ldots$&$\ldots$&0.02&$\ldots$\\
DB220&G062.818+03.144&62.819&3.144&19:37:55.70&27:53:33.3&$\ldots$&$\ldots$&$\ldots$&0.05&$\ldots$\\
DB086&G064.862+03.119&64.862&3.120&19:42:35.69&29:39:28.6&$\ldots$&$\ldots$&$\ldots$&0.05&$\ldots$\\
DB107&G066.607+02.060&66.608&2.061&19:50:52.81&30:38:08.0&$\ldots$&$\ldots$&$\ldots$&0.03&$\ldots$\\
DB087&G067.138+01.965&67.138&1.966&19:52:31.21&31:02:33.1&$\ldots$&$\ldots$&$\ldots$&0.05&$\ldots$\\
\enddata
\tablecomments{Parameters in this table are a result of the CASA routine imstat with user-defined masks.}
\end{deluxetable*}

\begin{figure}[htb!]
\centering
\includegraphics[width=0.73\columnwidth]{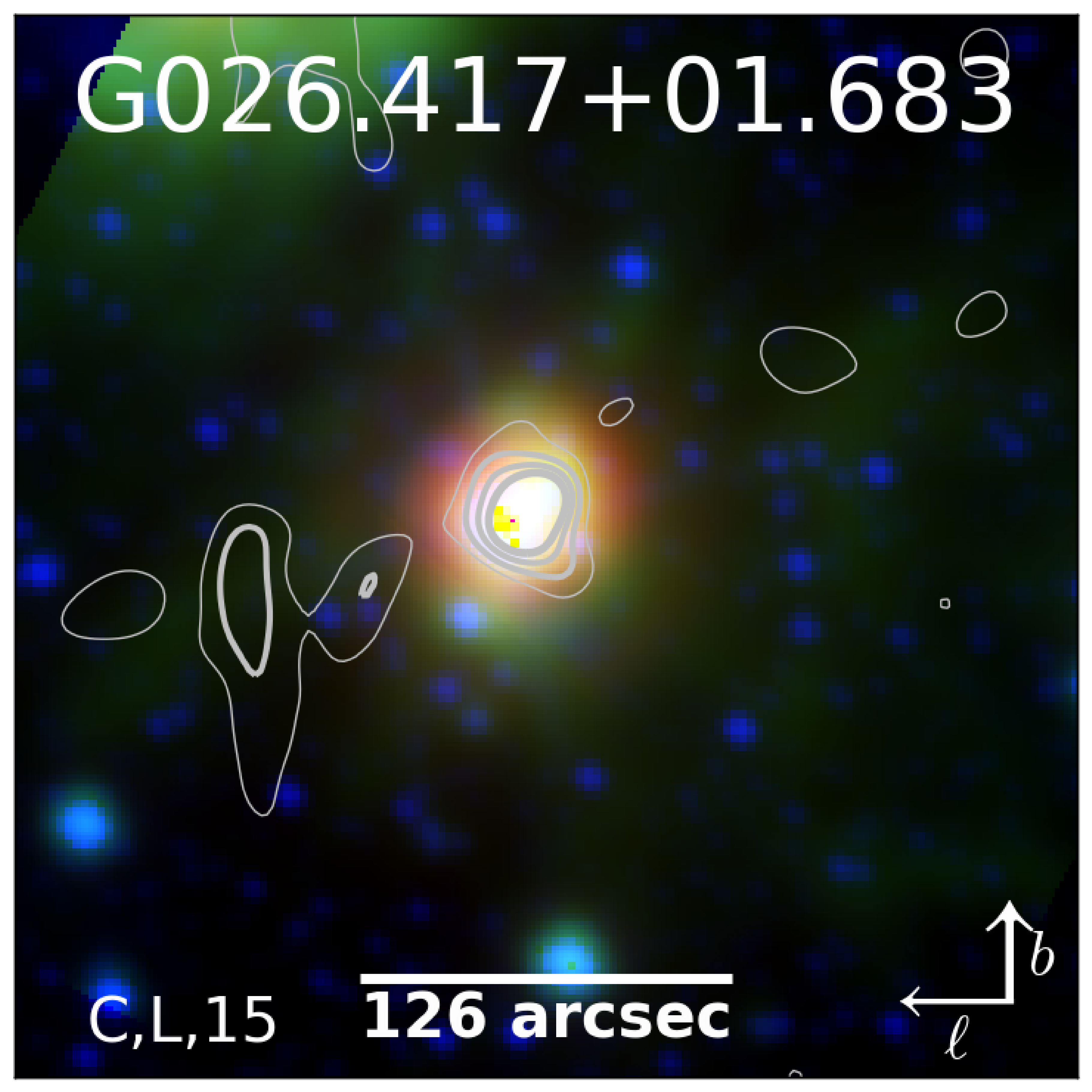}
\caption{\hii\ region snapshot showing both infrared and radio morphologies. Infrared WISE bands w2 (4.6 \microns\/), w3 (12 \microns\/), and w4 (22 \microns\/) are represented by blue, green, and red respectively. VLA continuum contours from 4 minute integrations at X-band in D configuration are overplotted, with contours at 30\%, 50\%, 70\%, and 90\% peak continuum flux (2.37 mJy beam$^{-1}$ here). The image is 6$^{\prime}$ on a side, and a scale bar represents the region's angular size as cataloged in the WISE Catalog of Galactic HII Regions \citep{anderson14}. This source was detected in radio continuum, water maser, and ammonia emission, so both a ``C" and ``L" are included on the lower left, indicating ``continuum" and ``line." Radio data for this source were smoothed by a 15$^{\prime\prime}$ tophat filter for display, and is marked by a ``15" to indicate the smoothing. Images of every source observed by the VLA are shown in Figure~\ref{fig:contSnapshots} in the Appendix. \label{fig:contSnapshotEx}}
\end{figure}

\section{Data Reduction \& Analysis\label{sec:analysis}}

\subsection{VLA Radio Continuum Images \label{sec:ContDetections}}
We used the CASA (Common Astronomy Software Applications) package to reduce and analyze our VLA data \citep{mcmullin07}. Version 1.2.2 of the CASA scripted pipeline automatically calibrated data for this project. The bulk of data analysis was done within CASA as well.

We formed images using the \textit{clean} routine in CASA where we defined masks interactively to contain all noticeable emission, repeating the process until we reached a typical image RMS of $<$0.1 mJy/beam. To extract radio continuum parameters for each source, we used the CASA routine \textit{imstat}, reporting peak and integrated fluxes within masks. These VLA continuum parameters are given in Table~\ref{tab:contParams}, showing integrated and peak fluxes, RMS noise (given as $\sigma$S$_{Peak}$), and the area of each mask. Where peak radio continuum emission was detected at the 3$\sigma$ level or greater, $\sigma$S$_{Peak}$ is the RMS noise outside of the masked region. Where no radio continuum emission was detected, $\sigma$S$_{Peak}$ is the RMS noise over the entire image. We detected radio continuum emission from 37 of the 65 sources observed with the VLA. Coupled with their common IR morphologies (described in Section~\ref{sec:targets}) detection of radio continuum \textit{strongly} suggests a bona fide \hii\ region.

We show a sample radio continuum snapshot in Figure~\ref{fig:contSnapshotEx}. Data from the VLA are shown as contours overlaid on WISE 4.6, 12, and 22-$\microns$ images. All sources observed with the VLA, whether with detected radio continuum emission or not, are shown in Figure~\ref{fig:contSnapshots}.

\subsection{GBT Spectra \label{sec:spectra}}
We analyzed the spectral line data using the NRAO software package GBTIDL. We used the \textit{getnod} routine. This was designed to handle data obtained by a multibeam receiver which nods beams between on and off-source positions. For each source, we combined the averaged spectra of both polarizations (LL and RR). In an effort to improve signal-to-noise ratios, we smoothed water maser observations with a Gaussian kernel to a spectral resolution of 0.3 \kms\/. We smoothed ammonia transitions to a coarser spectral resolution of 0.75 \kms\/ because the ammonia transitions tended to be fainter and broader than the observed water maser transitions. This smoothing method maintained 3-5 data points across any given spectral line, with typical widths of 1-1.5 \kms\/. We used a 3$\sigma$ detection threshold for the lowest energy detection for each source, typically $\mathrm{NH}_{3}\ (J,K)=(1,1)$. If a source had a bright ($>$3$\sigma$ detection) line for one ammonia transition and slightly weaker hyperfine lines or higher transitions at the correct velocity, those weaker transitions were also recorded. One exception is G053.580+01.387 which had a coincident CO detection at the same velocity by \citet{lundquist15}. This source falls just short of the $3\sigma$ criterion but was included as a detection based on prior knowledge. 

We estimate opacity at K-band for each observing epoch from weather forecasting data, taken as the average zenith opacity, $\tau_0$, of the nearby sites of Elkins, Lewisburg, and Hot Springs, WV. This is the same weather prediction used in the GBT's Dynamic Scheduling System (DSS), and has been shown to match observing conditions quite well at K-band \citep{maddalena10}.  Zenith opacity did not vary significantly throughout one observing epoch or across the total bandwidth of observations ($<$10\%), as observing conditions were generally good. We averaged the opacity centered on the NH$_{3}$ (J,K) = (1,1) transition (i.e. 23.6945 GHz) throughout one night. This averaged opacity, $\tau_0$, was used to determine the antenna temperature corrected for atmospheric attenuation:
\begin{equation}
T'_A = T_A \, e^{A \, \tau_0}
\label{eq:temp_opacity}
\end{equation}


\noindent
Here, the antenna temperature, $T_A$, was measured by the telescope. $T'_A$ is the antenna temperature corrected for atmospheric opacity using the average opacity and source airmass, A. This was calculated for each observation according to the following formula in which $z$ is the zenith angle of the source \citep{rozenberg66}.
\begin{equation}
A^{-1}=\cos{z}+0.025\, e^{-11\, \cos{z}}
\end{equation}

\noindent
We corrected each observation for atmospheric opacity based on the source zenith angle at the start of each two-minute scan.  As an additional correction, persistent RFI confined to a single channel at 23.876 GHz in the NH$_{3}$ (J,K) = (3,3) band was manually excised.

\begin{figure}[bh]
\includegraphics[width=\columnwidth]{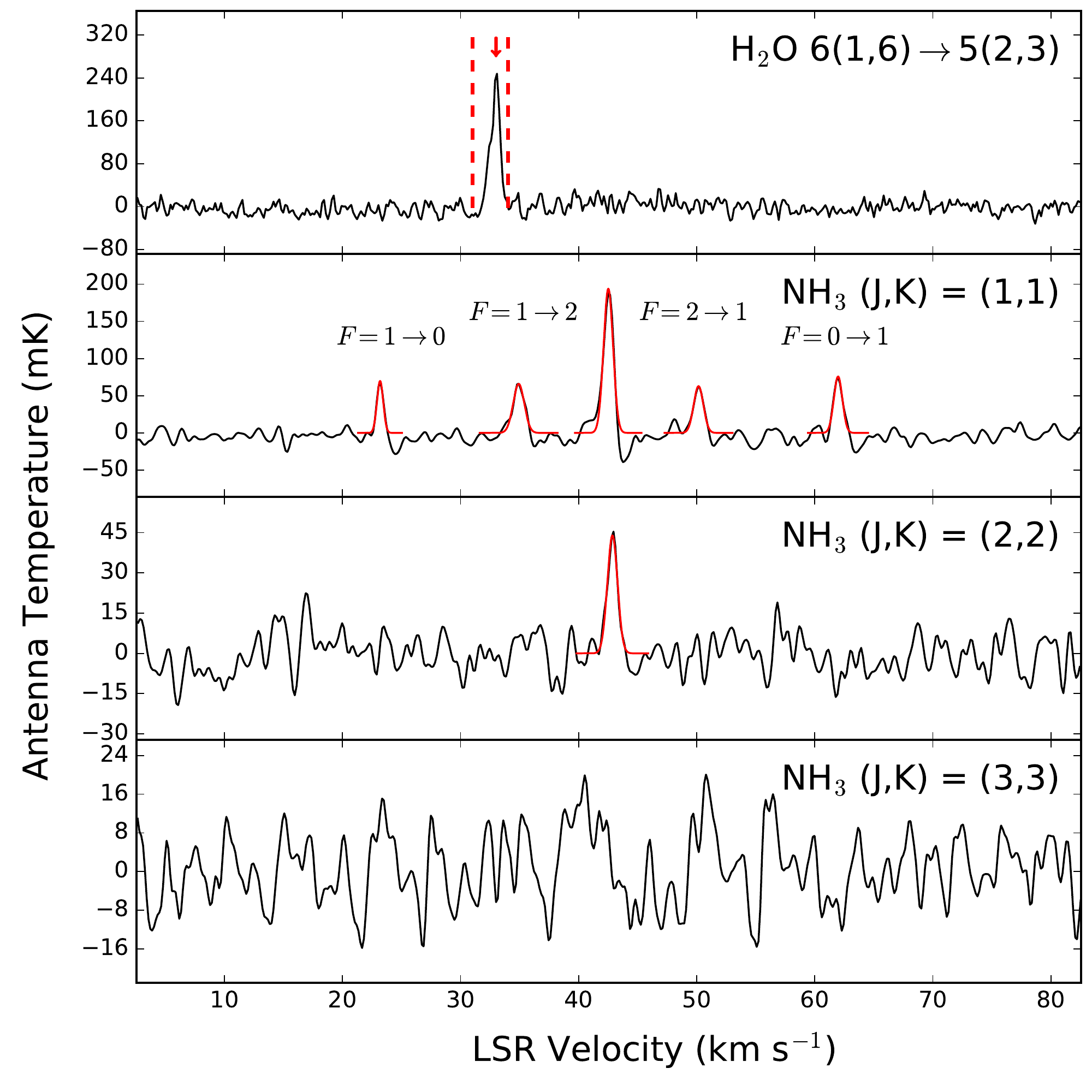}
\caption{Sample \hii\ region molecular spectra. Shown here are spectra from G026.417+01.683 of water maser (H$_2$O 6(1,6)$\rightarrow$5(2,3)) and ammonia (NH$_{3}$ (J,K)=(1,1), (2,2), (3,3)) transitions. Water maser detections (top panel) are indicated by red dashed lines demarking the upper and lower ranges of emission with a red arrow indicating the maser peak velocity. Ammonia detections (bottom three panels) are overlaid with red gaussians, including central peaks and auxiliary lines where detected. Each NH$_{3}$ (J,K)=(1,1) hyperfine transition is labeled to assist in interpretation of Table~\ref{tab:lineParams}. Maser emission is smoothed to 0.3 \kms\ whereas ammonia emission is smoothed to 0.75 \kms\/. \label{fig:NH3SampleSpectra}}
\end{figure}

\setlength{\tabcolsep}{3pt}
\begin{deluxetable*}{crrRrRrRrRrr}
\tabletypesize{\scriptsize} 
\tablecaption{Molecular Line Parameters \label{tab:lineParams}}
\tablecolumns{12}
\tablenum{3}
\tablewidth{4pt}
\tablehead{
\colhead{Name} & \colhead{$\ra_{J2000}$} & \colhead{$\dec_{J2000}$} & \colhead{Transition\tablenotemark{a}} & \colhead{T$_{L}$\tablenotemark{b}} & \colhead{$\sigma$T$_{L}$} & \colhead{$\Delta$V} & \colhead{$\sigma \Delta$V\tablenotemark{c}} & \colhead{V$_{LSR}$} & \colhead{$\sigma$V$_{LSR}$} & \colhead{rms} & \colhead{Note\tablenotemark{d}}\\
& \colhead{(hh:mm:ss)} & \colhead{(dd:mm:ss)} & & \colhead{(mK)} & \colhead{(mK)} & \colhead{(km s$^{-1}$)} & \colhead{(km s$^{-1}$)} & \colhead{(km s$^{-1}$)} & \colhead{(km s$^{-1}$)} & \colhead{(mK)}&
}
\startdata
G026.417+01.683&18:33:30.61&$-$05:01:17.2&$\mathrm{NH}_{3}$\ (J,K)=(1,1)&194&6&0.94&0.04&42.51&0.02&8&1a,1b,2a\\
&&&F=1$\rightarrow$0&70&8&0.62&0.08&23.18&0.03&$\ldots$&\\
&&&F=1$\rightarrow$2&66&6&1.10&0.12&34.92&0.05&$\ldots$&\\
&&&F=2$\rightarrow$1&63&6&0.96&0.11&50.15&0.05&$\ldots$&\\
&&&F=0$\rightarrow$1&76&7&0.85&0.09&61.97&0.04&$\ldots$&\\
&&&$\mathrm{NH}_{3}$\ (J,K)=(2,2)&44&2&1.01&0.05&42.86&0.02&7&\\
&&&$\mathrm{H}_2\mathrm{O}$\ 6(1,6)$\rightarrow$5(2,3)&248&$\ldots$&31,34&$\ldots$&33&$\ldots$&11&\\
G028.320+01.243&18:38:34.88&$-$03:32:04.8&$\mathrm{H}_2\mathrm{O}$\ 6(1,6)$\rightarrow$5(2,3)&3039&$\ldots$&$-$49, $-$38.5&$\ldots$&$-$46&$\ldots$&36&1a,3b\\
G029.138+02.218&18:36:36.47&$-$02:21:38.3&$\mathrm{NH}_{3}$\ (J,K)=(1,1)&377&9&1.62&0.05&34.72&0.02&24&1a\\
&&&F=1$\rightarrow$0&108&11&1.14&0.14&15.06&0.06&$\ldots$&\\
&&&F=1$\rightarrow$2&139&9&1.54&0.12&27.12&0.05&$\ldots$&\\
&&&F=2$\rightarrow$1&100&10&1.59&0.17&42.28&0.07&$\ldots$&\\
&&&F=0$\rightarrow$1&126&10&1.33&0.13&54.22&0.05&$\ldots$&\\
&&&$\mathrm{NH}_{3}$\ (J,K)=(2,2)&129&4&1.44&0.07&34.48&0.02&22&\\
&&&$\mathrm{H}_2\mathrm{O}$\ 6(1,6)$\rightarrow$5(2,3)&14505&$\ldots$&28, 46&$\ldots$&36.5&$\ldots$&24&\\
G034.133+00.471&02:16:31.92&00:28:15.6&$\mathrm{NH}_{3}$\ (J,K)=(1,1)&59&3&3.78&0.23&35.64&0.10&10&1b,2a,3a\\
&&&$\mathrm{NH}_{3}$\ (J,K)=(2,2)&39&3&3.44&0.26&35.97&0.11&10&\\
&&&$\mathrm{NH}_{3}$\ (J,K)=(3,3)&22&3&6.00&1.33&35.84&0.39&9&\\
G037.419+01.513&18:54:13.87&04:41:00.5&$\mathrm{NH}_{3}$\ (J,K)=(1,1)&113&5&1.86&0.09&42.45&0.04&11&1a,1c,4\\
G039.183$-$01.422&19:07:56.99&04:54:29.9&$\mathrm{H}_2\mathrm{O}$\ 6(1,6)$\rightarrow$5(2,3)&668&$\ldots$&$-$53, $-$50&$\ldots$&$-$51&$\ldots$&14&1a,3b\\
G039.536+00872&19:00:24.43&06:16:26.4&$\mathrm{H}_2\mathrm{O}$\ 6(1,6)$\rightarrow$5(2,3)&729&$\ldots$&$-$50, $-$36.5&$\ldots$&$-$38.5&$\ldots$&18&$\ldots$\\
G039.801+01.984&18:56:54.33&07:01:03.3&$\mathrm{NH}_{3}$\ (J,K)=(1,1)&33&1&1.33&0.03&29.33&0.01&7&$\ldots$\\
G040.954+02.473&18:57:16.22&08:15:59.2&$\mathrm{H}_2\mathrm{O}$\ 6(1,6)$\rightarrow$5(2,3)&247&$\ldots$&$-$62, $-$43&$\ldots$&$-$52.5&$\ldots$&40&1a\\
G048.589+01.125&19:16:28.10&14:25:26.4&$\mathrm{NH}_{3}$\ (J,K)=(1,1)&49&4&1.74&0.18&$-$34.24&0.08&11&$\ldots$\\
&&&F=1$\rightarrow$2&23&6&0.85&0.27&$-$42.50&0.12&$\ldots$&\\
&&&F=2$\rightarrow$1&26&6&1.07&0.27&$-$27.07&0.12&$\ldots$&\\
&&&$\mathrm{NH}_{3}$\ (J,K)=(2,2)&15&3&2.66&0.57&$-$33.97&0.24&8&\\
&&&$\mathrm{NH}_{3}$\ (J,K)=(3,3)&21&2&4.33&0.46&$-$34.33&0.20&10&\\
G050.900+01.055&19:21:12.59&16:26:04.8&$\mathrm{H}_2\mathrm{O}$\ 6(1,6)$\rightarrow$5(2,3)&290&$\ldots$&$-$61, $-$46&$\ldots$&$-$59&$\ldots$&17&$\ldots$\\
G053.580+01.387&19:25:17.44&18:57:12.8&$\mathrm{NH}_{3}$\ (J,K)=(1,1)&57&4&2.44&0.22&39.43&0.09&21&5a\\
G054.490+01.579&19:26:24.49&19:50:41.8&$\mathrm{NH}_{3}$\ (J,K)=(1,1)&52&3&1.42&0.08&$-$39.90&0.03&9&3b,5b\\
&&&$\mathrm{NH}_{3}$\ (J,K)=(3,3)&38&2&1.46&0.07&$-$39.96&0.03&9&\\
G055.114+02.422&19:24:29.93&20:47:33.2&$\mathrm{H}_2\mathrm{O}$\ 6(1,6)$\rightarrow$5(2,3)&380&$\ldots$&$-$63.5, $-$62&$\ldots$&$-$62.5&$\ldots$&20&3b\\
G064.151+01.282&04:16:36.24&01:16:55.2&$\mathrm{NH}_{3}$\ (J,K)=(1,1)&55&3&2.13&0.13&$-$55.92&0.05&8&2a\\
&&&$\mathrm{NH}_{3}$\ (J,K)=(2,2)&24&2&3.90&0.41&$-$56.81&0.17&8&\\
&&&$\mathrm{NH}_{3}$\ (J,K)=(3,3)&27&3&2.76&0.31&$-$56.55&0.13&8&\\
&&&$\mathrm{H}_2\mathrm{O}$\ 6(1,6)$\rightarrow$5(2,3)&90&$\ldots$&$-$68, $-$67&$\ldots$&$-$67.5&$\ldots$&12&\\
G066.607+02.060&19:50:52.81&30:38:08.0&$\mathrm{NH}_{3}$\ (J,K)=(1,1)&154&3&1.11&0.03&$-$69.16&0.01&21&4\\
&&&$\mathrm{H}_2\mathrm{O}$\ 6(1,6)$\rightarrow$5(2,3)&135&$\ldots$&$-$70.5, $-$66&$\ldots$&$-$68&$\ldots$&35&\\
\enddata
\setlength{\tabcolsep}{6pt}
\tablenotetext{a}{Transitions reported are the three lowest energy ammonia inversion transitions [i.e. (1,1), (2,2), and (3,3)], their hyperfine lines, and water maser \\ emission, where detected. Ammonia's hyperfine lines are labeled according to their transition, shown in Figure~\ref{fig:NH3SampleSpectra}. }
\tablenotetext{b}{For ammonia detections, values reported are the result of Gaussian fits to each component. For water maser detections, peak and center velocities\\ are read directly from the data and rounded to the nearest 0.5 km s$^{-1}$. }
\tablenotetext{c}{Water maser velocities are given as the range over which emission rises 3$\sigma$ above the noise.}
\tablenotetext{d}{The note column indicates previous line detections at comparable velocities, including maser species (1a : \citet{sunada07}, \\1b : \citet{urquhart12b}), ammonia (2 : \citet{urquhart12b}), radio recombination line (3a : \citet{anderson11}, \\3b : \citet{anderson15b}), CS (4 : \citet{bronfman96}), and CO (5a : \citet{lundquist15}, 5b : \citet{ao04}).}
\end{deluxetable*}

$ $

\clearpage

In a concurrent project at K-band with identical setup (GBT15B-356), we nodded on the bright calibrator source 3C147 as a calibration check. Comparing the 3C147 source intensity at the center of each bandpass with the expected values from \citet{peng00}, we found the LL and RR polarizations to agree within $\sim$6\% and $\sim$2\%, respectively, across the bandpass. As a secondary calibration check, we observed a well characterized ammonia and water maser source, G038.3453$-$00.9519 from \citet{urquhart11} and \citet{lumsden13}. The NH$_3$ (J,K) = (1,1) inversion transition for this source was previously measured with the GBT to have a peak main beam temperature of 4.1 K. Our observations of the source varied from this value by $\sim$10\% over year-long timescales. Thus, all of our calibration tests were consistent to within 10\%.

We report spectral line detections in Table~\ref{tab:lineParams} with example spectra shown in Figure~\ref{fig:NH3SampleSpectra}. The table contains line intensities T$_L$, FWHM line widths $\Delta$V, peak velocities V$_{LSR}$, and rms baseline noise. We also include notes on previous detections. All sources with ammonia or water maser detections are shown in Figure~\ref{fig:NH3spectra} in the Appendix. For ammonia detections, we report the results of Gaussian fits to each component. Our smoothing maintained $\sim$5 points across each spectral line. This allowed us to constrain Gaussian fit uncertainties for line width $\sigma\Delta$V and peak position $\sigma$V$_{LSR}$ to below our 0.75 \kms\ spectral resolution. In addition to the central peak velocity, intensity, and width, we also give parameters for hyperfine lines. These hyperfine lines are labeled \textit{F=1$\rightarrow$0}, etc. as indicated in Figure~\ref{fig:NH3SampleSpectra}.

For water maser emission, peak and center velocities are rounded to the nearest 0.5 km s$^{-1}$. The peak velocity and intensity of maser emission often varies with time, but over the course of our observations, we did not see them drift by more than 0.5 \kms\/.

We detected molecular line emission from $\sim$20\% of our targets. In total, sixteen sources were detected in ammonia or water maser emission. This includes ten water masers, ten NH$_3$ (J,K) = (1,1) lines, five (2,2) lines, and four (3,3) lines. Previous detections for each source are indicated in the Note column of Table~\ref{tab:lineParams}.


\section{Results \label{sec:results}}
\subsection{Detections \label{sec:detections}}

Of our 65 VLA candidate snapshots, 37 had detected radio continuum emission at X-band, giving us a detection rate of nearly 60\%. The majority of these targets had no previously detected radio continuum emission. Five of our radio continuum detections had previous radio recombination line measurements, and we are therefore certain of their status as \hii\ regions. These include OSC regions G033.007+01.150, G039.183$-$01.422, G054.093+01.748, and G055.114+02.422 (S83), as well as the Outer arm region G060.592+01.572 \citep{anderson15b}. All targets with previous RRL emission were detected by the VLA in radio continuum emission. These were reobserved because we knew of their status as extremely distant \hii\ regions but had no high quality radio continuum measurments. There are 6 additional OSC \hii\ regions that were not observed by the VLA because their classification as OSC \hii\ regions was unknown at the time of the VLA observations.

Our ammonia and water maser emission detection rates were markedly lower than radio continuum emission detections, with 16 out of 75 sources yielding molecular line detections, or a detection rate of $\sim$20\%. Four of these reported detections have no other molecular line detections in the literature. While we would also expect less CO at large Galactocentric radii than in the inner Galaxy, preliminary results from a CO survey with the 12 m Arizona Radio Observatory of the same targets shows a higher detection rate and will add significantly to the understanding of molecular gas around distant \hii\ region candidates in the first quadrant (Wenger et al. 2017, in preparation).

\subsection{Unusual Sources \label{sec:unusual}}

The source G039.183-01.422 has a velocity placing it at the distance of the OSC arm, but it is 1\degper\/4 below the Galactic plane. \citet{anderson15b} detected RRL emission from the region. It lies in a small finger of \hi\ emission apparently extending down from the arm, below the Galactic plane. This can be seen in the filled contours of Figure~\ref{fig:dameTargets} between Galactic longitudes of 38$^{\circ}$ and 40$^{\circ}$. While we have reobservered and detected a previously known water maser at this location, offset by 2 \kms\ from its original detection in \citep{sunada07}, we do not detect the source in ammonia. \citet{anderson15b} surmised that G039.183-01.422 is not likely to be a ``runaway star" or planetary nebula, but further observations are needed to fully characterize the source.

As noted earlier, the source G055.114+02.422 was first reported by \citet{sharpless59} as S83 and has a large negative velocity placing it within the OSC \citep{anderson15b}. It was by far our brightest radio continuum detection. In our first epoch of GBT observations, water maser emission was seen near the expected velocity of S83 with an antenna temperature less than 100 mK (a $\sim$1.5$\sigma$ detection). Upon reobservation of faint possible detections in January 2017, S83 exhibited strong water maser emission, a 380 mK detection. At $-$62.5 \kms\/, the peak velocity for this maser differed from the ionized gas velocity of $-$76.1 \kms\ by nearly 15 \kms\/. The drastic variability indicates that searches for only water maser emission are likely to miss OSC candidates.

The source G064.151+01.282 shows ammonia emission and water maser emission offset from each other by $\sim$10 \kms\/. This could be due to outflows from the \hii\ region's central core. Velocity offsets like this are not uncommon, as evidenced above for G055.114+02.422, but they indicate that \hii\ regions detected spectroscopically by water maser emission must have their kinematic distances refined with follow-up measurements of other gas species.

\subsection{Spectral Types \label{sec:stellarTypes}}

We derive spectral types by using the VLA radio continuum images to calculate the number of ionizing photons ($N_{Ly}$) required to maintain the \hii\ region. Here, we assume that each nebula is ionized by a single star. The initial mass function (IMF) is steep for high-mass stars. Because the number of Lyman-continuum photons emitted from a star increases sharply with mass, the highest mass star will contribute significantly more Lyman-continuum photons than the other members of its cluster. We then use stellar models to match spectral type with our derived $N_{Ly}$.

We calculate two separate source luminosities using kinematic distances from two different rotation curves, \citet{brand93} and \citet{reid14}. For these rotation curve models, we use the velocity of the NH$_{3}$ (J,K) = (1,1) inversion transition to determine distances. If no ammonia transition is available, we use peak water maser velocity or velocities previously cataloged in \citet{anderson15b}. For each rotation curve, we determine source luminosity with the following formula.

\begin{equation}
L_{\nu} = 4\pi~10^{-26}~\bigg[\frac{D}{\mathrm{m}}\bigg]^{2}~\bigg[\frac{S_{int}}{\mathrm{Jy}} \bigg]~\big[\mathrm{W~Hz^{-1}} \big]
\label{eq:luminosity}
\end{equation}

\citet{rubin68} first explored the relationship between \hii\ region luminosity and the spectral type of its most luminous central star. Using a derived form of Rubin's original equation from \citet{condon2016}, Equation~\ref{eq:lymanPhots}, we calculate the number of Lyman-continuum photons (in photons $s^{-1}$) ionizing each \hii\ region. Only sources with both kinematic and radio continuum measurements can be evaluated in this way as we need distance to determine luminosity.

\begin{equation}
N_{Ly} = 6.3\times10^{52} \bigg[ \frac{T_e}{10^4\:\mathrm{K}} \bigg] ^{-0.45} \bigg[ \frac{\nu}{\ghz\ } \bigg] ^{0.1} \bigg[ \frac{L_\nu}{10^{20}\:\mathrm{W}\:\mathrm{Hz}^{-1}} \bigg]  \big[\mathrm{s}^{-1} \big]
\label{eq:lymanPhots}
\end{equation}

Here we assume an electron temperature of $10^4$K and use the central frequency of our VLA observations, $\sim$9 \ghz\/.

We assign spectral types using a combination of the stellar models from \citet{martins05} and \citet{smith02}. Both models convert from Lyman-continuum photon count to spectral type using non-LTE line-blanketing, assuming an expanding atmosphere. The \citeauthor{martins05} models are reported for stellar type O9.5 through O3, whereas \citeauthor{smith02} give results for stars down to stellar type B1.5. In the range where these models overlap, they are consistent with each other for solar metallicity and luminosity class V to within 50\% on average, differing at most by $\sim$150\%. As \citeauthor{smith02} modeled lower-mass stars (down to B1.5), we used Smith for B1.5-B0 and Martins for O9.5-O3, shown in Table~\ref{tab:spectralTypeValues}. This provided the finest gridding of spectral types from B1.5 to O3 type stars.

\begin{deluxetable}{lcccclcc}
\tabletypesize{\footnotesize} 
\tablecaption{Single-Star \hii\ Region Parameters \label{tab:spectralTypeValues}}
\tablecolumns{8}
\tablenum{4}
\tablewidth{0pt}
\tablehead{\colhead{Spectral} & \multicolumn{2}{c}{Log$_{10}$(N$_{Ly}$)\,\,(s$^{-1}$)}& && \colhead{Spectral} & \multicolumn{2}{c}{Log$_{10}$(N$_{Ly}$)\,\,(s$^{-1}$)}  \\ \cline{2-3} \cline{7-8}
\colhead{Type} & \colhead{Smith} &\colhead{Martins} &&&\colhead{Type}  &  \colhead{Smith} &\colhead{Martins} }
\startdata
B1.5&\textbf{46.10}&$-$&&&O7.5&48.70&\textbf{48.44}\\
B1&\textbf{46.50}&$-$&&&O7&49.00&\textbf{48.63}\\
B0.5&\textbf{47.00}&$-$&&&O6.5&$-$&\textbf{48.80}\\
B0&\textbf{47.40}&$-$&&&O6&$-$&\textbf{48.96}\\
O9.5&$-$&\textbf{47.56}&&&O5.5&$-$&\textbf{49.11}\\
O9&47.90&\textbf{47.90}&&&O5&49.20&\textbf{49.26}\\
O8.5&$-$&\textbf{48.10}&&&O4&49.40&\textbf{49.47}\\
O8&48.50&\textbf{48.29}&&&O3&49.50&\textbf{49.63}\\
\enddata
\end{deluxetable}

We compile results from Equation~\ref{eq:lymanPhots} and parameters derived using both rotation curves in Table~\ref{tab:derivedParams}.  This table gives derived \hii\ region parameters including Galactocentric radius $R_{Gal}$, Solar distance $d_{\sun}$, number of Lyman photon emitted $Log_{10}(\mathrm N_{Ly})$, and spectral type. Comparing results between the Brand and Reid rotation curves, we see that derived spectral types never differ by more than a 0.5 spectral type. For sources within the Solar circle, there is a well-known ambiguity in their distance measurement, known as the ``kinematic distance ambiguity," or KDA, in which one velocity corresponds to two potential distances. Here, we analyze both near and far distances, not resolving the KDA. Spectral types for near versus far distances never differ by more than one spectral type. Sources within the Solar circle have values for both distances given in the same column of Table~\ref{tab:derivedParams} with near parameters reported first. We also include \hii\ regions without detected radio continuum but with kinematic information; for these sources, we assign no Lyman photon flux.

\setlength{\tabcolsep}{3pt}
\begin{deluxetable*}{hlhhrrrchccrrchccr}
\tabletypesize{\scriptsize} 
\tablecaption{Derived Region Parameters \label{tab:derivedParams}}
\tablecolumns{19}
\tablenum{5}
\tablewidth{.8\textwidth}
\tablehead{
& &  & & & & \multicolumn{5}{c}{Brand} && \multicolumn{5}{c}{Reid} &\\ \cline{7-11} \cline{13-17}
&\colhead{Name}  &  & & \colhead{V$_{LSR}$}  & & \colhead{R$_{Gal}$} &  \colhead{d$_{\sun}$} &  & \colhead{Log$_{10}$(N$_{Ly}$)\tablenotemark{a}} & \colhead{Spectral\tablenotemark{b}} & &\colhead{R$_{Gal}$} &  \colhead{d$_{\sun}$} & & \colhead{Log$_{10}$(N$_{Ly}$)\tablenotemark{a}} & \colhead{Spectral\tablenotemark{b}} & \colhead{Note\tablenotemark{c}} \\ 
& &  & & \colhead{(km s$^{-1}$)} && \colhead{(kpc)} & \colhead{(kpc)} & & \colhead{(s$^{-1}$)} & \colhead{Type}& & \colhead{(kpc)} & \colhead{(kpc)} &  & \colhead{(s$^{-1}$)} & \colhead{Type} &
}
\startdata
DB005&G026.417+01.683&26.418&1.683&42.55&&5.93&3.05/12.18&0.09/0.36&45.77/46.98&B1.5/B0.5&&6.57&2.23/12.99&0.07/0.38&45.50/47.03&B1.5/B0&\\
DB100&G028.320+01.243&28.320&1.243&$-$44.00&&15.25&22.18&0.48&47.97&O8.5&&16.43&23.41&0.51&48.02&O8.5&OSC\\
DB009&G029.138+02.218&29.138&2.219&34.71&&6.44&2.48/12.36&0.10/0.48&$\ldots$&$\ldots$&&7.15&1.59/13.26&0.06/0.51&$\ldots$&$\ldots$&\\
DB015&G033.007+01.150&33.008&1.151&$-$57.60&&17.05&23.54&0.47&48.18&O8&&18.31&24.84&0.50&48.23&O8&W,OSC\\
LA055&G034.133+00.471&34.133&0.471&35.65&&6.63&2.44/11.63&0.02/0.10&$\ldots$&$\ldots$&&7.36&1.44/12.64&0.01/0.10&$\ldots$&$\ldots$&\\
DB018&G037.419+01.513&37.419&1.514&42.13&&6.49&2.83/10.68&0.07/0.28&45.77/46.92&B1.5/B0.5&&7.20&1.73/11.78&0.05/0.31&45.34/47.01&B1.5/B0&\\
FQ009&G039.183$-$01.422&39.183&-1.422&$-$51.00&&13.88&19.39&-0.48&48.28&O8&&15.01&20.61&-0.51&48.33&O7.5&W,OSC\\
FQ438&G039.536+00.872&39.536&0.872&$-$38.50&&12.05&17.32&0.26&47.33&B0&&13.10&18.49&0.28&47.39&B0&\\
DB021&G039.801+01.984&39.801&1.985&29.31&&7.09&1.99/11.07&0.07/0.38&44.91/46.40&B1.5/B1&&7.86&0.85/12.21&0.03/0.42&44.18/46.48&B1.5/B1&\\
DB201&G040.954+02.473&40.955&2.473&$-$52.50&&13.82&19.07&0.82&$\ldots$&$\ldots$&&14.95&20.29&0.88&$\ldots$&$\ldots$&OSC\\
DB048&G048.589+01.125&48.589&1.125&$-$34.25&&10.99&14.57&0.29&46.93&B0.5&&11.99&15.78&0.31&47.00&B0.5&\\
LA802&G050.900+01.055&50.901&1.056&$-$59.00&&13.40&17.03&0.31&47.78&O9&&14.52&18.30&0.34&47.84&O9&\\
DB065&G053.580+01.387&53.581&1.388&39.44&&7.00&3.56/6.53&0.09/0.16&$\ldots$&$\ldots$&&7.77&1.37/8.72&0.03/0.21&$\ldots$&$\ldots$&\\
DB067&G054.093+01.748&54.094&1.748&$-$85.30&&16.99&20.52&0.63&48.06&O8.5&&18.25&21.89&0.67&48.11&O8&W,OSC\\
DB070&G054.490+01.579&54.491&1.579&$-$39.94&&11.22&13.77&0.38&47.83&O9&&12.23&15.03&0.41&47.91&O8.5&\\
S83&G055.114+02.422&55.114&2.422&$-$76.10&&15.26&18.43&0.78&49.28&O4&&16.44&19.75&0.84&49.35&O4&W,OSC\\
DB077&G060.592+01.572&60.592&1.572&$-$49.40&&11.76&13.31&0.37&48.21&O8&&12.80&14.62&0.40&48.29&O7.5&W\\
&G064.151+01.282&64.151&1.282&$-$55.93&&12.18&13.19&0.30&$\ldots$&$\ldots$&&13.24&14.52&0.32&$\ldots$&$\ldots$&\\
DB107&G066.607+02.060&66.608&2.061&$-$69.10&&13.34&14.20&0.51&$\ldots$&$\ldots$&&14.45&15.54&0.56&$\ldots$&$\ldots$
\enddata
\tablecomments{Sources within the Solar circle have two values for the Distance (d$_{\sun}$), N$_{Ly}$, and Spectral Type columns since we did not resolve the \\kinematic distance ambiguity. Sources with data blanked by ``$\ldots$" in the N$_{Ly}$ and Spectral Types column have measured ammonia or maser \\velocities from our GBT pointings, but no detected radio continuum.}
\tablenotetext{a}{Log$_{10}$ of the number of Lyman continuum photons emitted per second is determined through Equation~\ref{eq:lymanPhots}.}
\tablenotetext{b}{Spectral type is assigned based on a combination of models from \citet{martins05} and \citet{smith02}, described in Section~\ref{sec:stellarTypes}.}
\tablenotetext{c}{Sources whose velocity places them within the Outer Scutum-Centaurus Arm are indicated by "OSC" in the note column. Two HII \\regions, G028.320+01.243 and G040.954+02.473, were discovered to be part of the OSC as part of this work. Additionally, sources previously \\identified as HII regions via RRLs in the WISE catalog are marked by a ``W."}
\end{deluxetable*}

\section{Discussion \label{sec:discussYes}}

We added two new sources to our OSC catalog based on their water maser velocities, G028.320+01.243 and G040.954+02.473; neither had ammonia detections. The source G028.320+01.243 had a previous radio recombination line measurement placing it on the edge of the OSC \citep[V$_{LSR}$ = $-$39.2 \kms\/,][]{anderson15b}. This \hii\ region also showed coincident radio continuum emission and had a water maser detection first reported in \citet{brand94}. The source G040.954+02.473 had water maser emission first detected in \citet{sunada07} and had no detected radio continuum emission. Both G028.320+01.243 and G040.954+02.473 showed the same characteristic infrared morphology of a 22-\microns\ core of emission surrounded by a more diffuse 12-\microns\ envelope. While this morphology has been demonstrated to correlate well with \hii\ regions, more follow-up observations will be conducted to strengthen the evidence that these are \hii\ regions within the OSC. All known OSC \hii\ regions to date are included in Table~\ref{tab:OSChiiRegions}, including regions not observed as part of this work.

\begin{figure*}[t!]
\graphicspath{ {Figures/} }
\includegraphics[width=\textwidth]{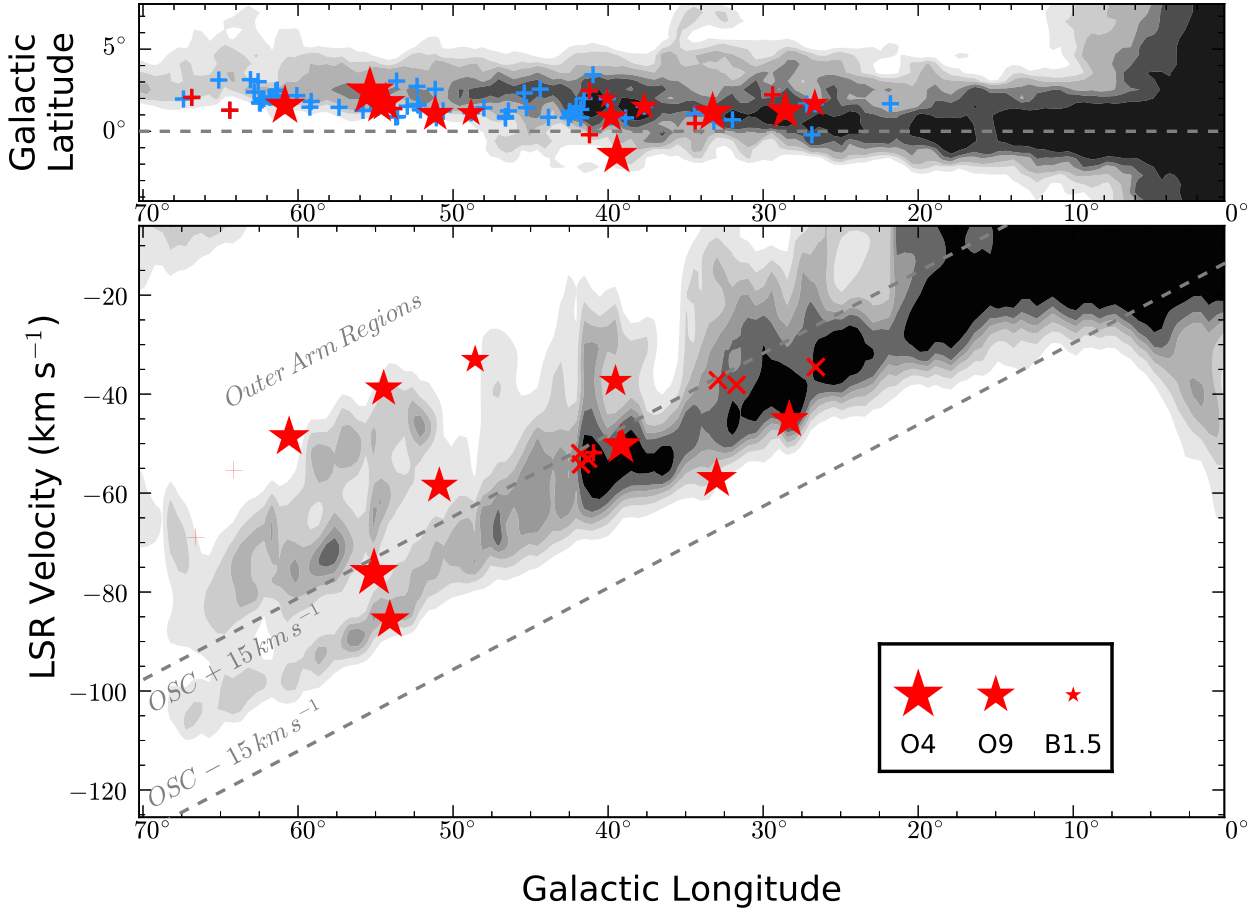}
\vspace{-15pt}
\caption{The Outer Scutum-Centaurus Arm as traced by integrated HI emission. All 12 detected OSC \hii\ regions are shown here. \textbf{Top:} Velocity-integrated \hi\ emission tracing the OSC arm, summed over a 14 \kms\ wide window following the center velocity given by $V_{LSR}$ = $-$1.6 km~s$^{-1}$ deg$^{-1}\times\gl\/$. Overplotted is a distribution of observed \hii\ region targets identified in WISE 12- and 22-\microns\ images at the \lb\ loci of the OSC arm. We show candidates without known velocities in blue crosses, wheras confirmed \hii\ regions are shown in red crosses and stars. Star markers indicate sources with detected radio continuum in addition to spectral lines; these markers are scaled by stellar type as described in Section~\ref{sec:stellarTypes}. \textbf{Bottom:} Longitude-velocity diagram of HI emission, summed over a 3.5$^{\circ}$ window following the arm in latitude according to $b = 0.375^{\circ} + 0.075\times$\gl\/. Overlaid are detected OSC \hii\ regions with the same symbols as the top panel as well as 6 OSC \hii\ regions detected previously but not observed as part of this survey, marked with x's. Dashed lines indicate the central locus of the OSC in \lv\ space as indicated by \citet{dame11}, $V_{LSR}$ = $-$1.6 km~s$^{-1}$ deg$^{-1}\times\gl\/ \pm 15 \kms\,$. This is a modified version of Fig. 3 from \citet{dame11}.  \label{fig:dameReplica}}
\end{figure*}

\begin{figure*}[tb]
\begin{center}
\graphicspath{ {Figures/} }
\includegraphics[width=0.8\textwidth]{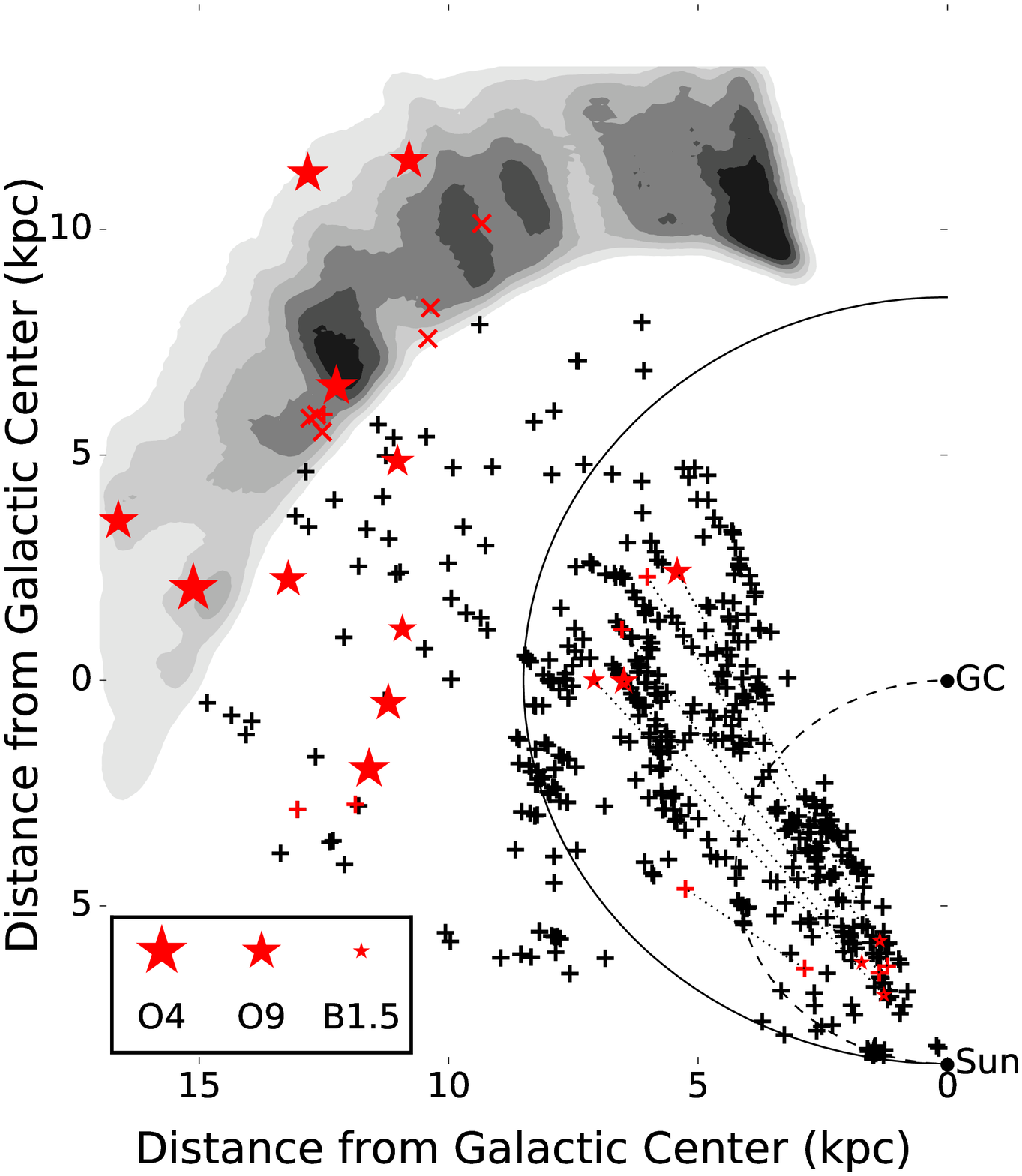}
\caption{Face-on map of \hii\ regions in the first Galactic quadrant. The Solar circle and tangent points are marked by solid and dashed lines, respectively. Black crosses mark \hii\ regions from the WISE Catalog of Galactic HII Regions with known distances \citep{anderson14}. Red x's represent \hii\ regions detected previously within the OSC. Red crosses and stars mark \hii\ regions examined or discovered by this study. Stars indicate sources with detected radio continuum emission in addition to ammonia emission; these markers are scaled in size by stellar spectral type. Stars within the Solar circle suffer from the kinematic distance ambiguity (KDA). For each source, both possible distances are shown connected by a dotted line. We also show \hi\ emission from the LAB survey following the arm in latitude and velocity: $V_{LSR}$ = $-$1.6 km~s$^{-1}$ deg$^{-1}\times\gl\/$, $b = 0.375^{\circ} + 0.075\times$\gl\/. This is the same emission as shown between the dashed lines in Figure~\ref{fig:dameReplica}, bottom panel, but transformed assuming a Brand rotation curve. \label{fig:faceOn}}
\end{center}
\end{figure*}

While our detection rate for molecular line emission was rather low (20\%), this could be due to both decreasing metallicity and decreasing star formation efficiency at high Galactocentric radii. Metallicity decreases with distance from the Galactic Center, which might mean we are simply running out of observable tracer molecules in these most distant \hii\ regions. Additionally, studies of external galaxies have found a rapid decline in star formation efficiency with increasing Galactocentric radius \citep{bigiel10}. Thus, our low detection rate for molecular gas emission is perhaps not surprising given the fact we are probing \hii\ regions in the far outer Galaxy. In the Green Bank Telescope RMS survey by \citet{urquhart11}, $\sim$80\% of observed \hii\ regions had detected ammonia emission while $\sim$50\% had detected water maser emission. Over 80\% of the RMS detections were within 1$^{\prime}$ of a source from the WISE Catalog of Galactic \hii\ Regions \citep{anderson15b}. The RMS detection rates were markedly higher than ours, likely because the compact RMS targets were younger and closer to the Sun on average. Simulations of star formation in the Galaxy which include the decreasing metallicity gradient could help us interpret our lack of molecular gas detections, but they are beyond the scope of this work.

In contrast, we detected nearly 60\% of our targeted ``radio quiet" candidates though their radio continuum emission, indicating that there is a significant population of faint, undetected \hii\ regions still to discover within the Milky Way.

We show our targets overlaid on the \hi\ content of the OSC in Figure~\ref{fig:dameReplica}. Sources with derived stellar types are shown with star markers, increasing in size with earlier stellar types. \hii\ regions with velocities placing them in the far outer Galaxy (i.e. large negative velocities) are mainly O-stars, though we do see two B-stars outside of the Solar Circle (G039.536+00.872 and G048.589+01.125). The most luminous source observed (G055.114+02.422) is found to have a type O4 star. This is a well known \hii\ region from the Sharpless catalog, S83 \citep{sharpless59}.

A face-on view of the first Galactic quadrant is shown in Figure~\ref{fig:faceOn}. Using a \citet{brand93} rotation curve to convert between velocities and distances, previous detections reported in the WISE Catalog of Galactic \hii\ Regions are plotted together with detections from this work in Figure~\ref{fig:faceOn}. Since the KDA was not resolved for our new detections, we have included both the near and far distances, with sources connected radially by dotted lines.

This was the first systematic survey targeting radio quiet candidates from the WISE Catalog of Galactic \hii\ Regions. Since over 60\% of these candidates were detected in radio continuum, we might expect similar detection rates for the other radio quiet candidates in the catalog, nearly 4000 in total. Radio continuum observations with the VLA of every radio quiet candidate from the WISE Catalog in the second and third Galactic quadrants will probe this further (Armentrout et al. 2017b, in prep).

\iftrue
\setlength{\tabcolsep}{3pt}
\begin{deluxetable*}{hlhhrrrchccrrchccc}
\tabletypesize{\scriptsize} 
\tablecaption{OSC \hii\ Regions \label{tab:OSChiiRegions}}
\tablecolumns{19}
\tablenum{6}
\tablewidth{.8\textwidth}
\tablehead{
& &  & & & & \multicolumn{5}{c}{Brand} && \multicolumn{5}{c}{Reid} &\\ \cline{7-11} \cline{13-17}
&\colhead{Name}  &  & & \colhead{V$_{LSR}$}  & & \colhead{R$_{Gal}$} &  \colhead{d$_{\sun}$} &  & \colhead{Log$_{10}$(N$_{Ly}$)\tablenotemark{a}} & \colhead{Spectral\tablenotemark{b}} & &\colhead{R$_{Gal}$} &  \colhead{d$_{\sun}$} & & \colhead{Log$_{10}$(N$_{Ly}$)\tablenotemark{a}} & \colhead{Spectral\tablenotemark{b}} & \colhead{Note\tablenotemark{c}} \\ 
& &  & & \colhead{(km s$^{-1}$)} && \colhead{(kpc)} & \colhead{(kpc)} & & \colhead{(s$^{-1}$)} & \colhead{Type}& & \colhead{(kpc)} & \colhead{(kpc)} &  & \colhead{(s$^{-1}$)} & \colhead{Type} &
}
\startdata
&G026.610$-$00.212&26.610&$-$0.212&$-$35.70&&13.77&20.84&-0.08&$\ldots$&$\ldots$&&14.90&22.00&-0.08&$\ldots$&$\ldots$&RRL\\
DB100&G028.320+01.243&28.320&1.243&$-$44.00&&15.25&22.18&0.48&47.97&O8.5&&16.43&23.41&0.51&48.02&O8.5&VLA,H$_{2}$O\\
&G031.727+00.698&31.727&0.698&$-$39.20&&13.26&19.71&0.24&$\ldots$&$\ldots$&&14.37&20.89&0.25&$\ldots$&$\ldots$&RRL\\
&G032.928+00.606&32.928&0.606&$-$38.30&&12.89&19.16&0.20&$\ldots$&$\ldots$&&13.98&20.32&0.21&$\ldots$&$\ldots$&RRL\\
DB015&G033.007+01.150&33.008&1.151&$-$57.60&&17.05&23.54&0.47&48.18&O8&&18.31&24.84&0.50&48.23&O8&RRL,VLA\\
FQ009&G039.183$-$01.422&39.183&-1.422&$-$51.00&&13.88&19.39&-0.48&48.28&O8&&15.01&20.61&-0.51&48.33&O7.5&RRL,VLA,H$_{2}$O\\
DB201&G040.954+02.473&40.955&2.473&$-$52.50&&13.82&19.07&0.82&$\ldots$&$\ldots$&&14.95&20.29&0.88&$\ldots$&$\ldots$&H$_{2}$O\\
&G041.304+01.997&41.304&1.997&$-$53.70&&13.95&19.16&0.67&$\ldots$&$\ldots$&&15.08&20.39&0.71&$\ldots$&$\ldots$&RRL\\
&G041.755+01.451\tablenotemark{d}&41.755&1.451&$-$54.80&&14.04&19.19&0.49&$\ldots$&$\ldots$&&15.18&20.43&0.52&$\ldots$&$\ldots$&RRL\\
&G041.804+01.503\tablenotemark{d}&41.804&1.503&$-$52.60&&13.69&18.80&0.49&$\ldots$&$\ldots$&&14.82&20.03&0.53&$\ldots$&$\ldots$&RRL\\
DB067&G054.093+01.748&54.094&1.748&$-$85.30&&16.99&20.52&0.63&48.06&O8.5&&18.25&21.89&0.67&48.11&O8&RRL,VLA\\
S83&G055.114+02.422&55.114&2.422&$-$76.10&&15.26&18.43&0.78&49.28&O4&&16.44&19.75&0.84&49.35&O4&RRL,VLA,H$_{2}$O\\
\enddata
\tablecomments{All known OSC HII regions to date. Sources with data blanked by ``$\ldots$" in the N$_{Ly}$ and Spectral Types column have measured \\ammonia or maser velocities from our GBT pointings, but no detected radio continuum.}
\tablenotetext{a}{Log$_{10}$ of the number of Lyman-continuum photons emitted per second is determined through Equation~\ref{eq:lymanPhots}.}
\tablenotetext{b}{Spectral type is assigned based on a combination of the \citet{martins05} and \citet{smith02} models of Lyman-continuum emission \\from high-mass stars to give the finest gridding of spectral types and cover B1.5 to O3 type stars, as described in Section~\ref{sec:stellarTypes}.}
\tablenotetext{c}{Sources are labeled by RRL, VLA, and H$_{2}$O indicating detected emission from radio recombination lines, VLA radio continuum emission, and water \\maser emission respectively. Two HII regions, G028.320+01.243 and G040.954+02.473, were discovered to be part of the OSC as part of this work \\and only have detected water maser emission.}
\tablenotetext{d}{The HII regions G041.755+01.451 and G041.804+01.503 are part of the same complex but have distinct morphologies.}
\end{deluxetable*}
\fi

\section{Summary \label{sec:summary}}

We observed \hii\ region candidates coincident with the Outer Scutum-Centaurus spiral arm. At $\sim$15 kpc from the Galactic Center, the OSC is affected by the Galactic warp and bends out of the Galactic plane to positive Galactic latitudes, leaving our targets unobstructed from the bulk of in-plane gas and dust. If the Galaxy is symmetric, the OSC is the first quadrant counterpart to the third quadrant Perseus Arm. We searched for thermal radio continuum emission with the Very Large Array and both water maser and ammonia emission with the Green Bank Telescope. By combining integrated luminosity information from radio continuum observations, kinematic information from spectroscopy, and stellar spectral models, we were able to calculate the underlying spectral types of stars within these distant \hii\ regions.

The motivation for this survey was to detect especially distant \hii\ regions in the first Galactic quadrant, coincident with the OSC. Our observations add two \hii\ regions to the census of high-mass star formation in the OSC, bringing the total known population to 12 \hii\ regions, detailed in Table~\ref{tab:OSChiiRegions}. One of these \hii\ regions (G055.114+02.422) was observed as early as 1953 by \citet{sharpless53}, but it was not recognized as extremely distant star formation at the time. This study shows that the OSC is forming high-mass stars with types as early as O4. With Galactocentric radii in excess of 15 kpc, the population of detected OSC \hii\ regions could represent an outer boundary for star formation in the Milky Way.

\begin{acknowledgments}
\nraoblurb\ This work was supported by NASA ADAP grant NNX12AI59G and NSF grant AST1516021. We thank West Virginia University for its financial support of GBT operations, which enabled some of the observations for this project. We also thank the staff at the Green Bank Telescope for their helpfulness and hospitality during the observations and data reduction and the technical staff at the Very Large Array for quickly addressing any issues we encountered throughout the data reduction process. The National Radio Astronomy Observatory is a facility of the National Science Foundation operated under cooperative agreement by Associated Universities, Inc. This research has made use of NASA's Astrophysics Data System Bibliographic Services, Astropy, a community-developed core Python package for Astronomy \citep{astropy13}, and also APLpy, an open-source plotting package for Python hosted at http://aplpy.github.com.
\end{acknowledgments}

\appendix

Snapshot images of each source observed with the VLA are shown in Figure~\ref{fig:contSnapshots}, while spectra for GBT detections are shown in Figure~\ref{fig:NH3spectra}.

The \textit{WISE} Catalog of Galactic \hii\ Regions website\footnote{http://astro.phys.wvu.edu/wise/} now includes these results. It contains an interactive map of the Galactic plane, showing all detected and candidate \hii\ regions cataloged to date, along with detected source velocities and known parameters.

\facility{Green Bank Telescope, Very Large Array}
\software{Astropy, APLpy, CASA, GBTIDL}


\bibliographystyle{aasjournal}
\bibliography{ref}

\renewcommand{\thefigure}{A\arabic{figure}}
\newcommand{\figSize}{0.33\textwidth}
\begin{figure*}[!ht]
\caption{Infrared and radio images of all candidate \hii\ regions observed with the VLA. WISE bands w2 (4.6 \microns\/), w3 (12 \microns\/), and w4 (22 \microns\/) are represented by blue, green, and red respectively. VLA radio continuum contours from 4 minute integrations at X-band in D configuration are overplotted where available. Contours are placed at 30\%, 50\%, 70\%, and 90\% peak radio continuum flux for sources detected in radio continuum.  Each image is 6$^{\prime}$ on a side, and scale bars represent the regions' angular sizes as cataloged in the WISE Catalog of Galactic HII Regions \citep{anderson14}. If an image was detected by the VLA in radio continuum, a ``C" appears in the bottom left corner. Ammonia and maser detections from this work are additionally denoted with an ``L". Sources marked ``C" or ``L" have further details in Tables~\ref{tab:contParams} \& \ref{tab:lineParams} respectively. Sources marked ``OSC" have been identified as part of the Outer Scutum-Centaurus spiral arm. We smoothed especially diffuse radio continuum emission by a 15$^{\prime\prime}$ tophat filter for display; these are marked by ``15." \label{fig:contSnapshots}}
\includegraphics[width=\figSize]{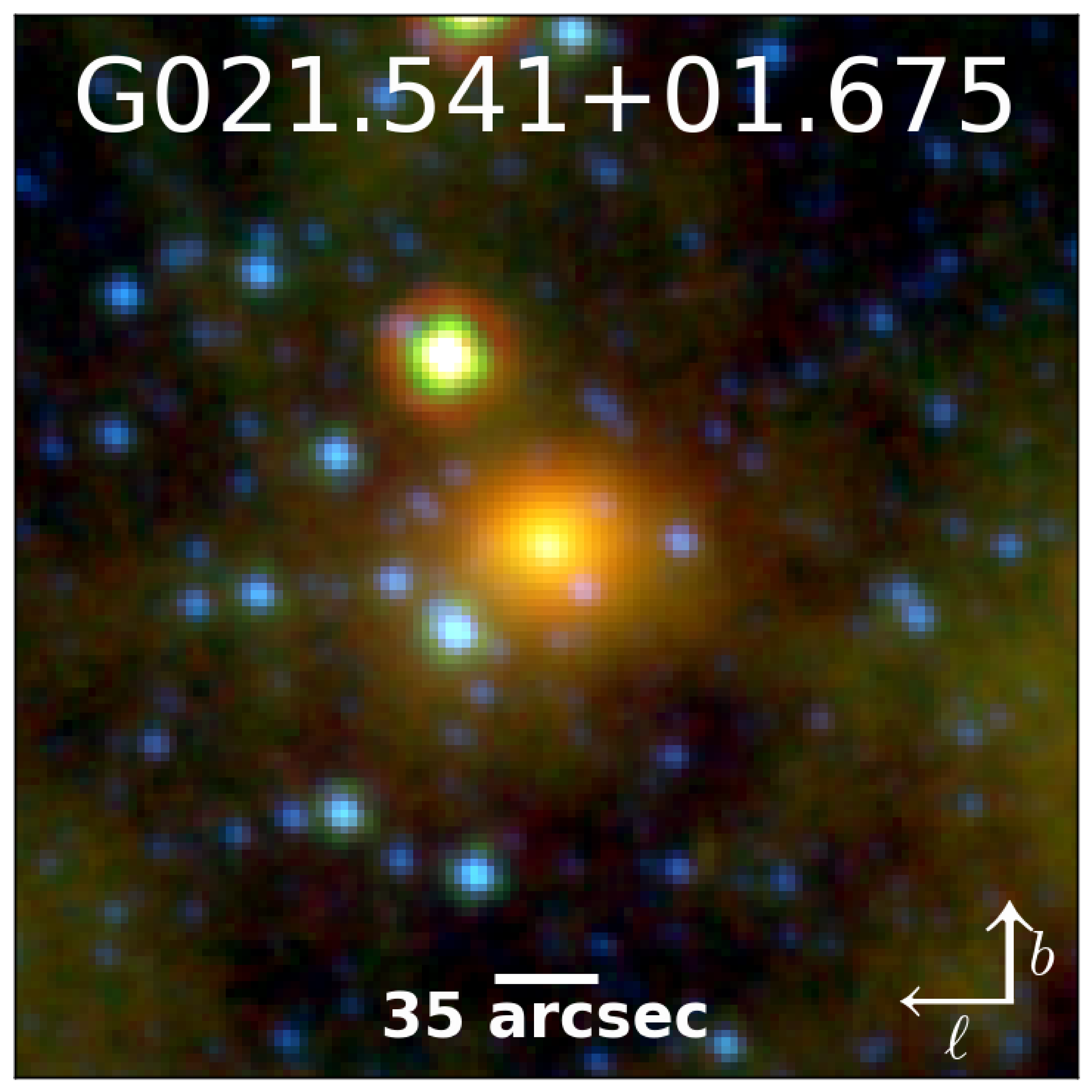}
\includegraphics[width=\figSize]{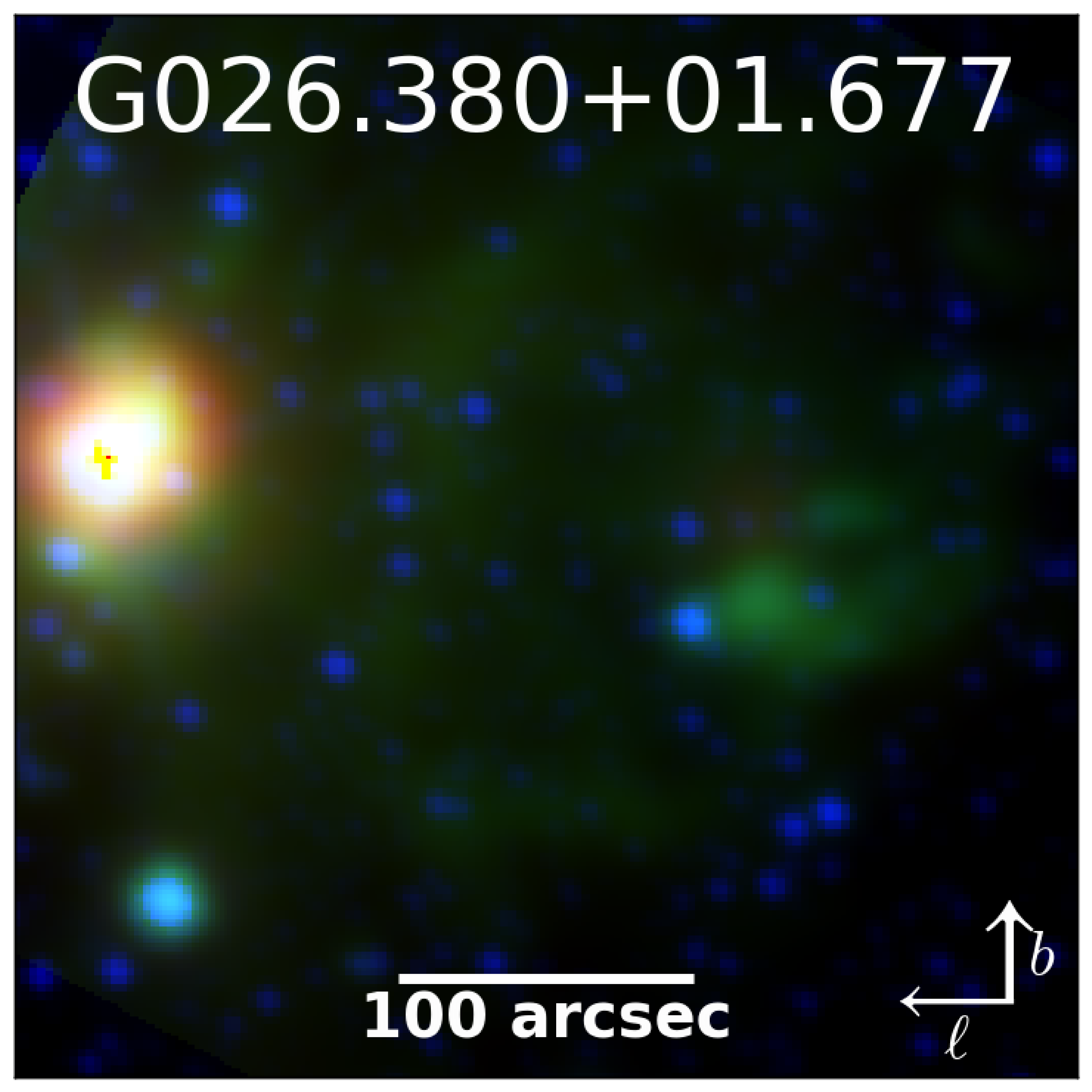}
\includegraphics[width=\figSize]{DB005.pdf}\\
\includegraphics[width=\figSize]{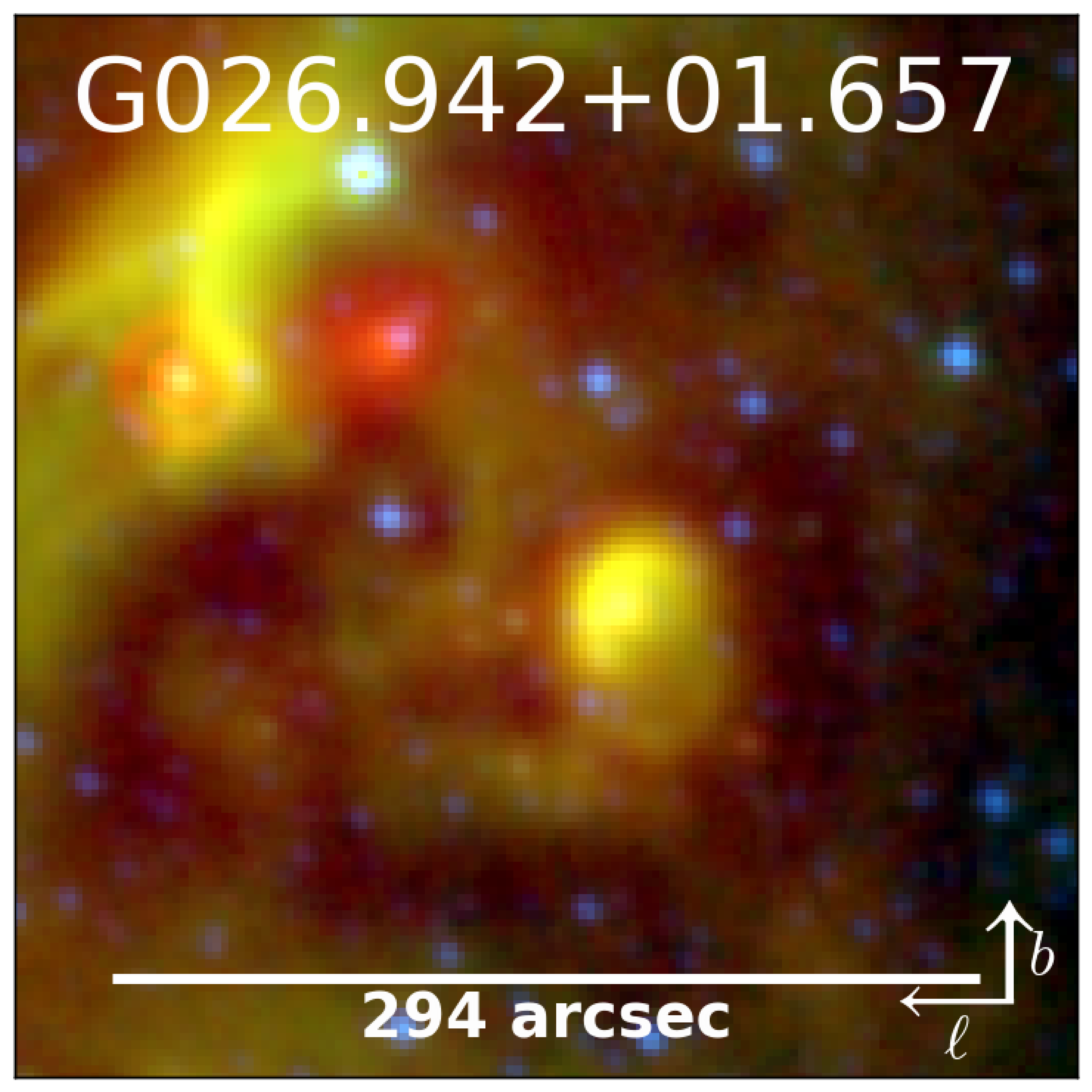}
\includegraphics[width=\figSize]{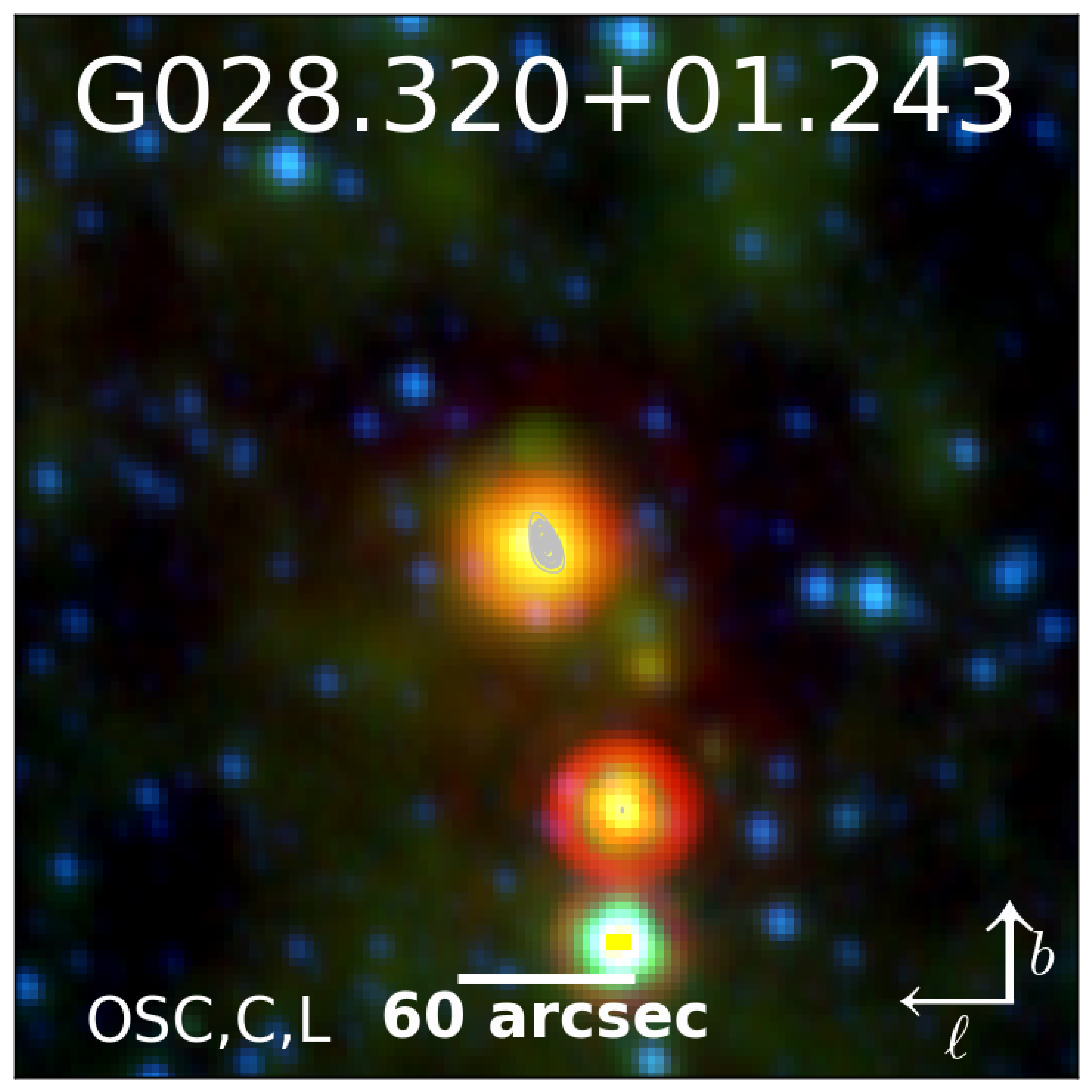}
\includegraphics[width=\figSize]{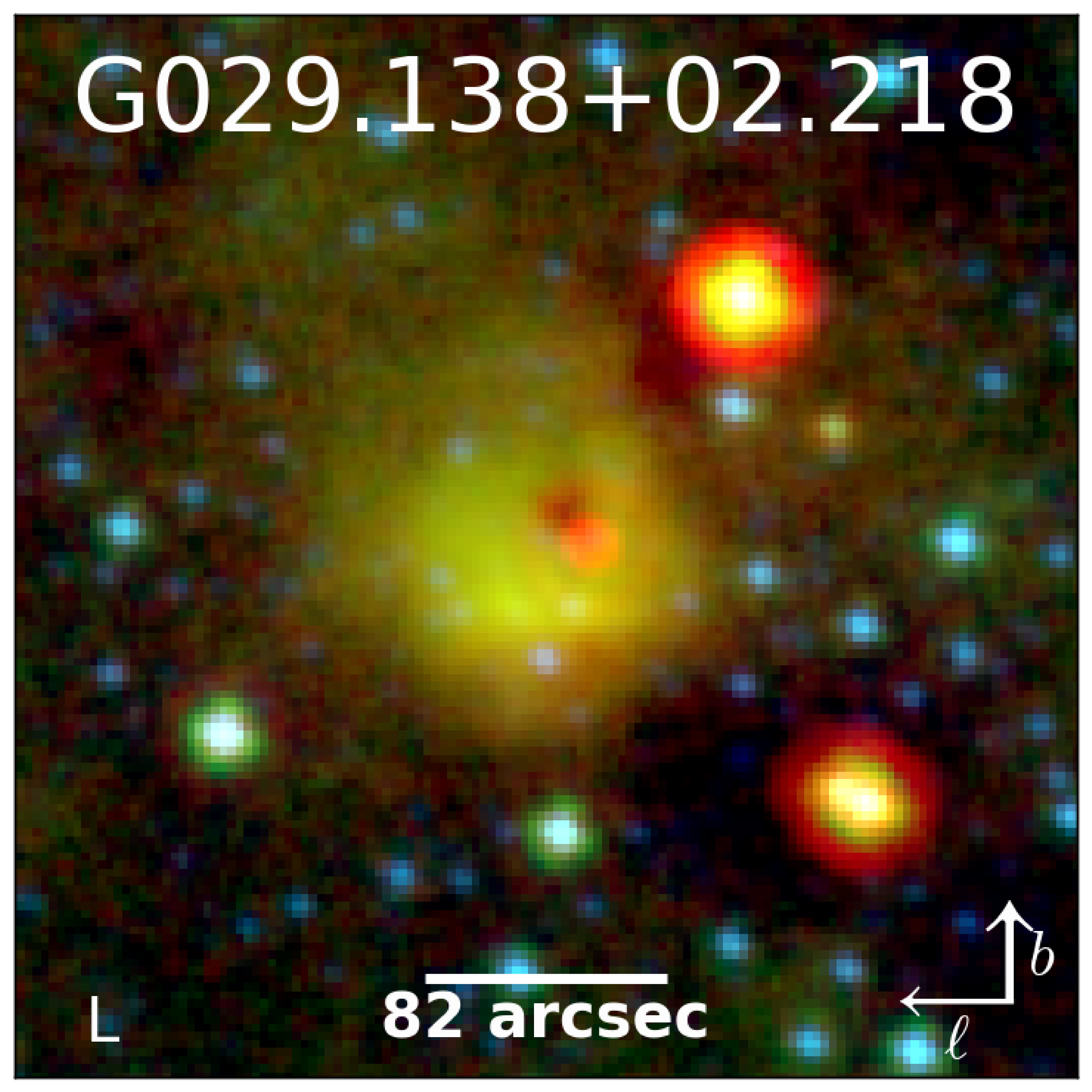}\\
\includegraphics[width=\figSize]{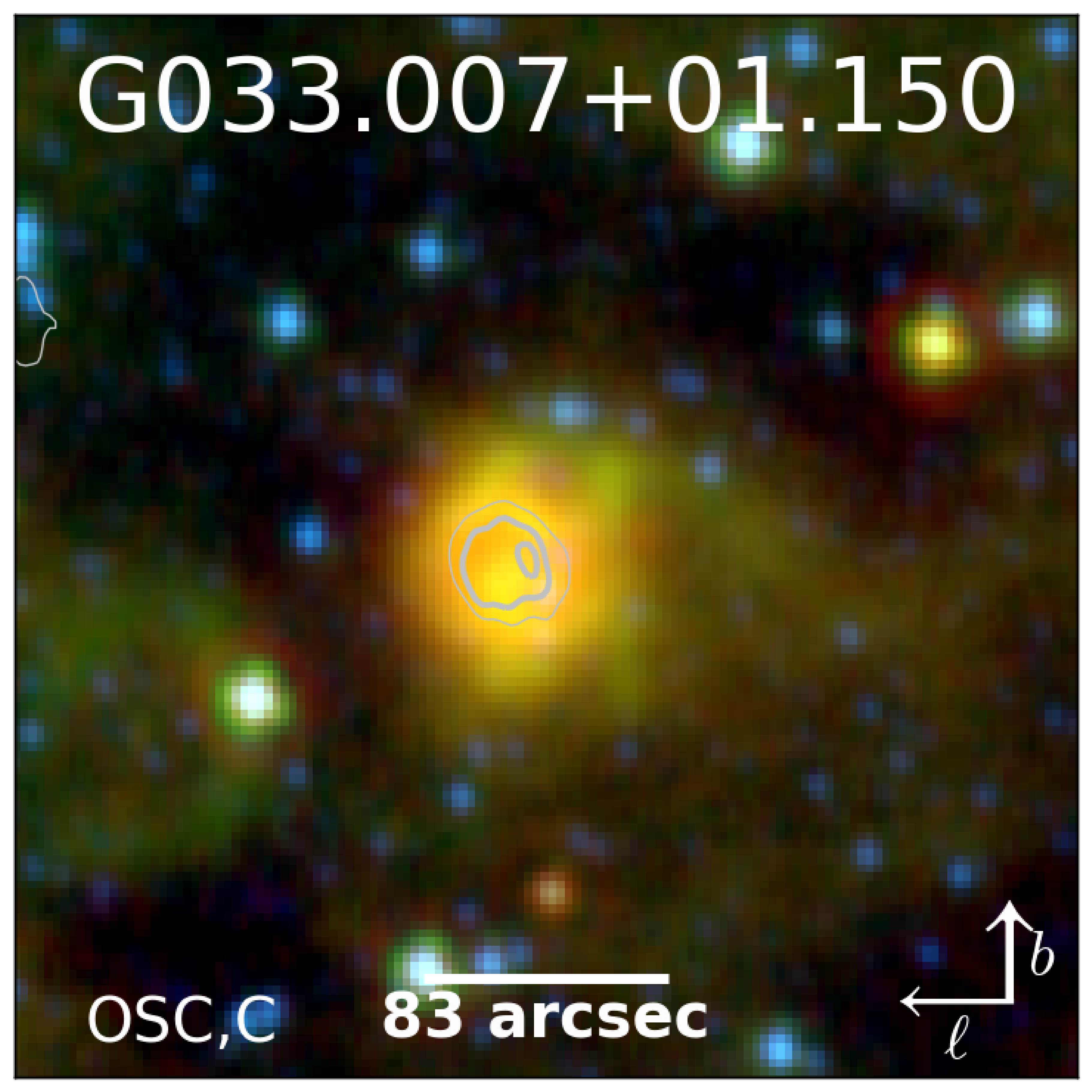}
\includegraphics[width=\figSize]{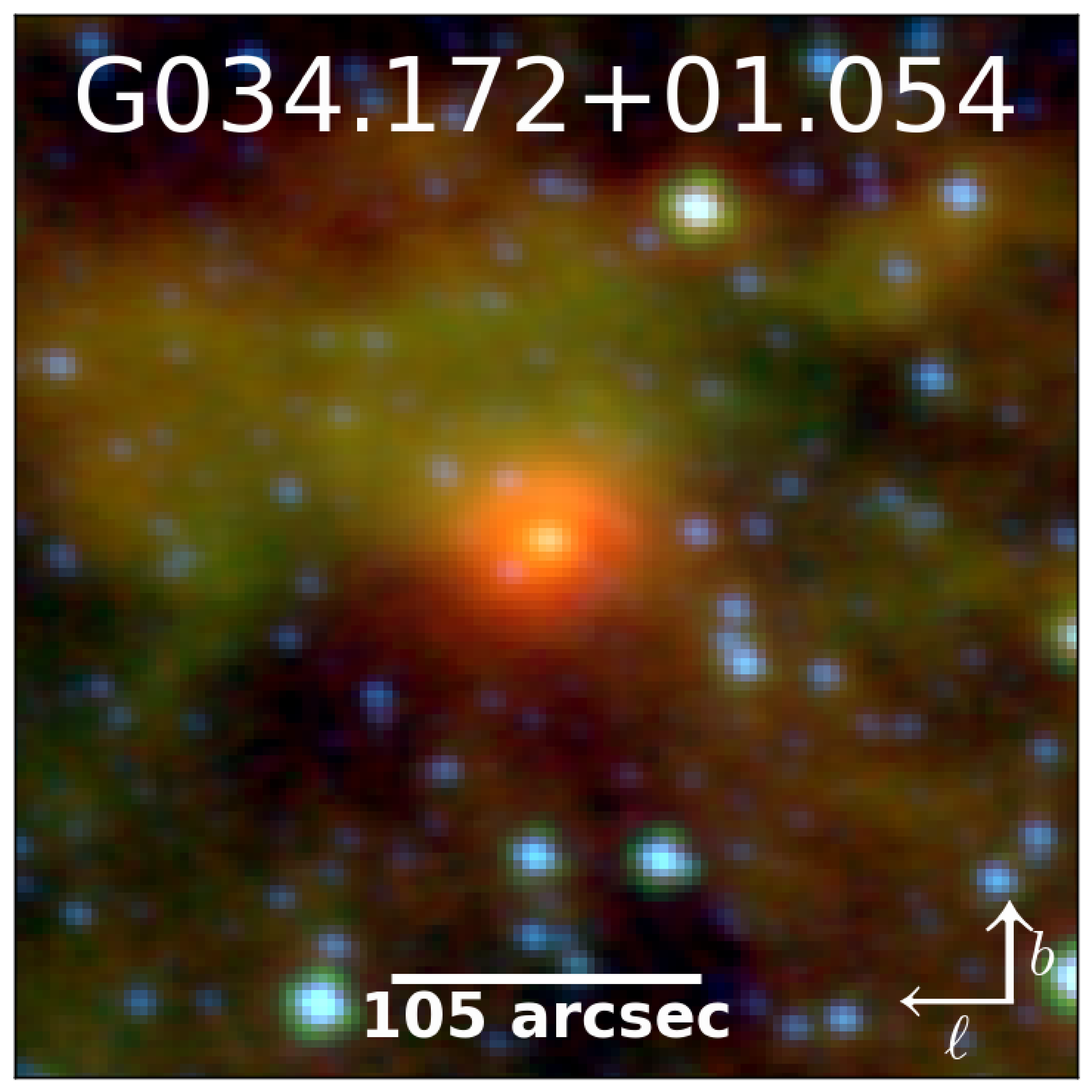}
\includegraphics[width=\figSize]{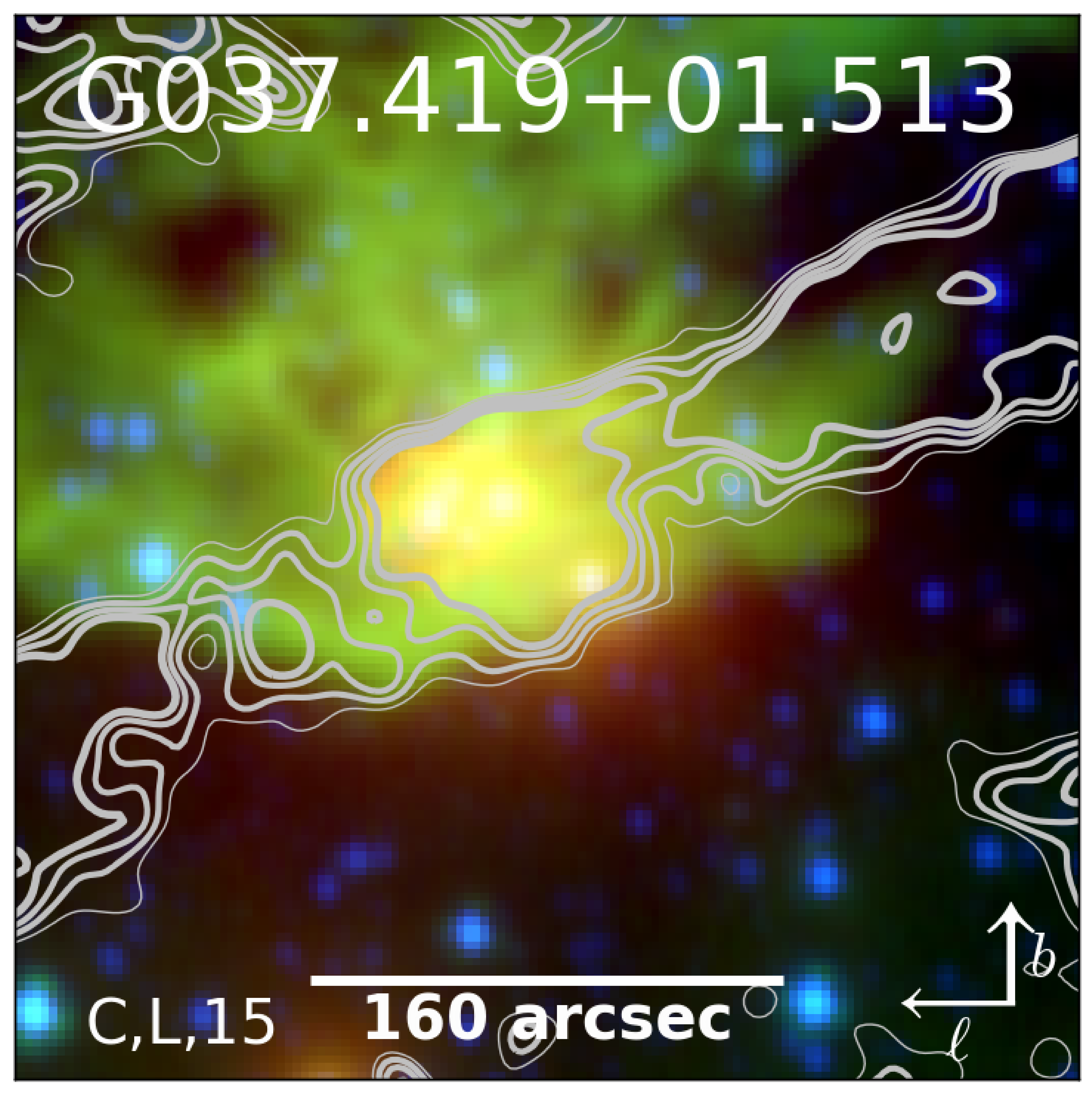}
\end{figure*}
\begin{figure*}[!htb]
\includegraphics[width=\figSize]{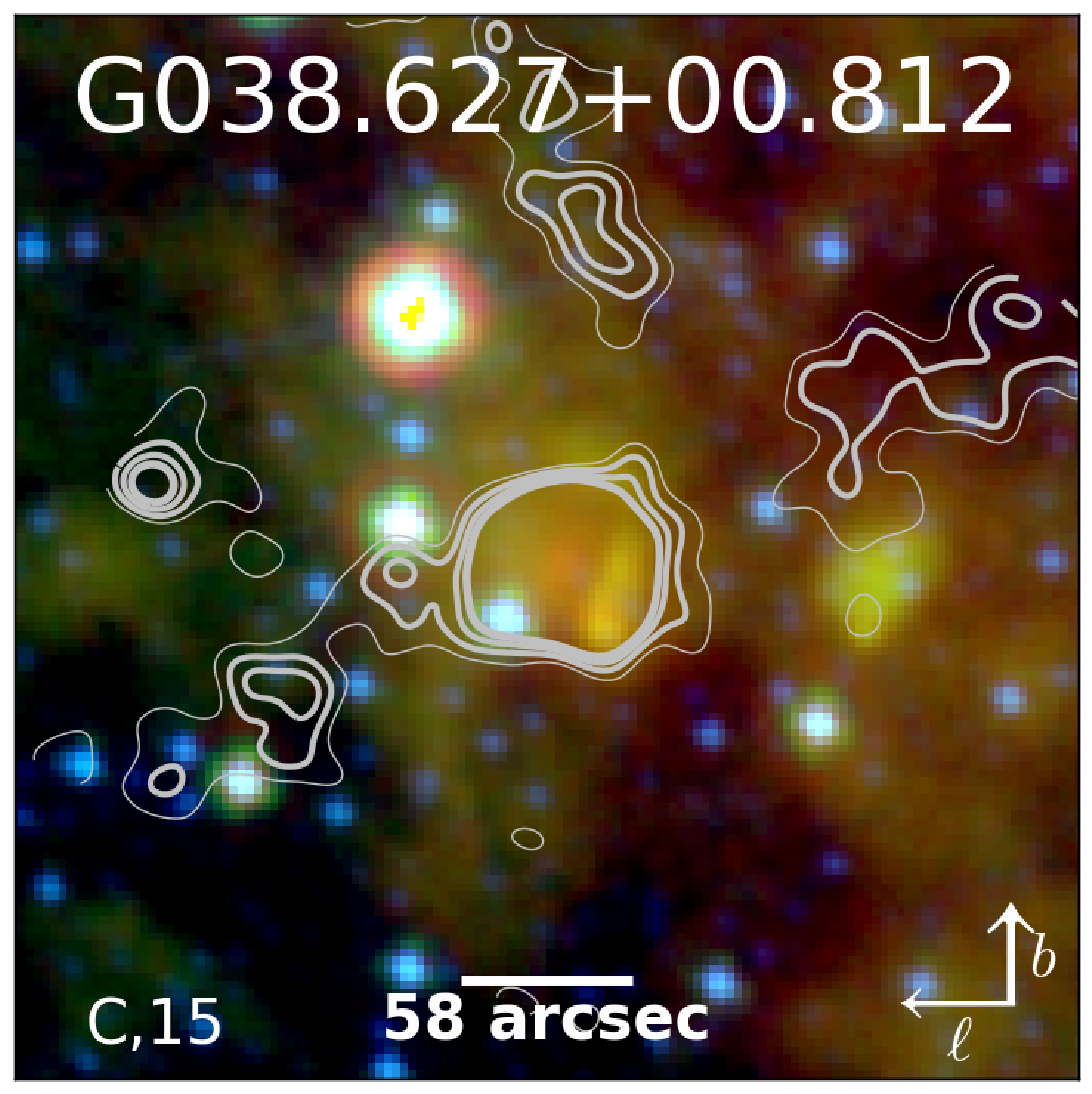}
\includegraphics[width=\figSize]{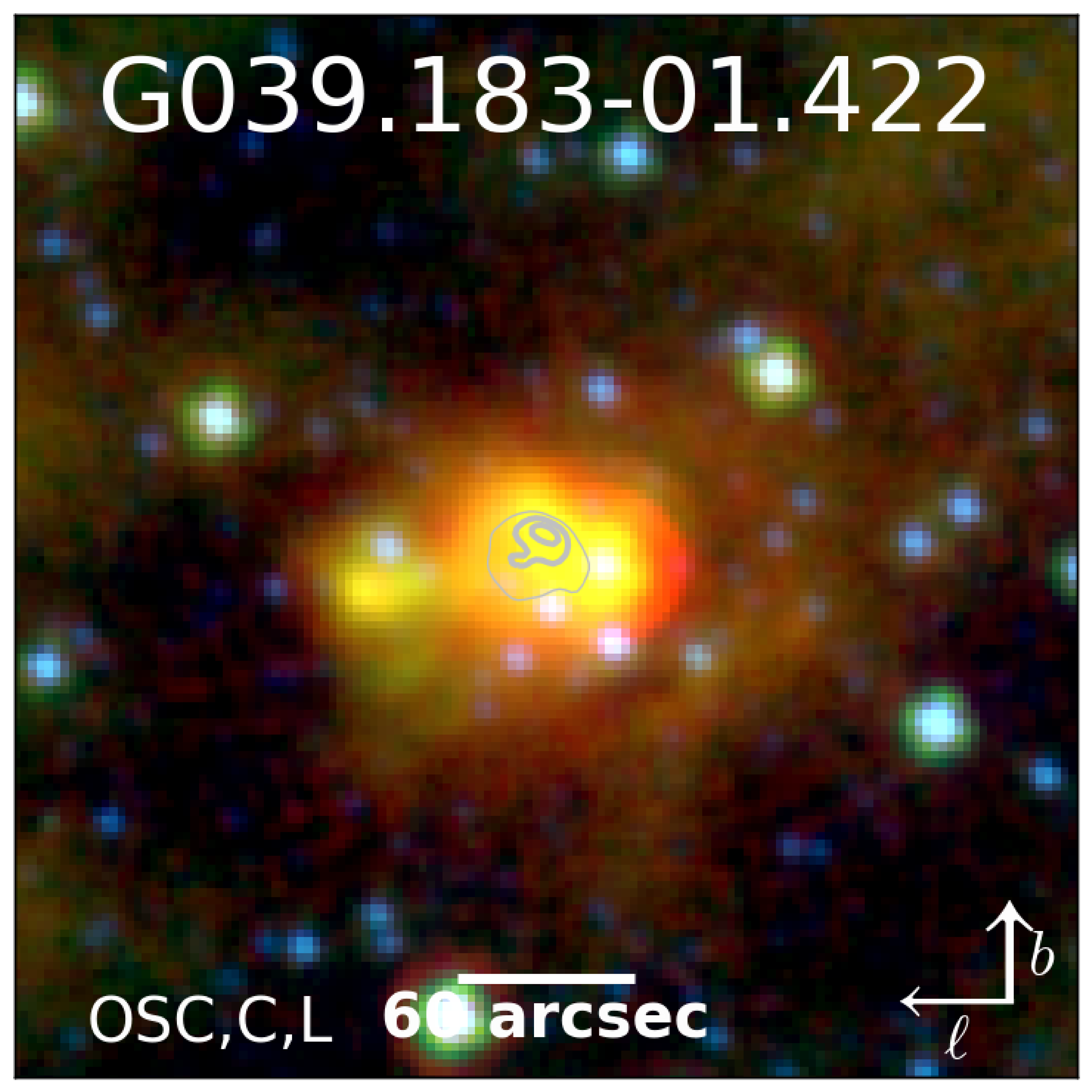}
\includegraphics[width=\figSize]{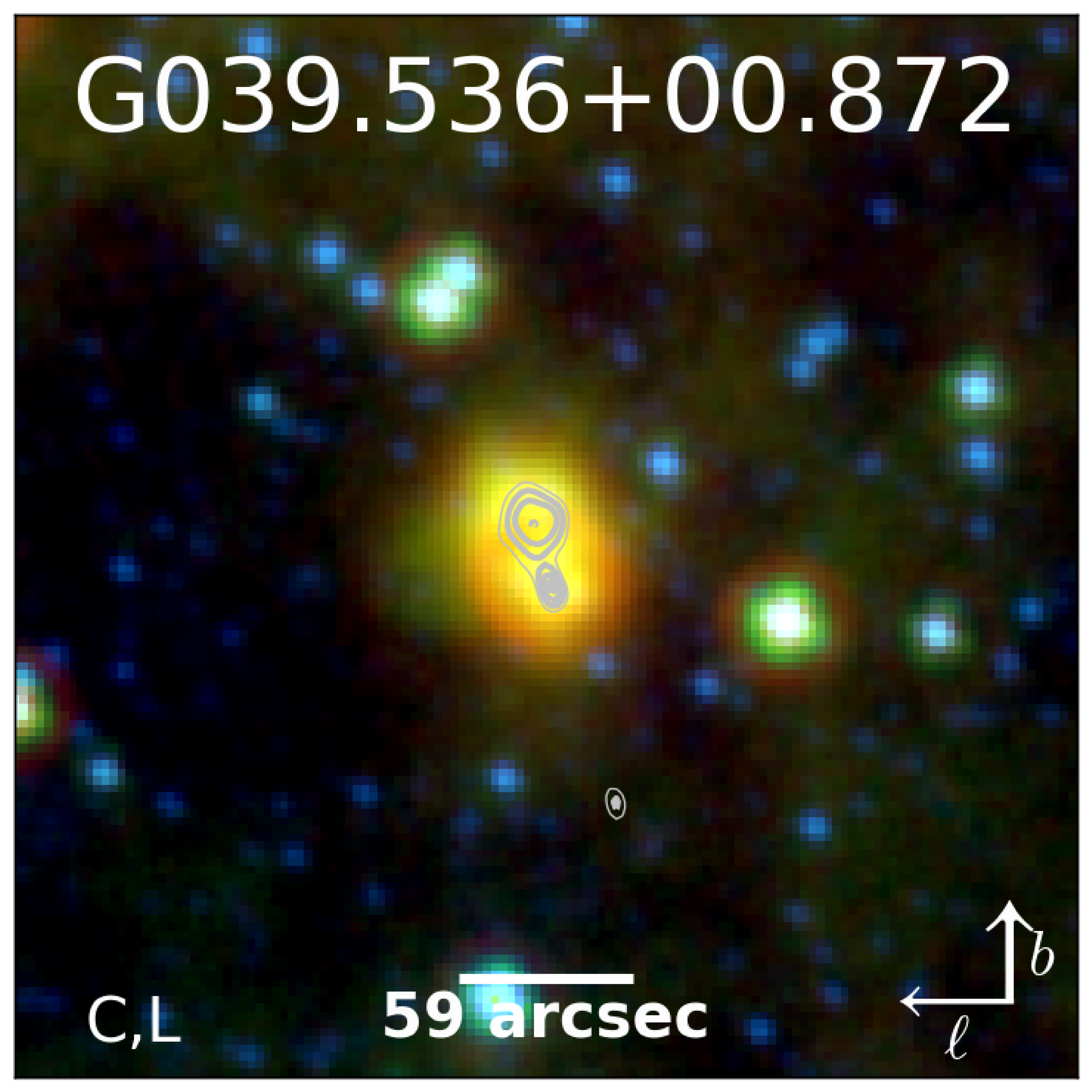}\\
\includegraphics[width=\figSize]{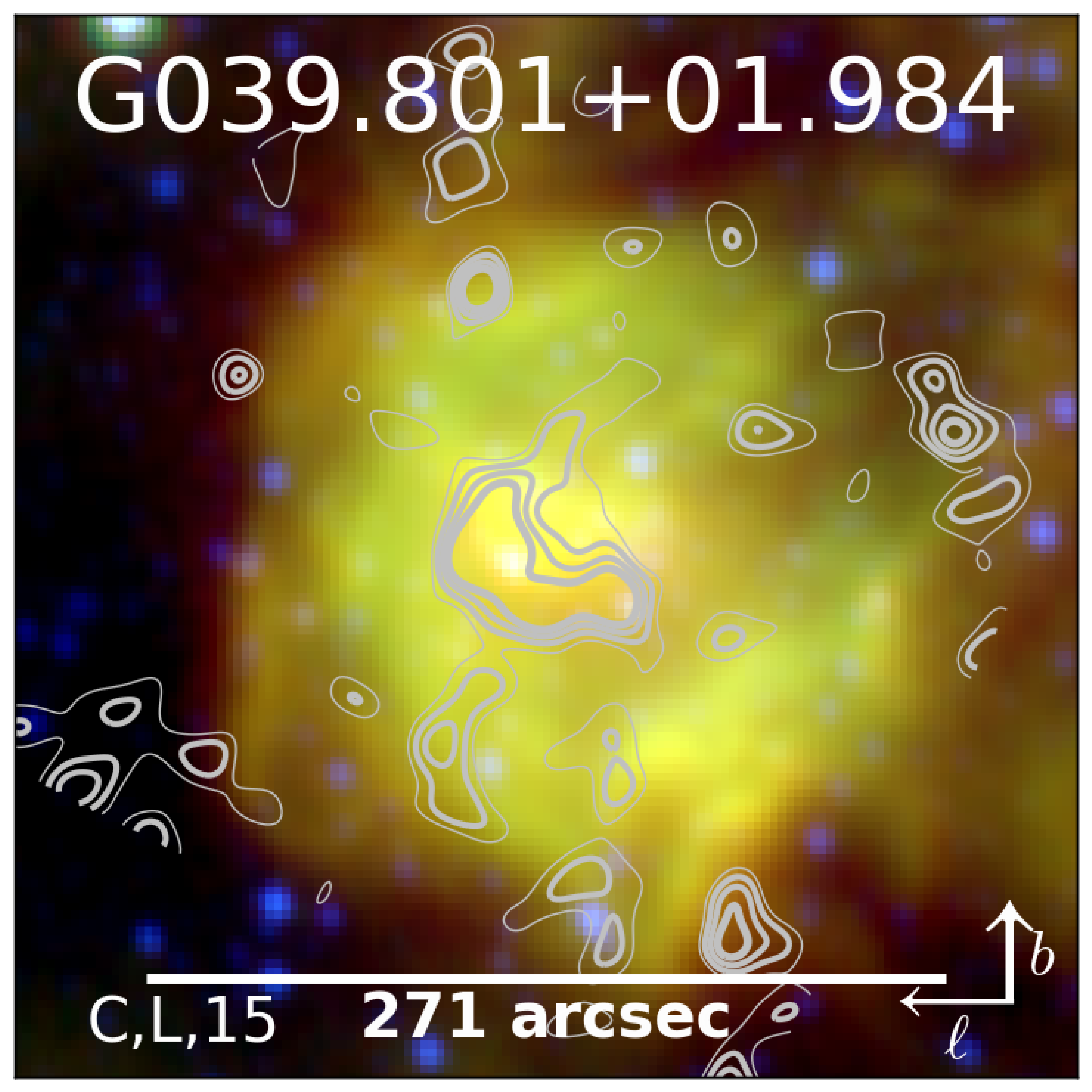}
\includegraphics[width=\figSize]{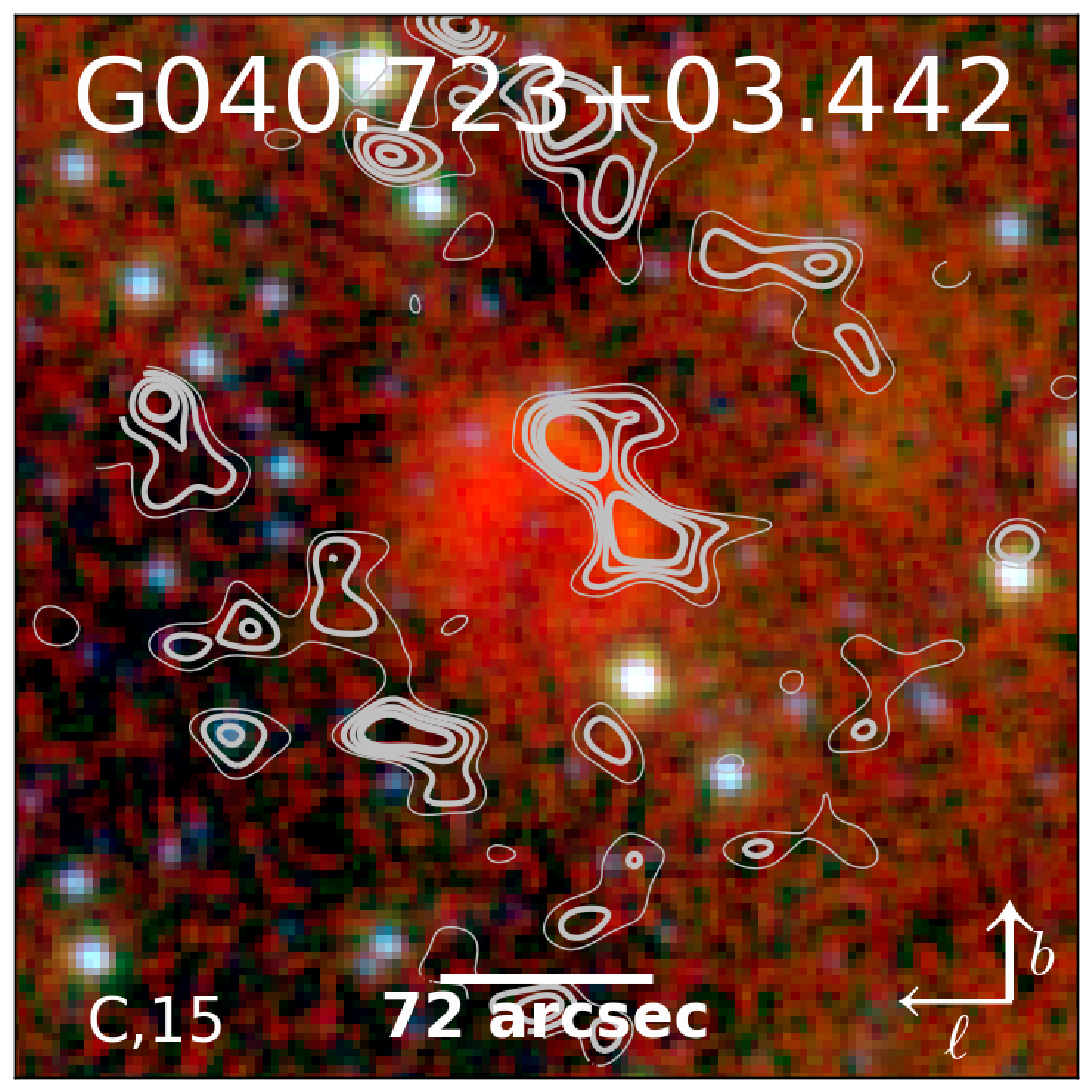}
\includegraphics[width=\figSize]{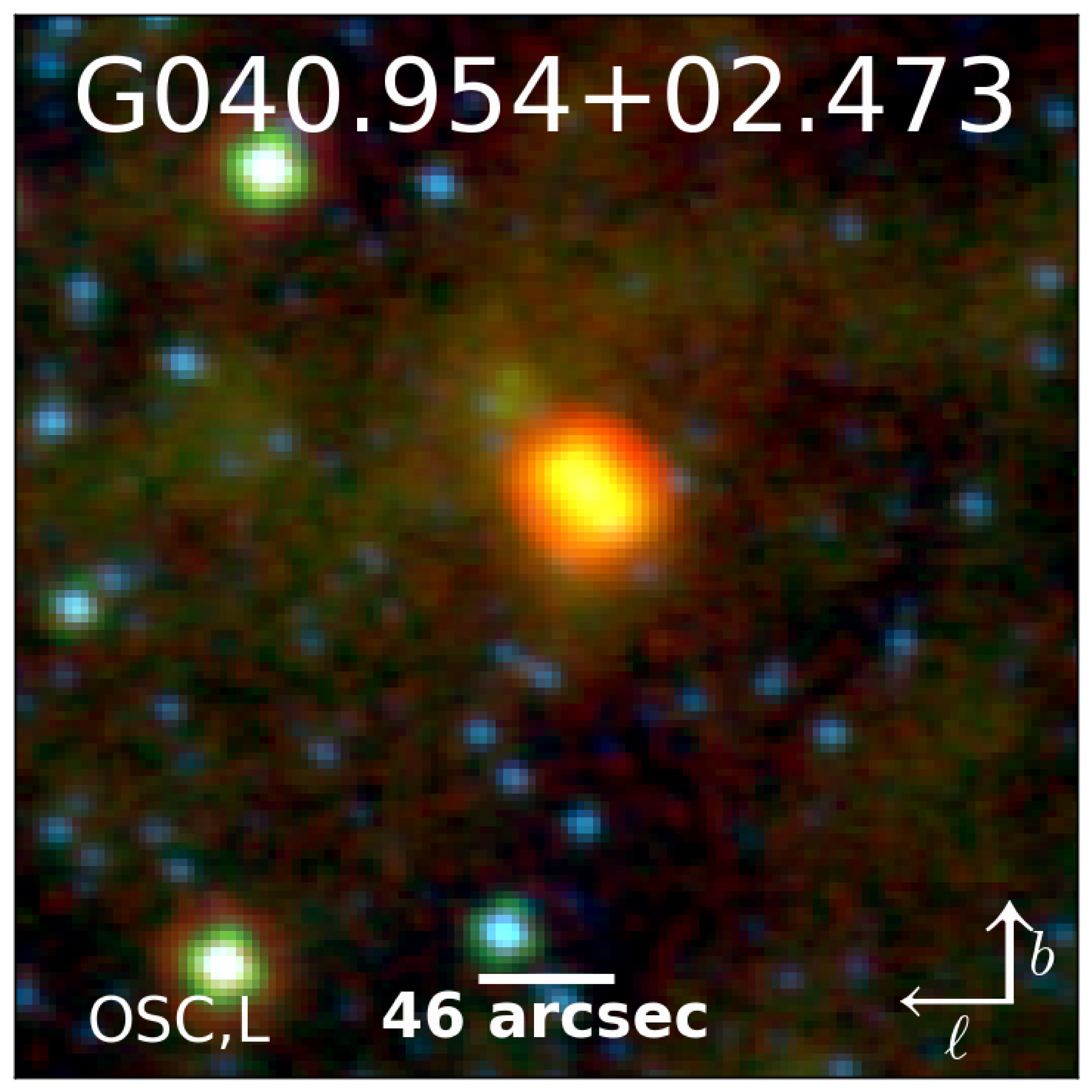}\\
\includegraphics[width=\figSize]{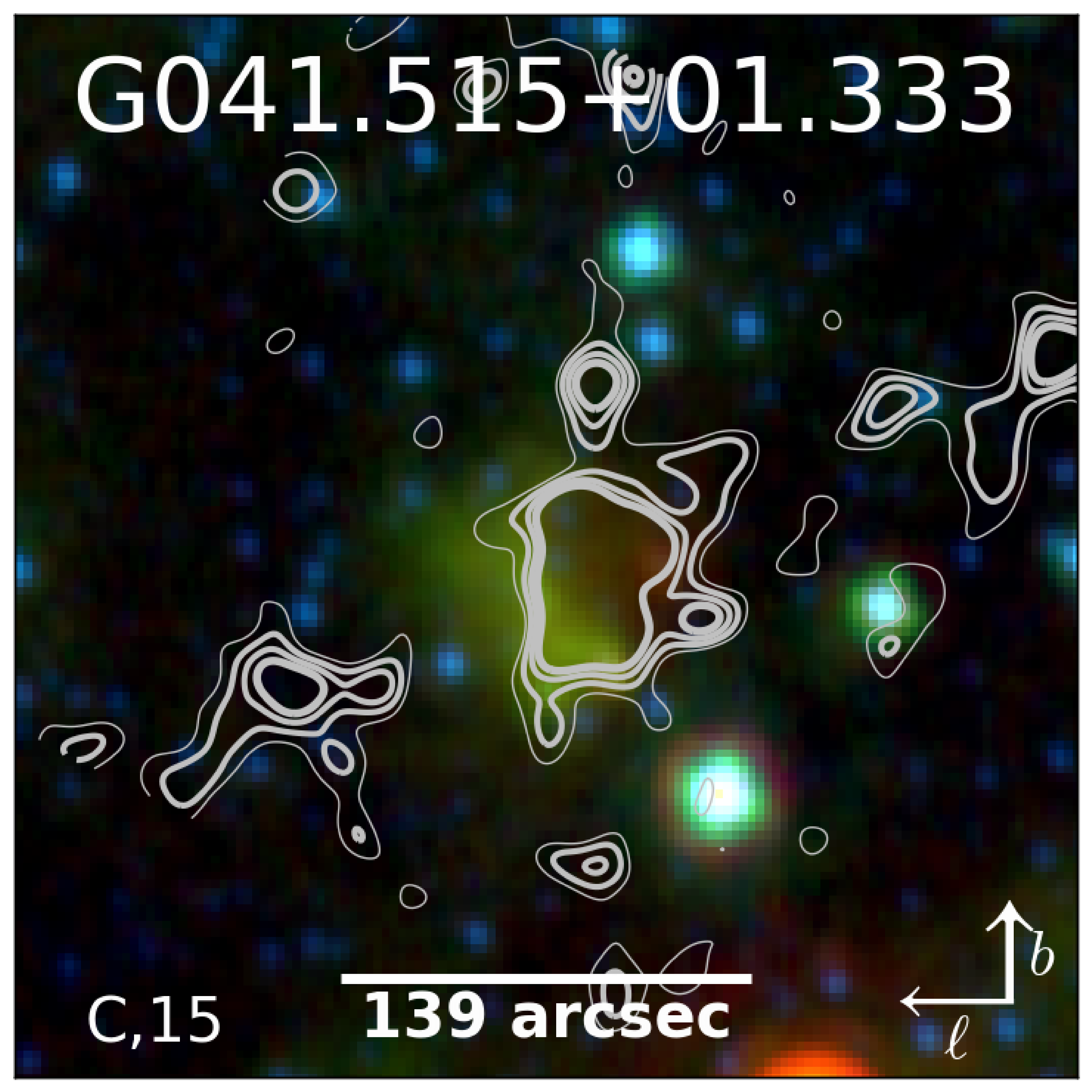}
\includegraphics[width=\figSize]{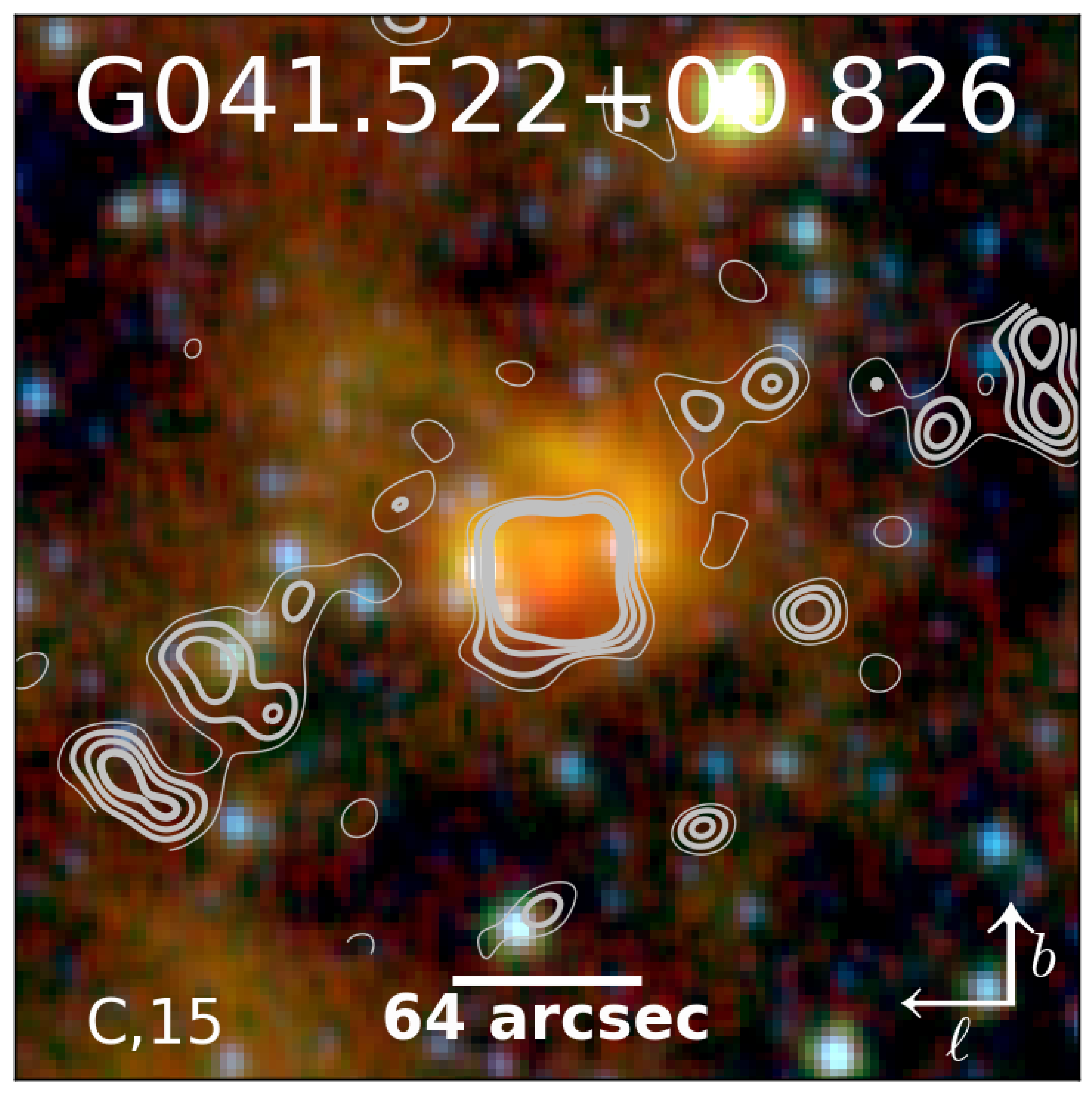}
\includegraphics[width=\figSize]{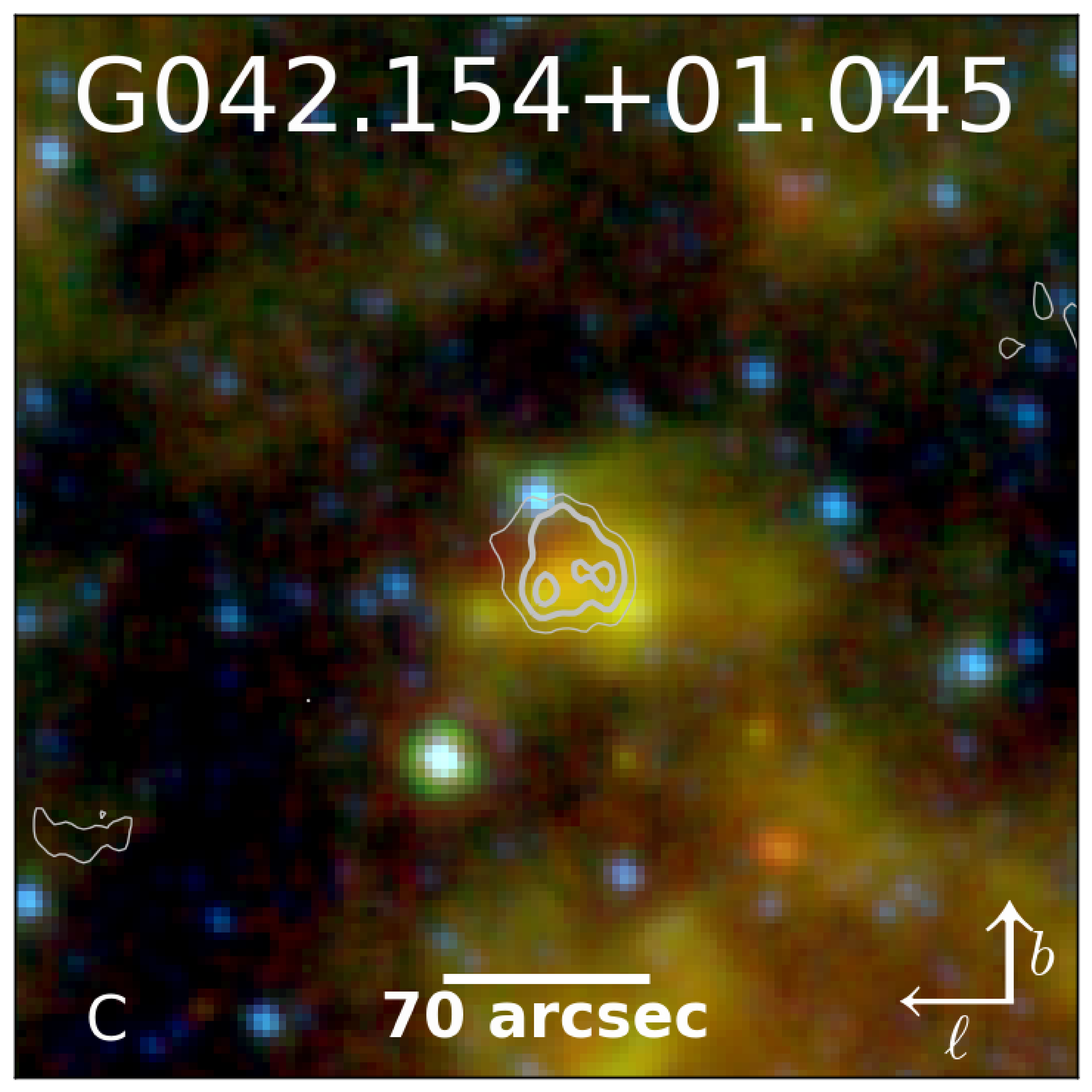}\\
\includegraphics[width=\figSize]{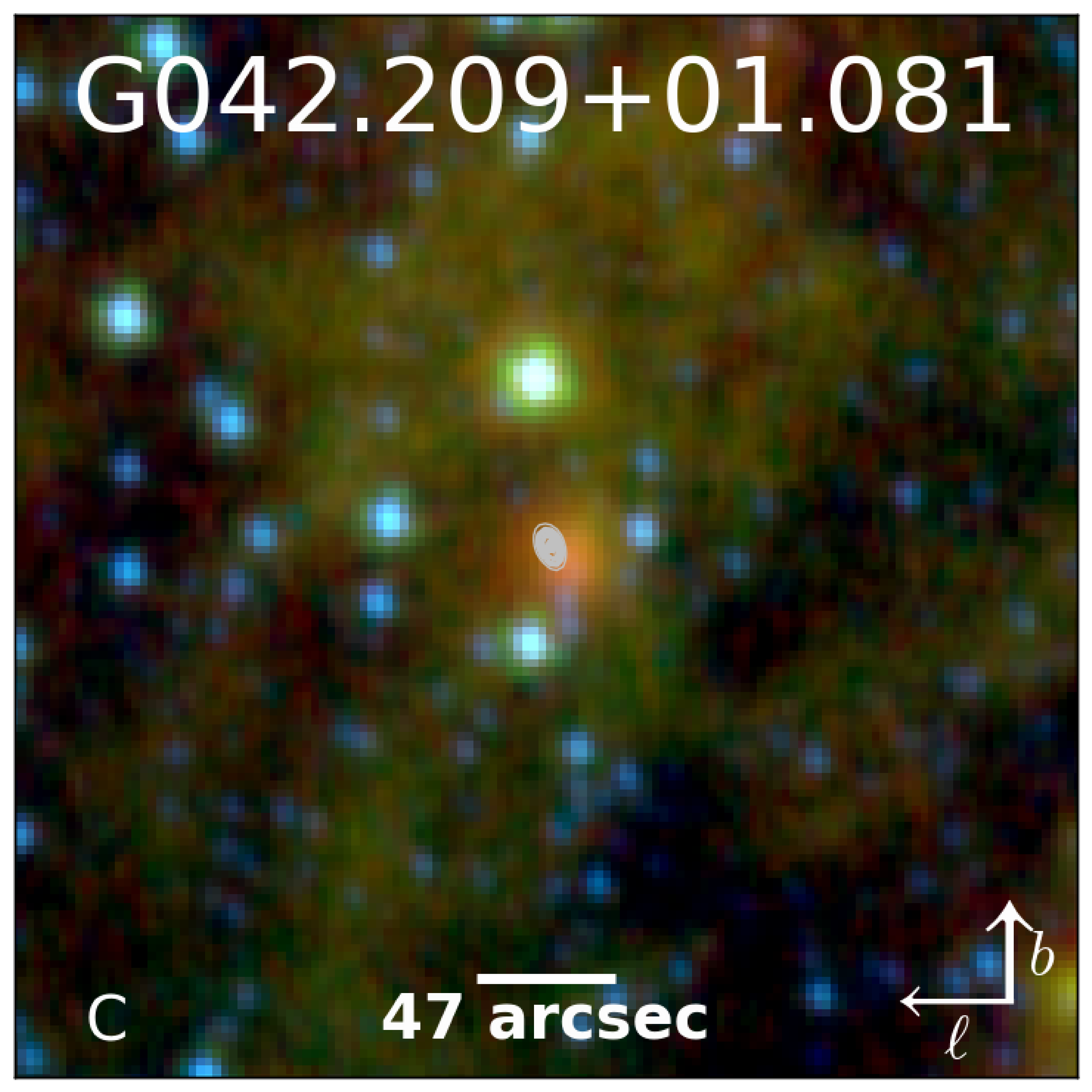}
\includegraphics[width=\figSize]{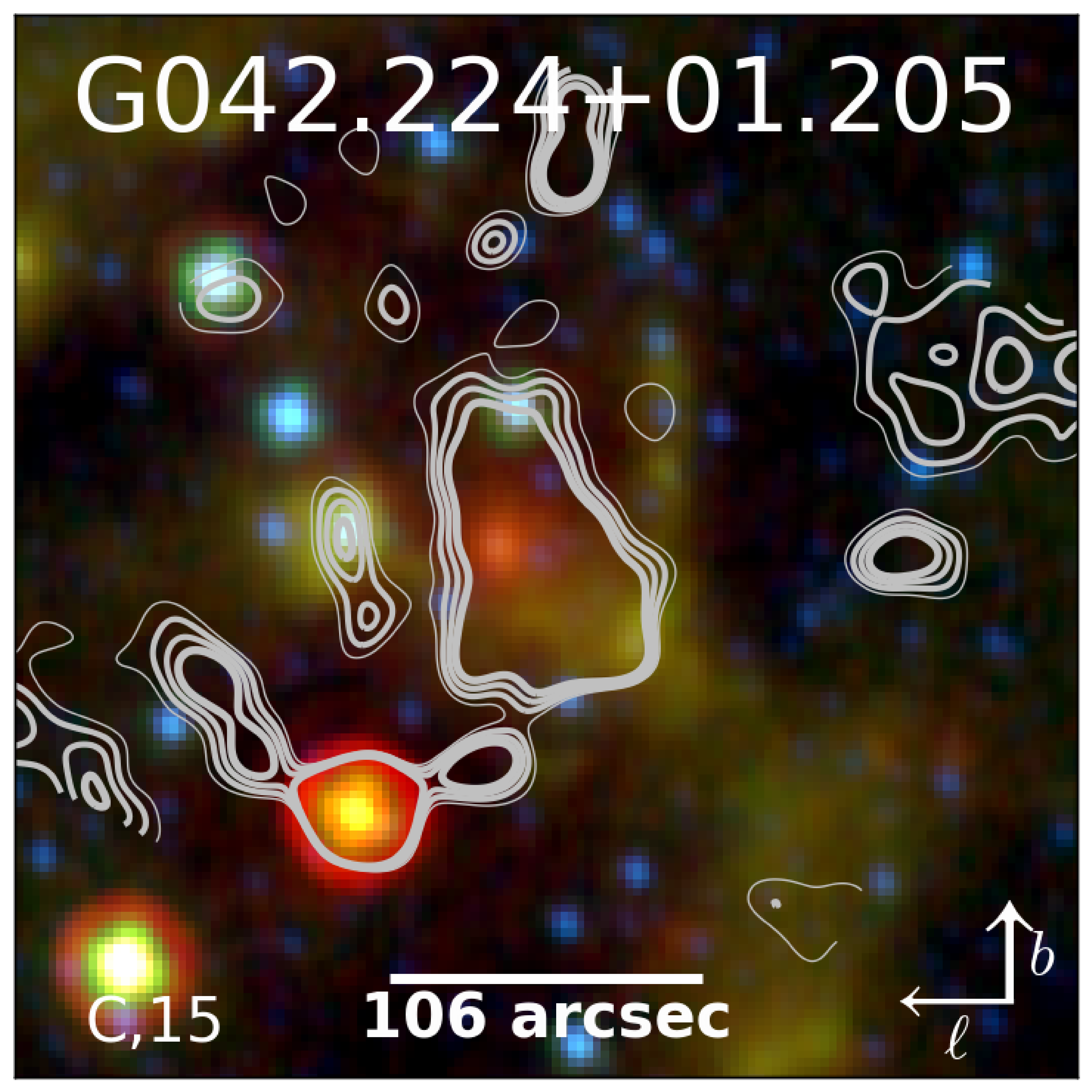}
\includegraphics[width=\figSize]{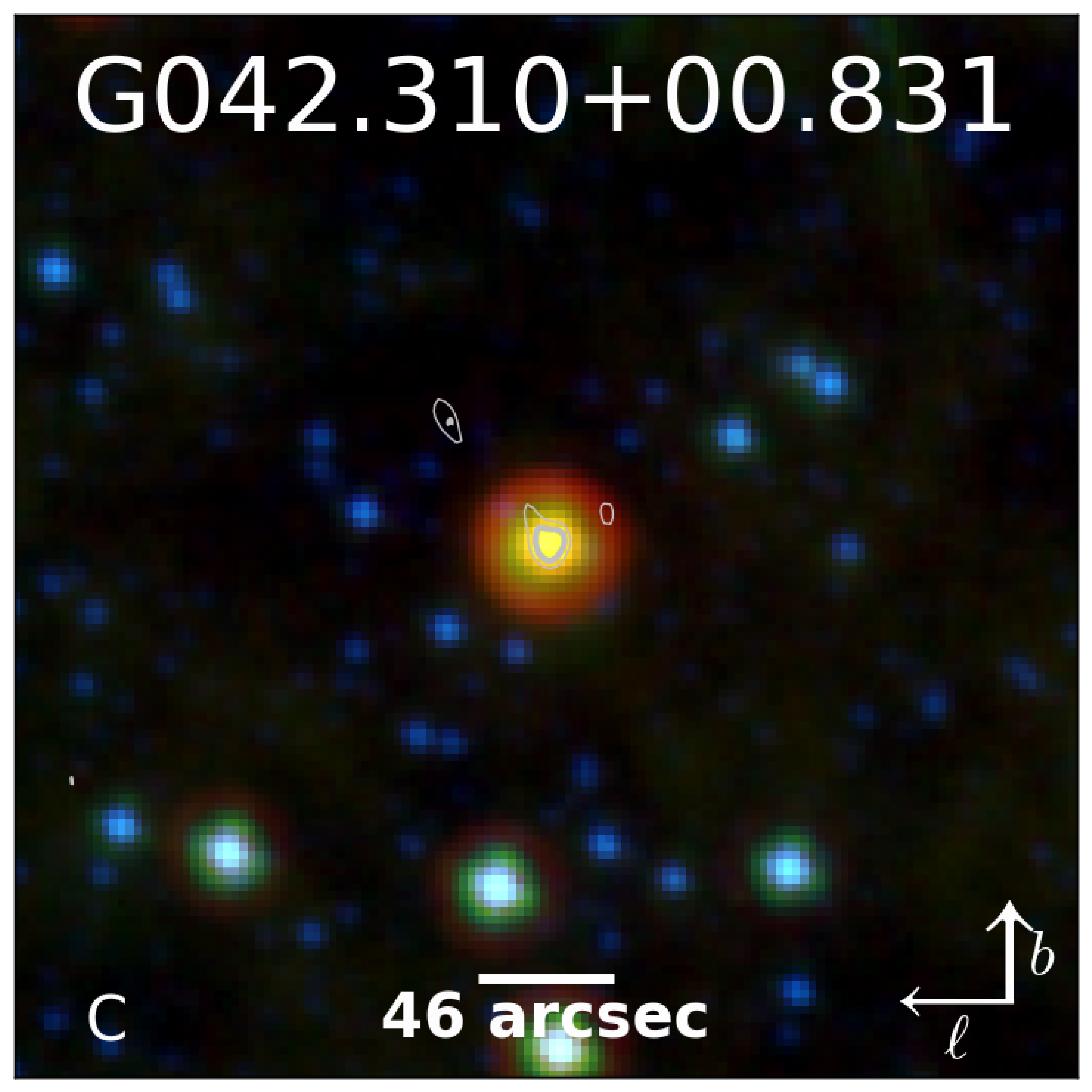}
\end{figure*}
\begin{figure*}[!htb]
\includegraphics[width=\figSize]{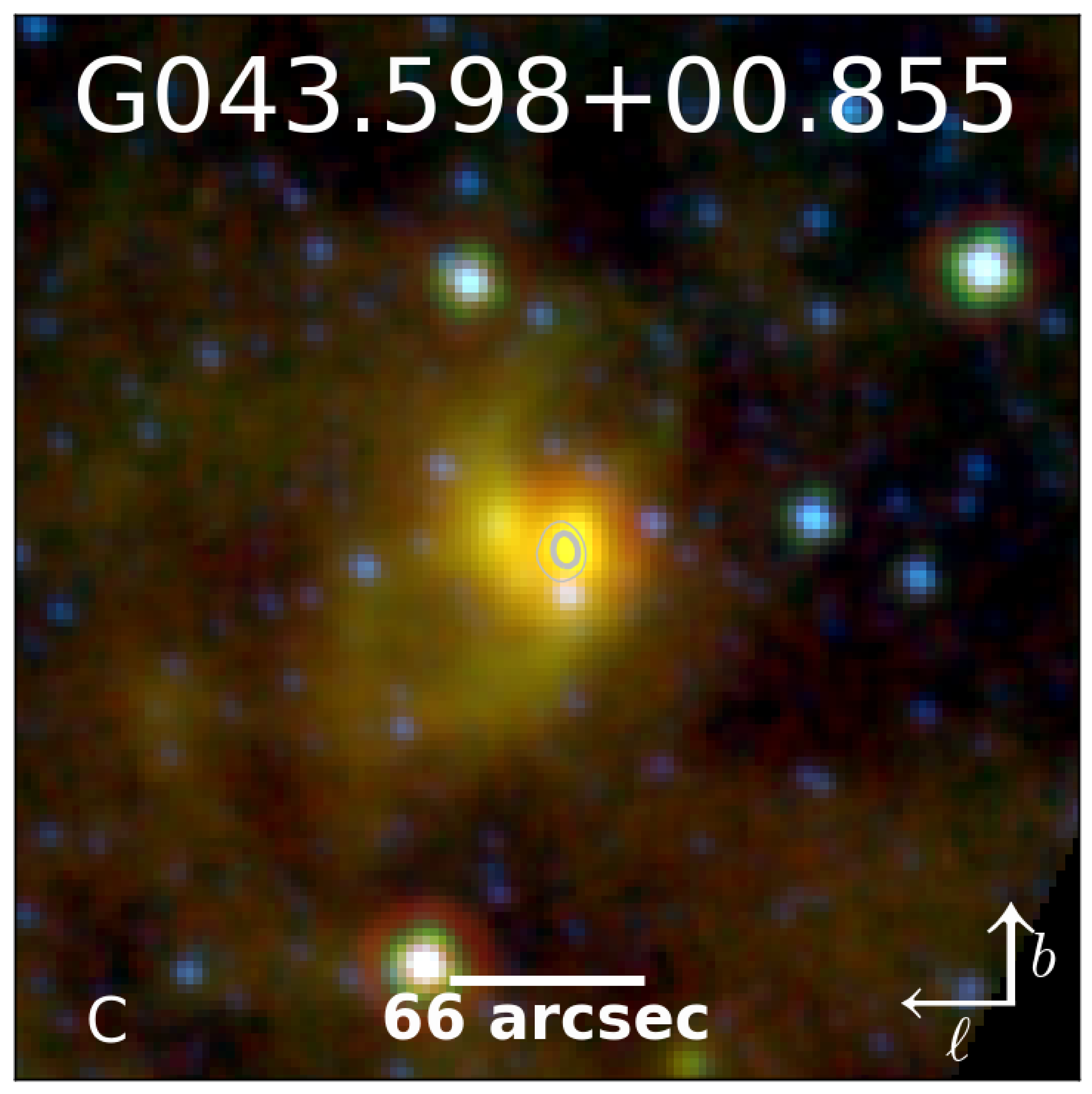}
\includegraphics[width=\figSize]{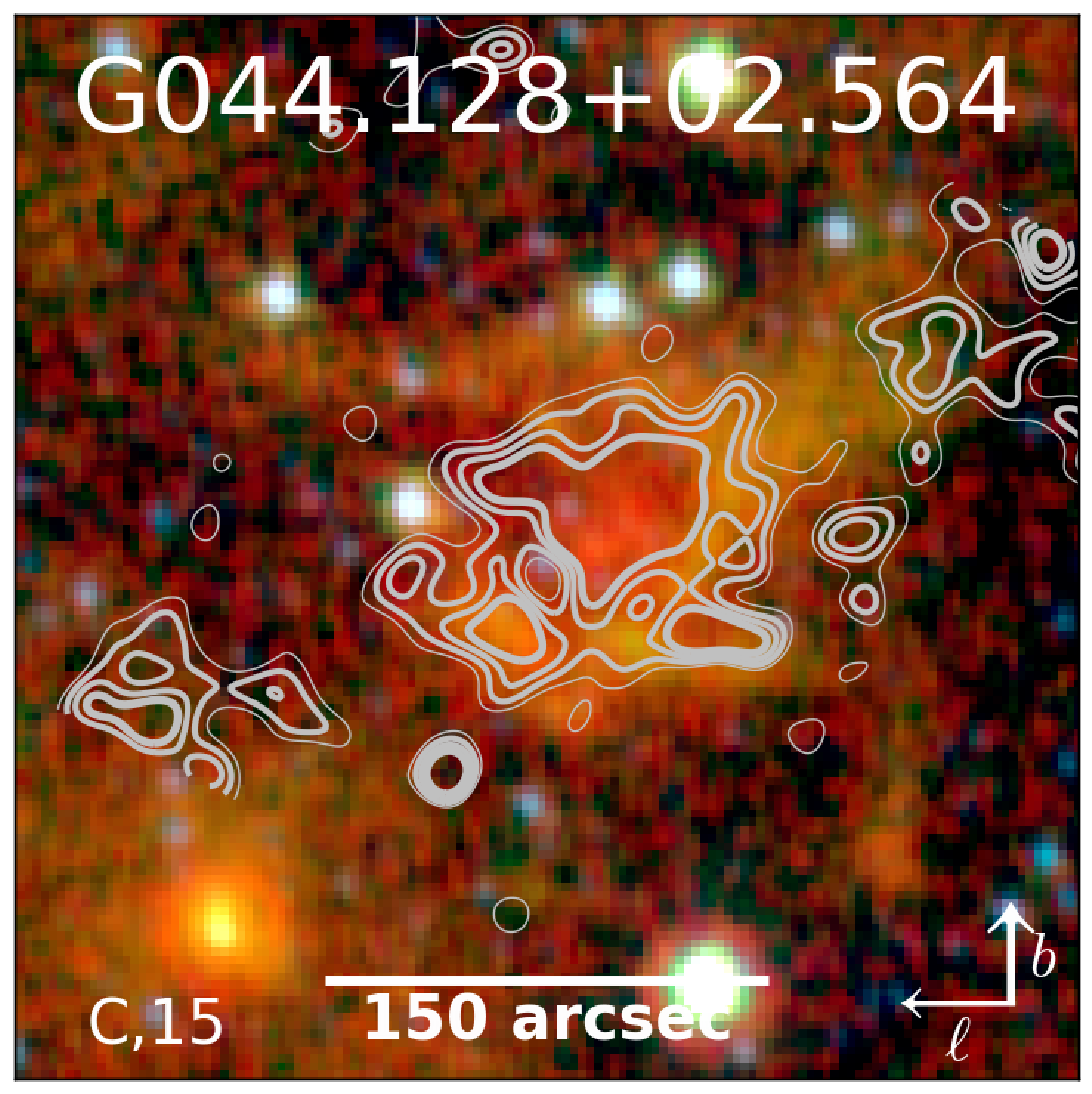}
\includegraphics[width=\figSize]{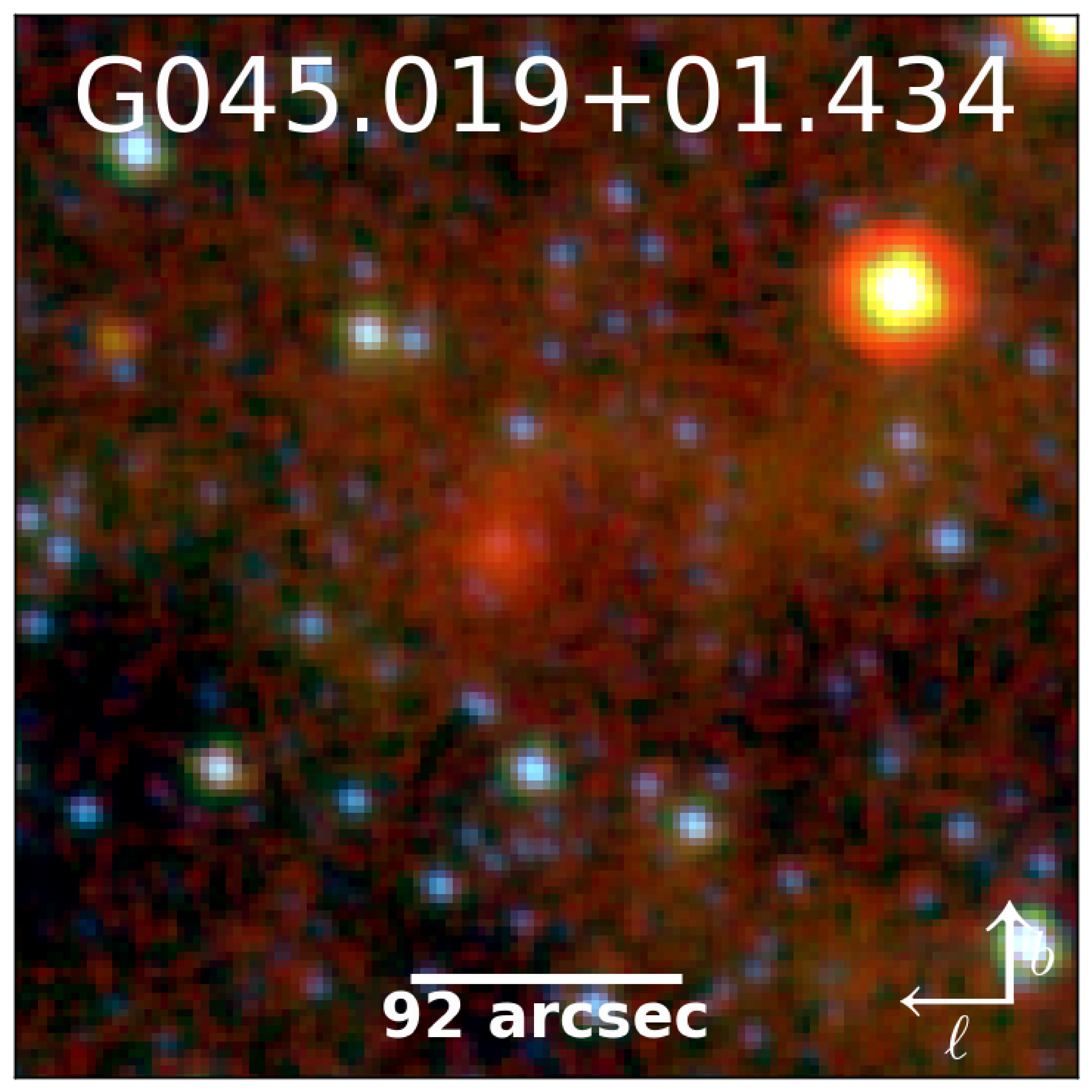}\\
\includegraphics[width=\figSize]{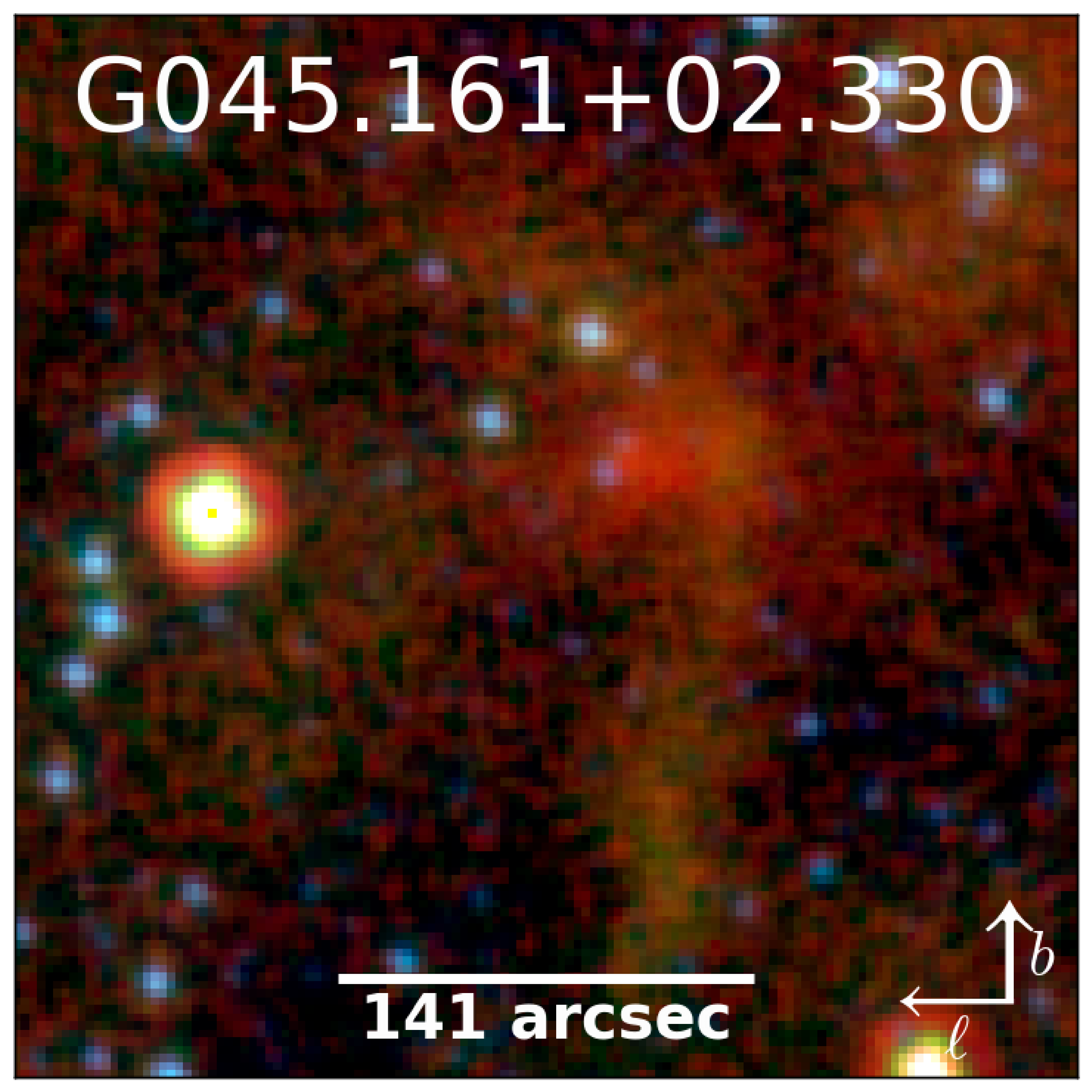}
\includegraphics[width=\figSize]{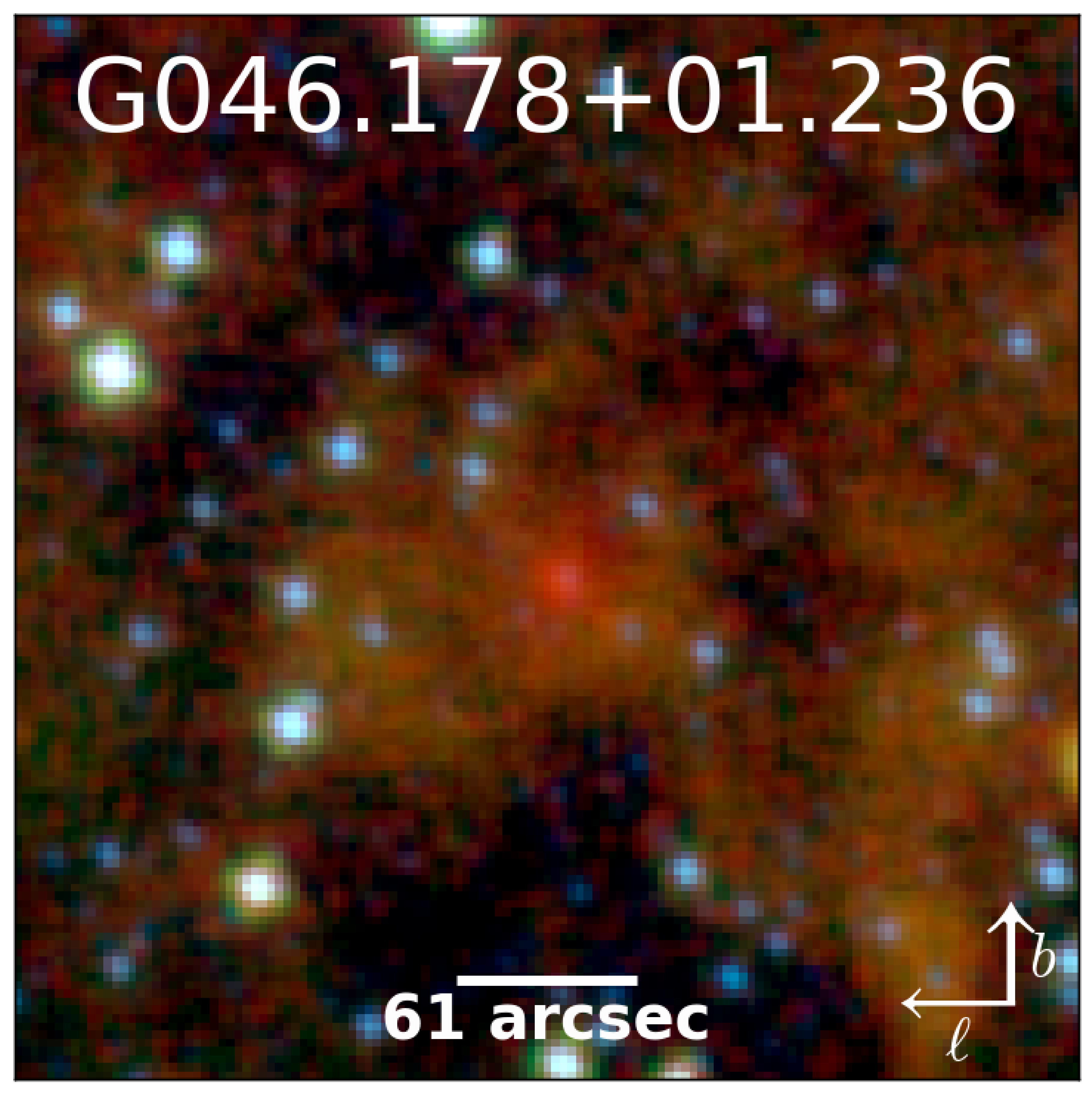}
\includegraphics[width=\figSize]{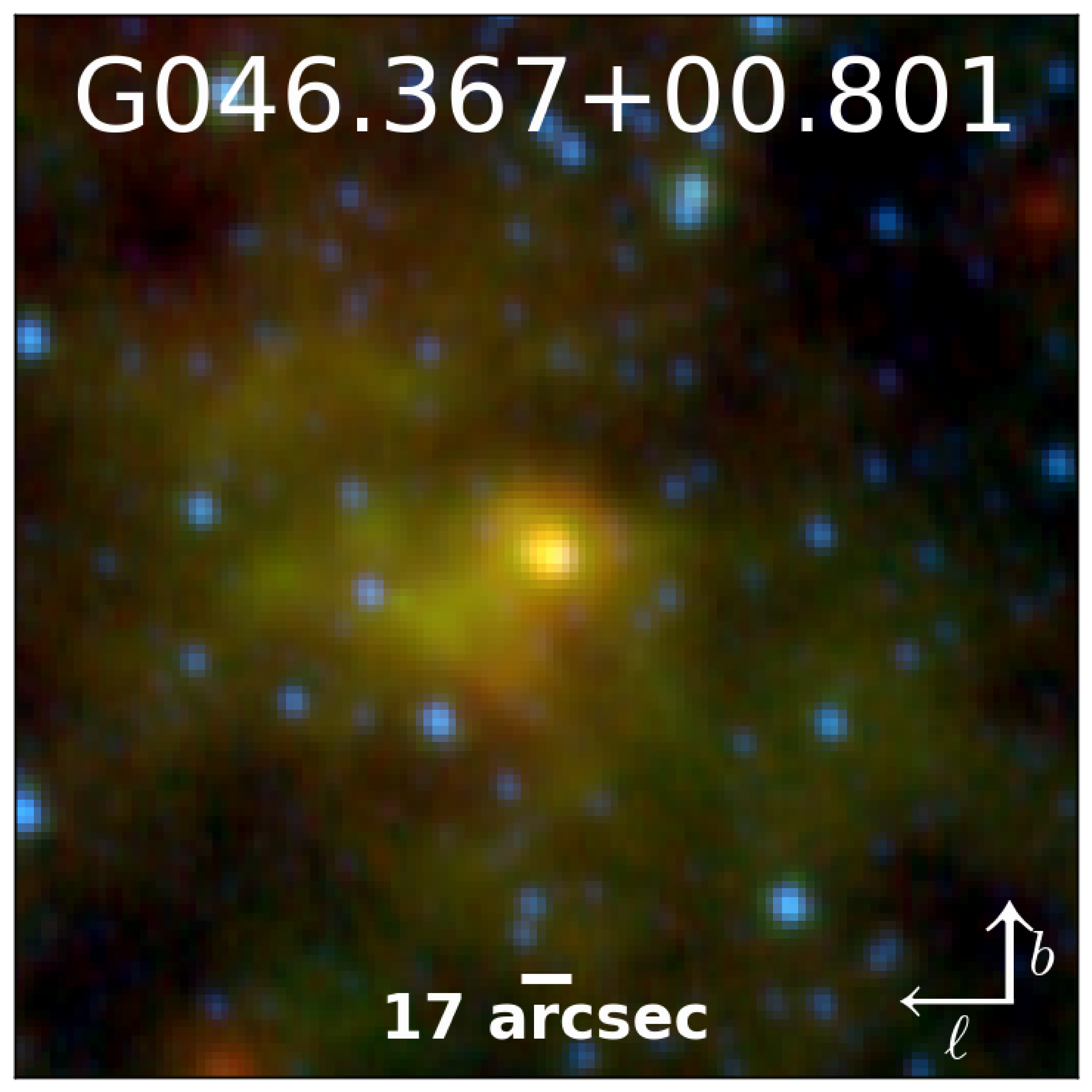}\\
\includegraphics[width=\figSize]{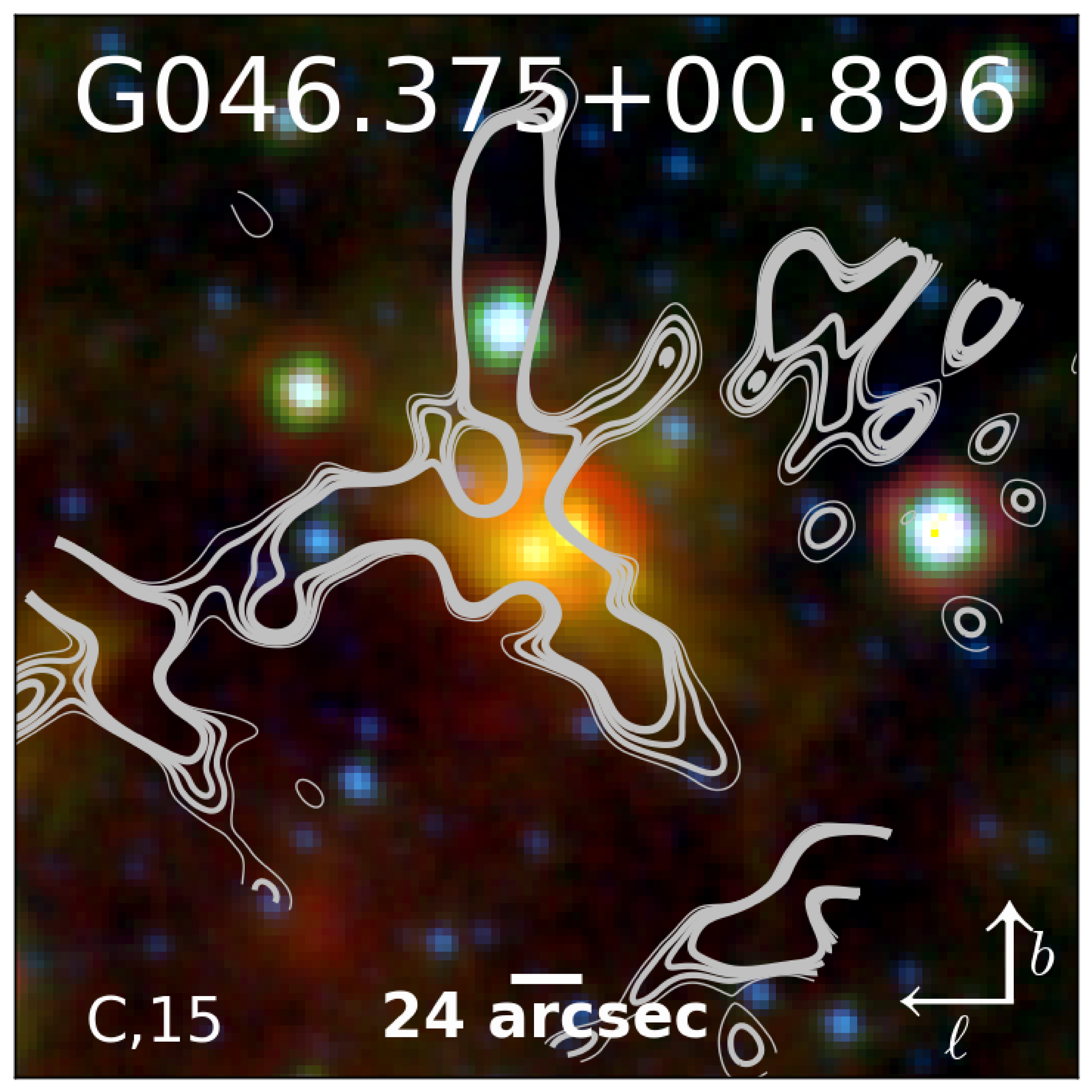}
\includegraphics[width=\figSize]{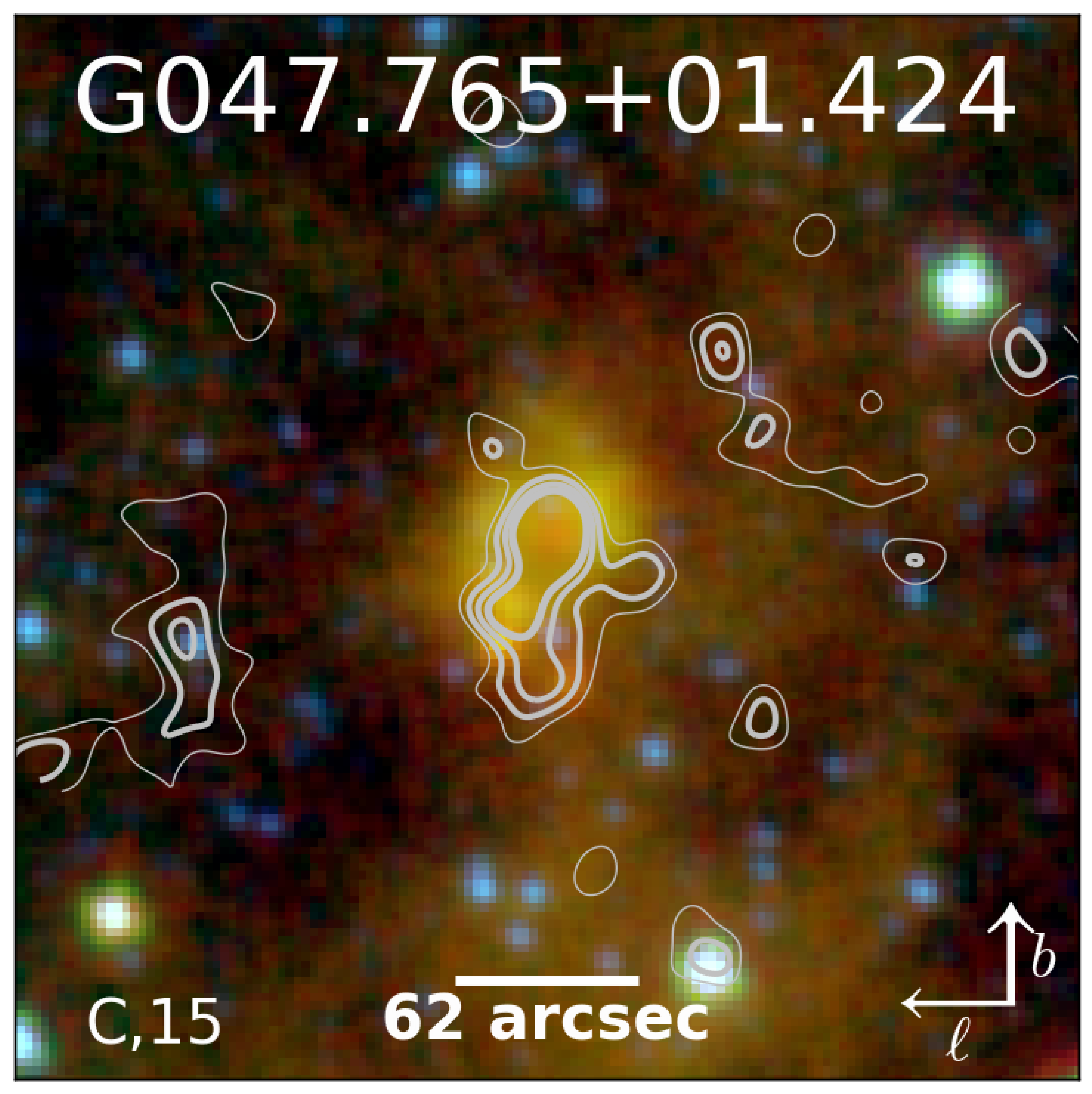}
\includegraphics[width=\figSize]{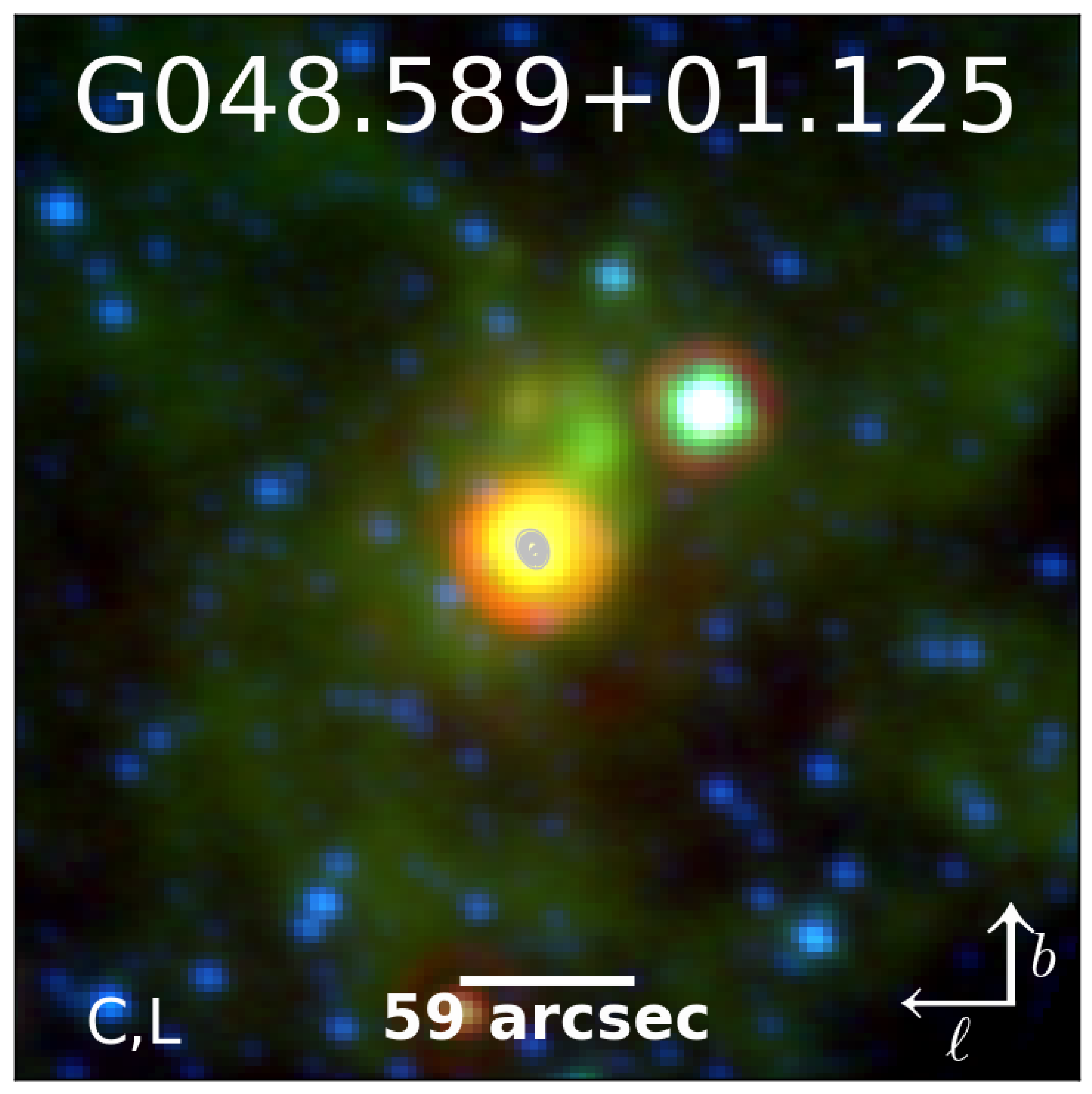}\\
\includegraphics[width=\figSize]{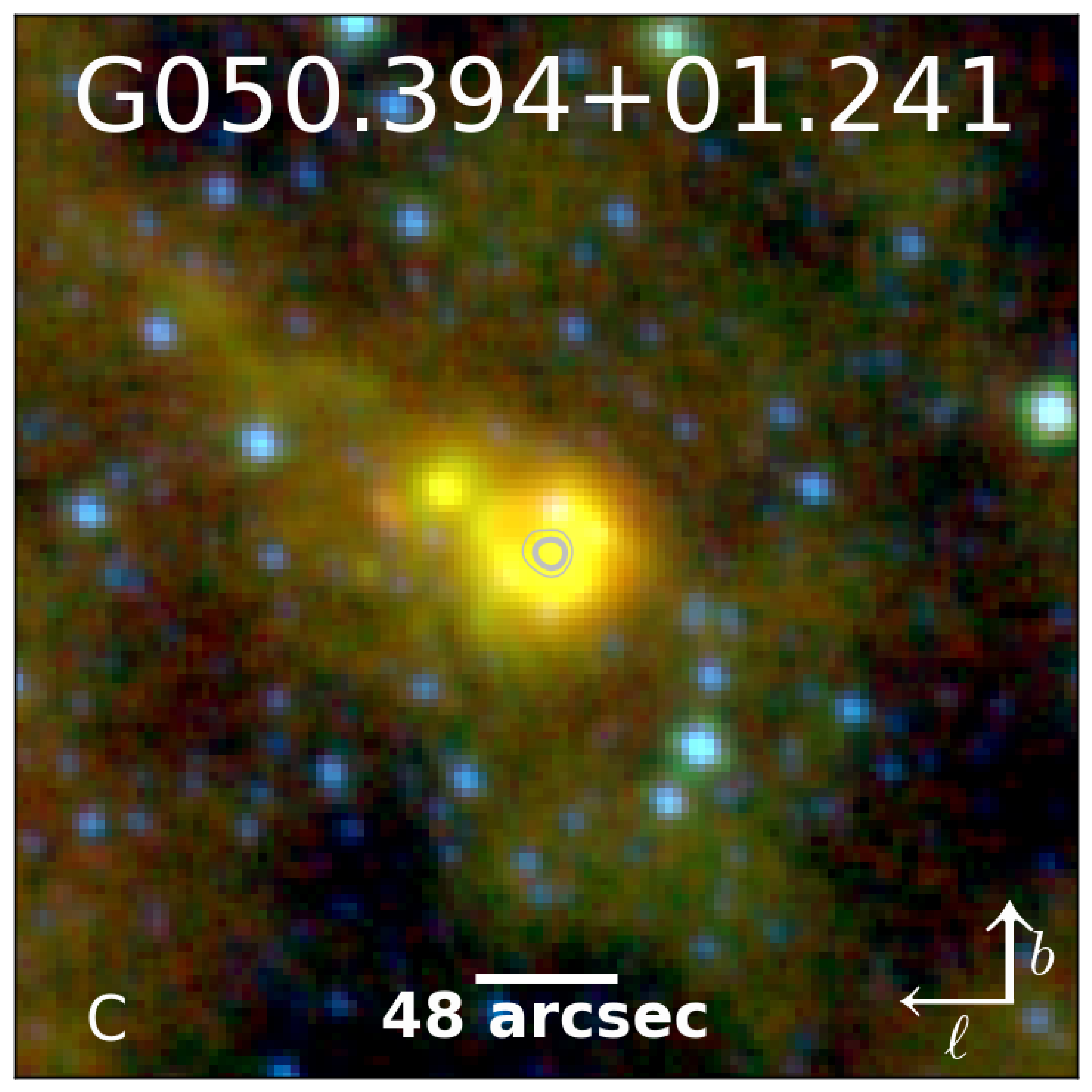}
\includegraphics[width=\figSize]{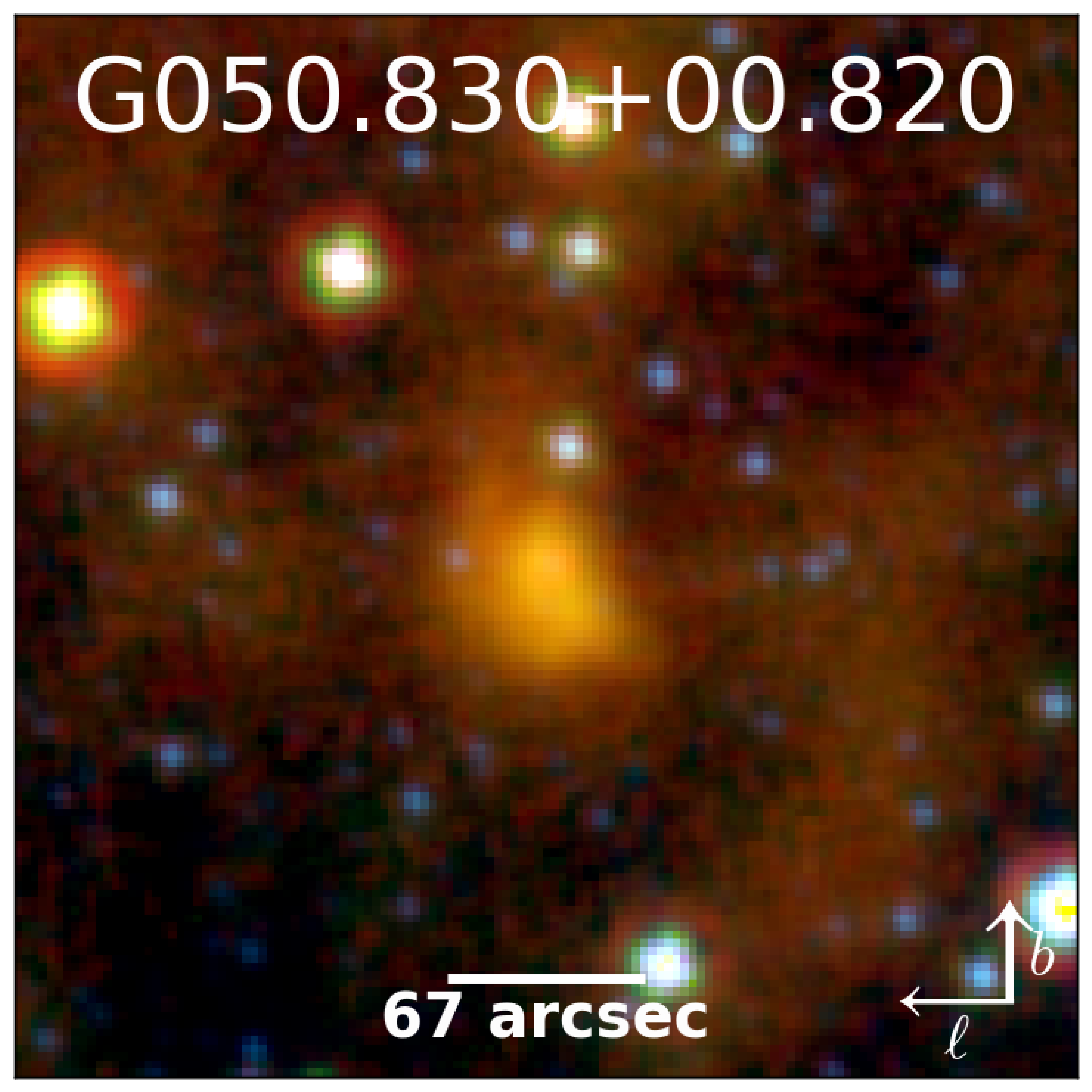}
\includegraphics[width=\figSize]{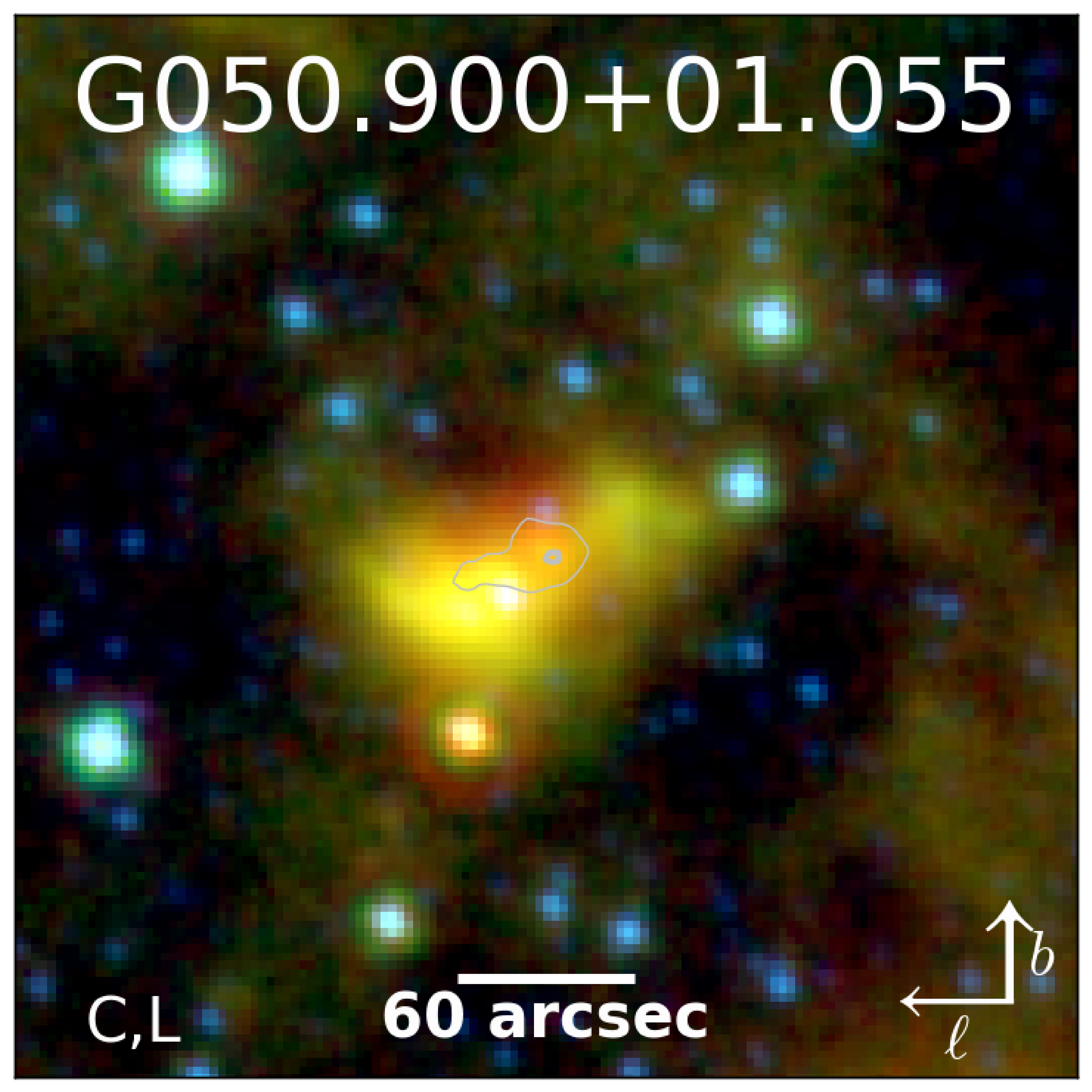}
\end{figure*}
\begin{figure*}[!htb]
\includegraphics[width=\figSize]{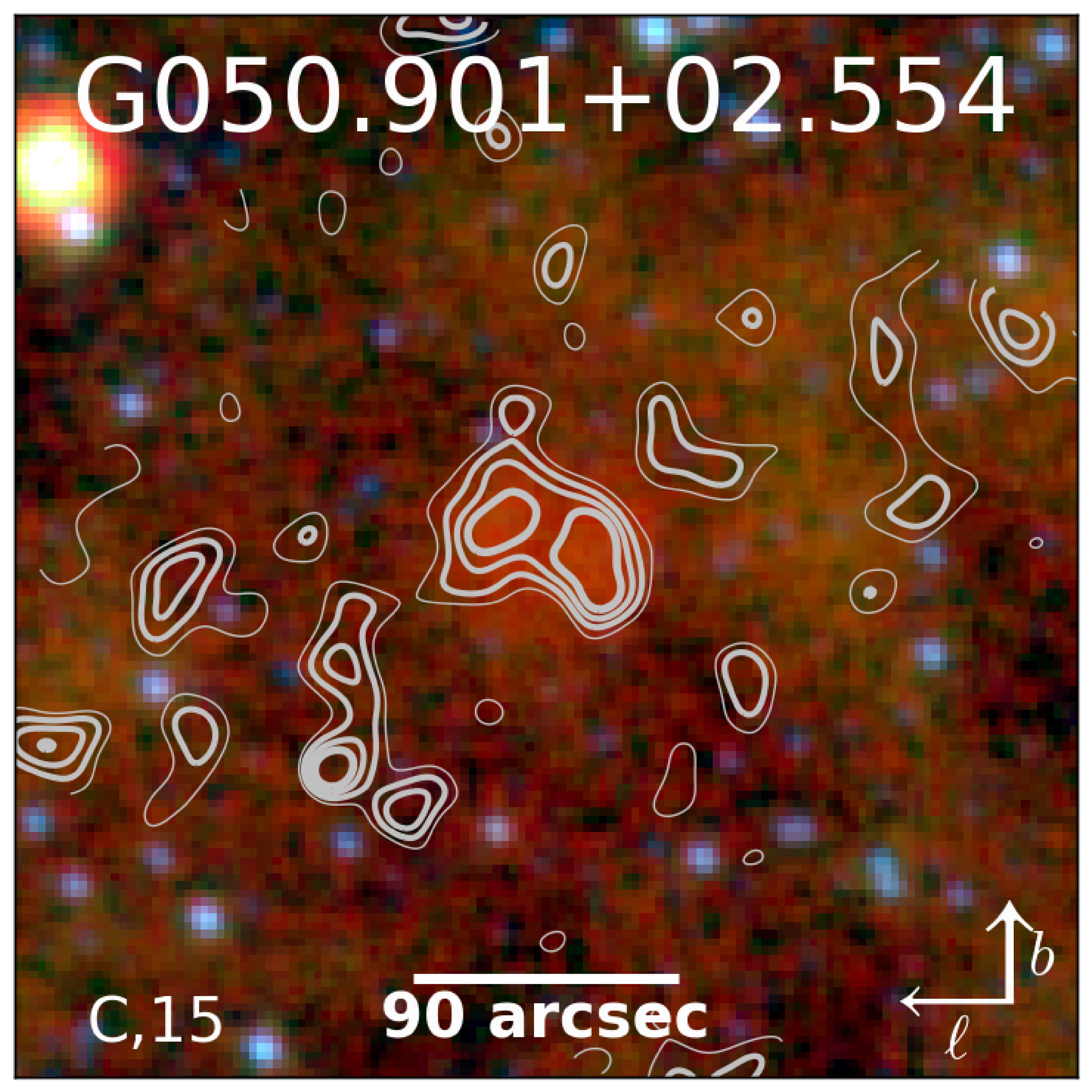}
\includegraphics[width=\figSize]{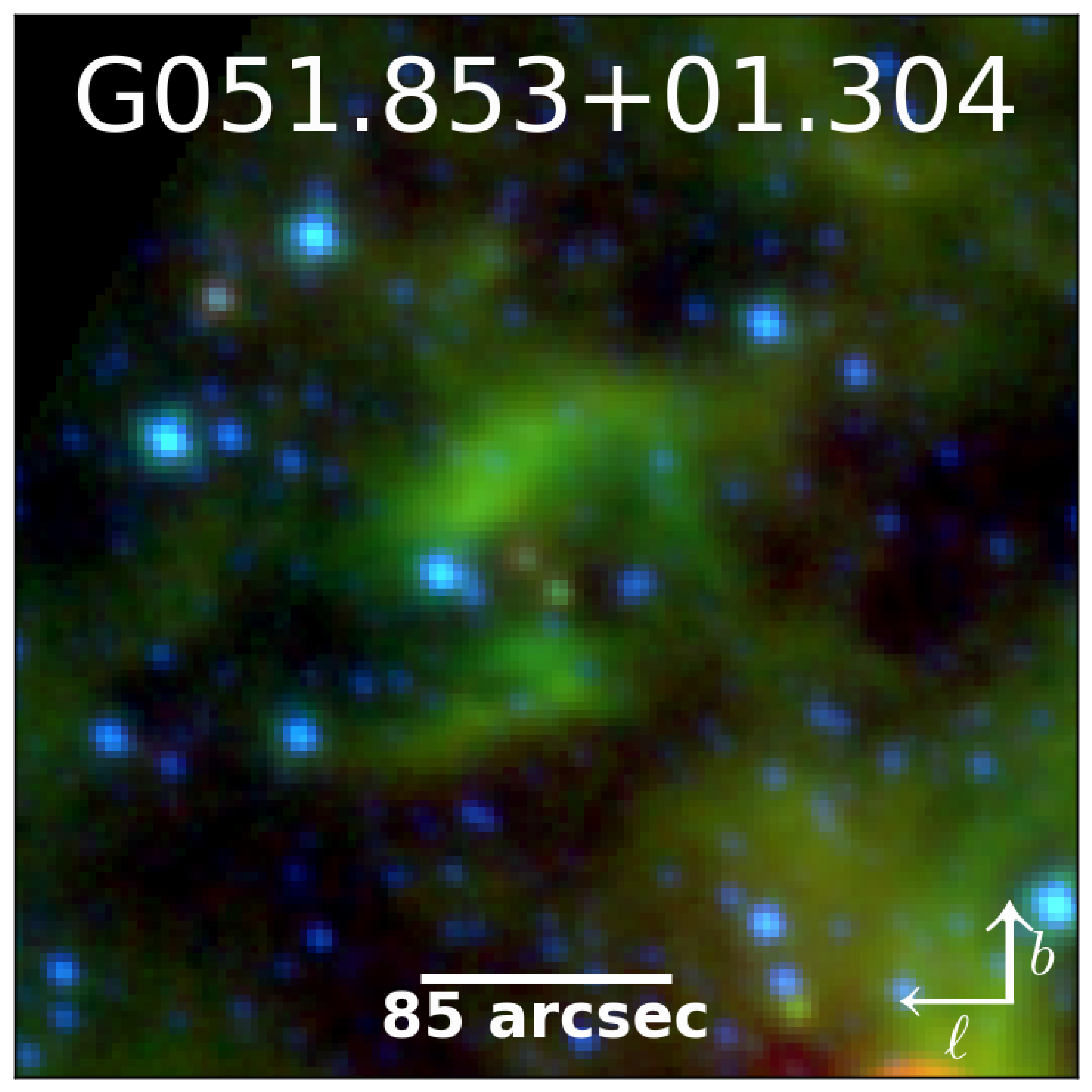}
\includegraphics[width=\figSize]{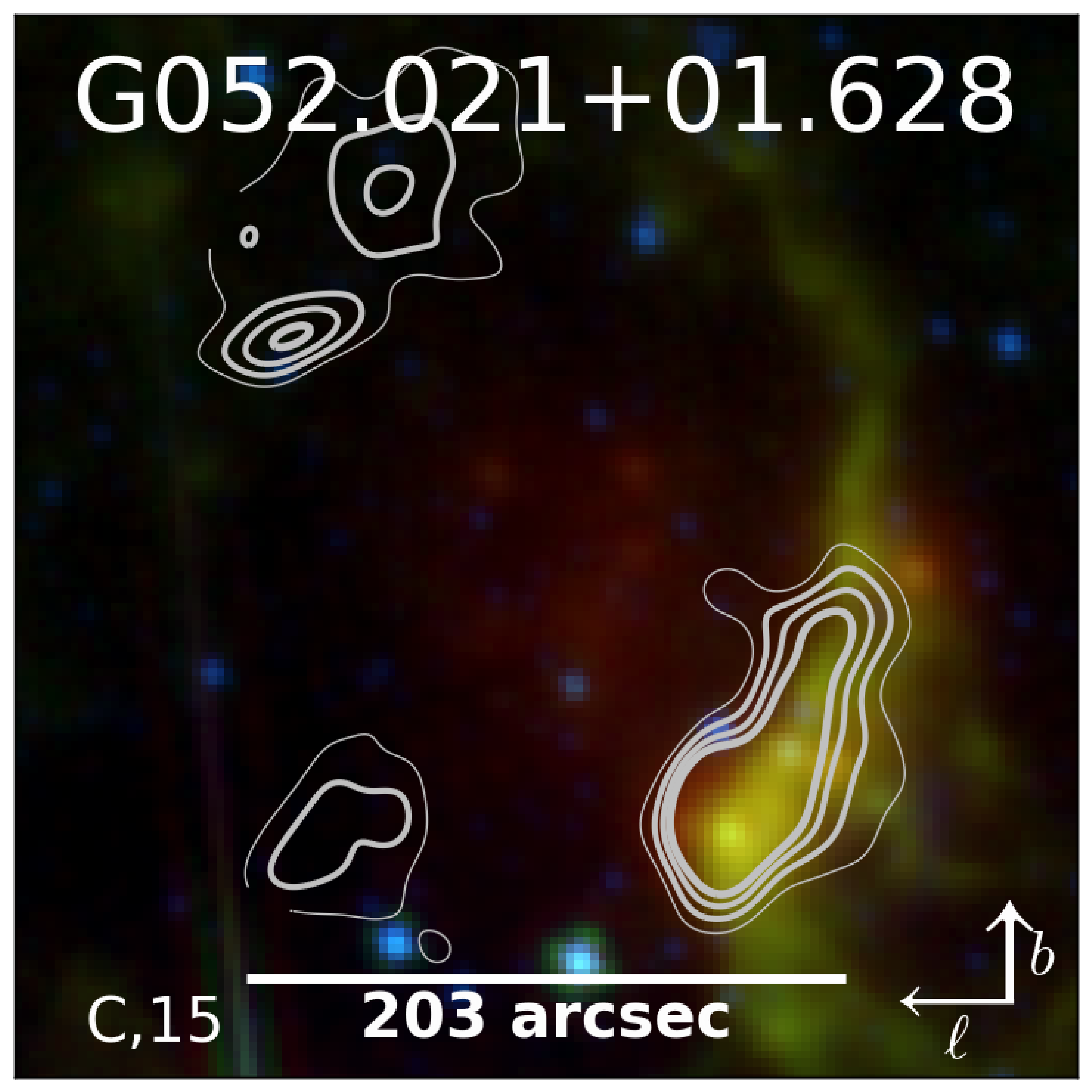}\\
\includegraphics[width=\figSize]{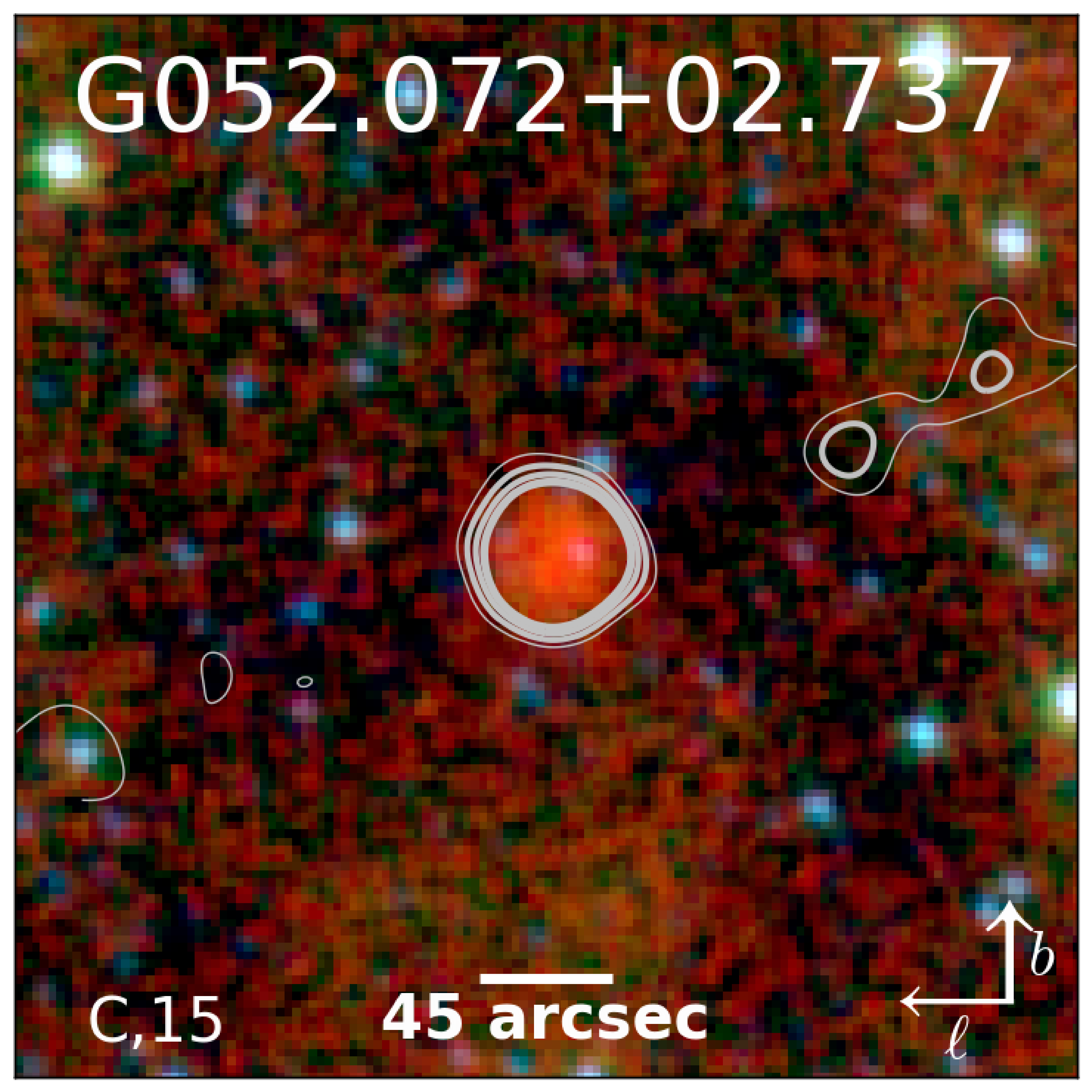}
\includegraphics[width=\figSize]{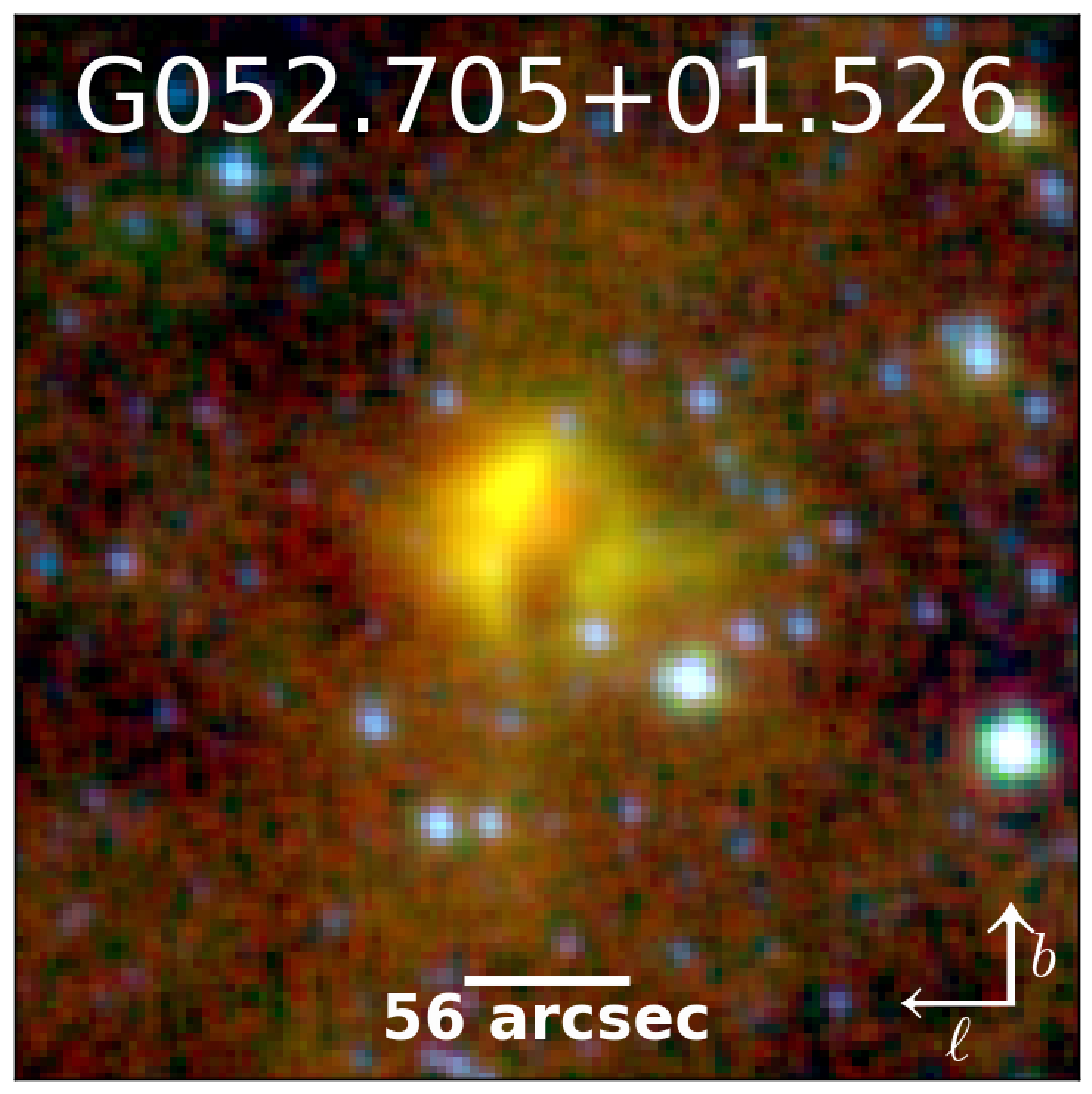}
\includegraphics[width=\figSize]{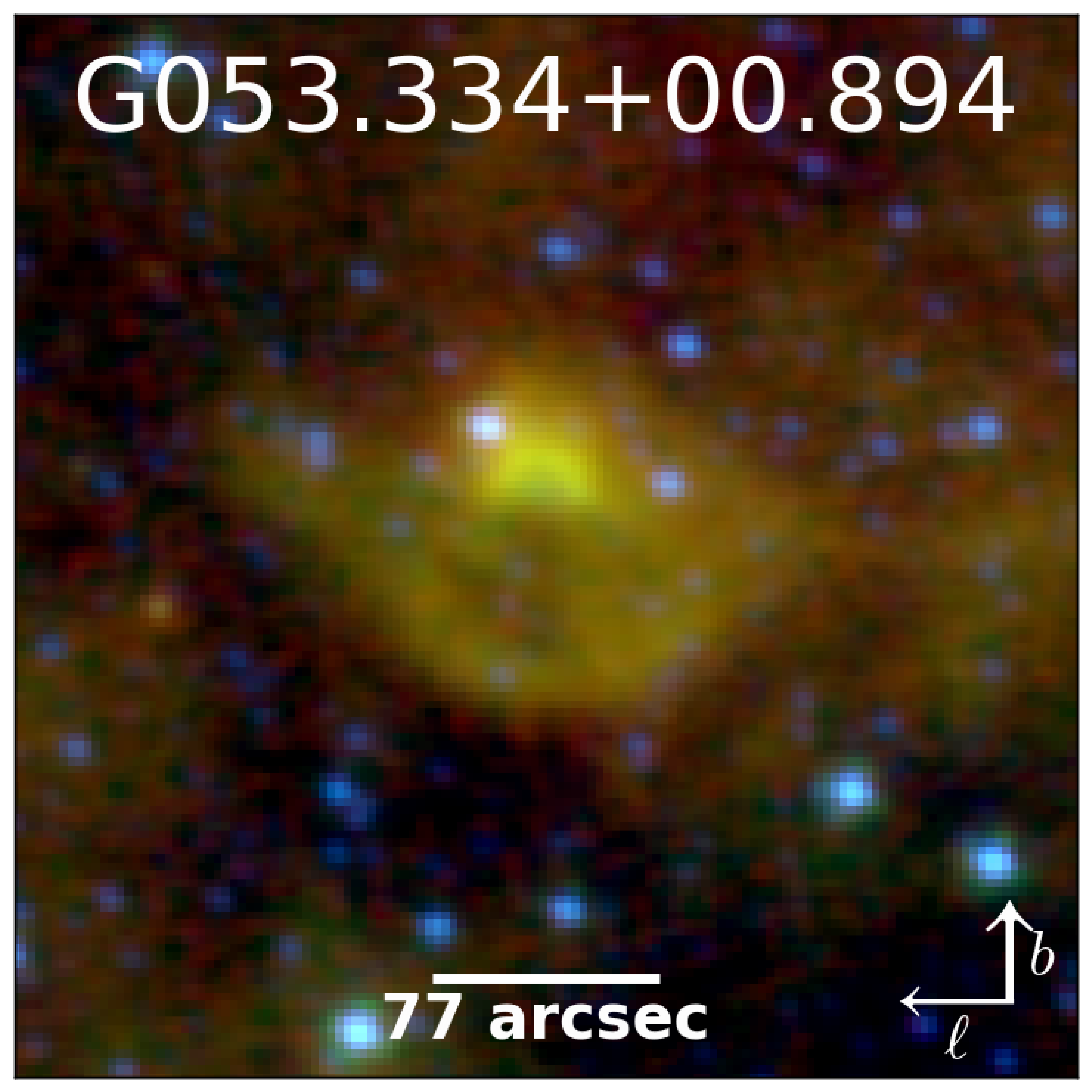}\\
\includegraphics[width=\figSize]{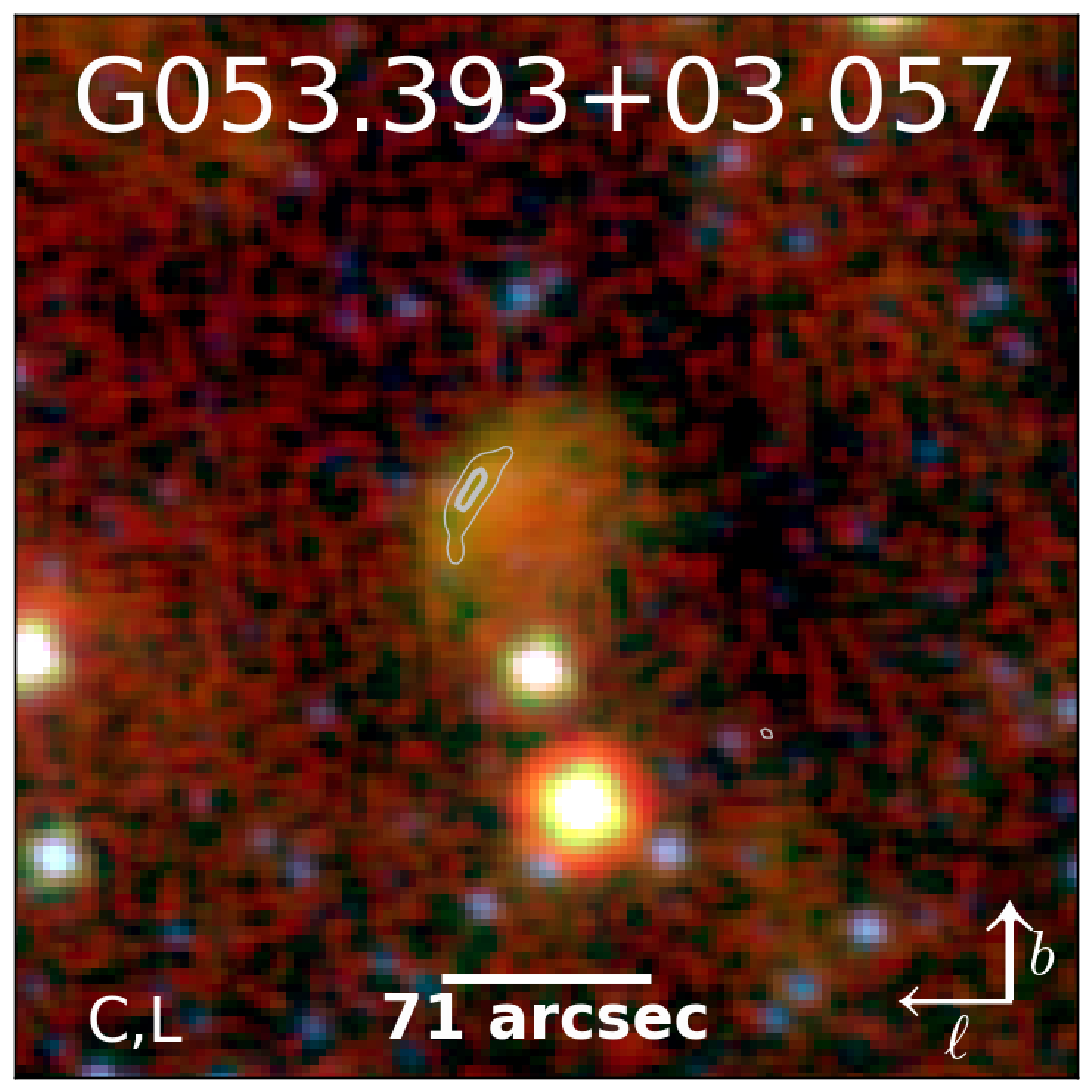}
\includegraphics[width=\figSize]{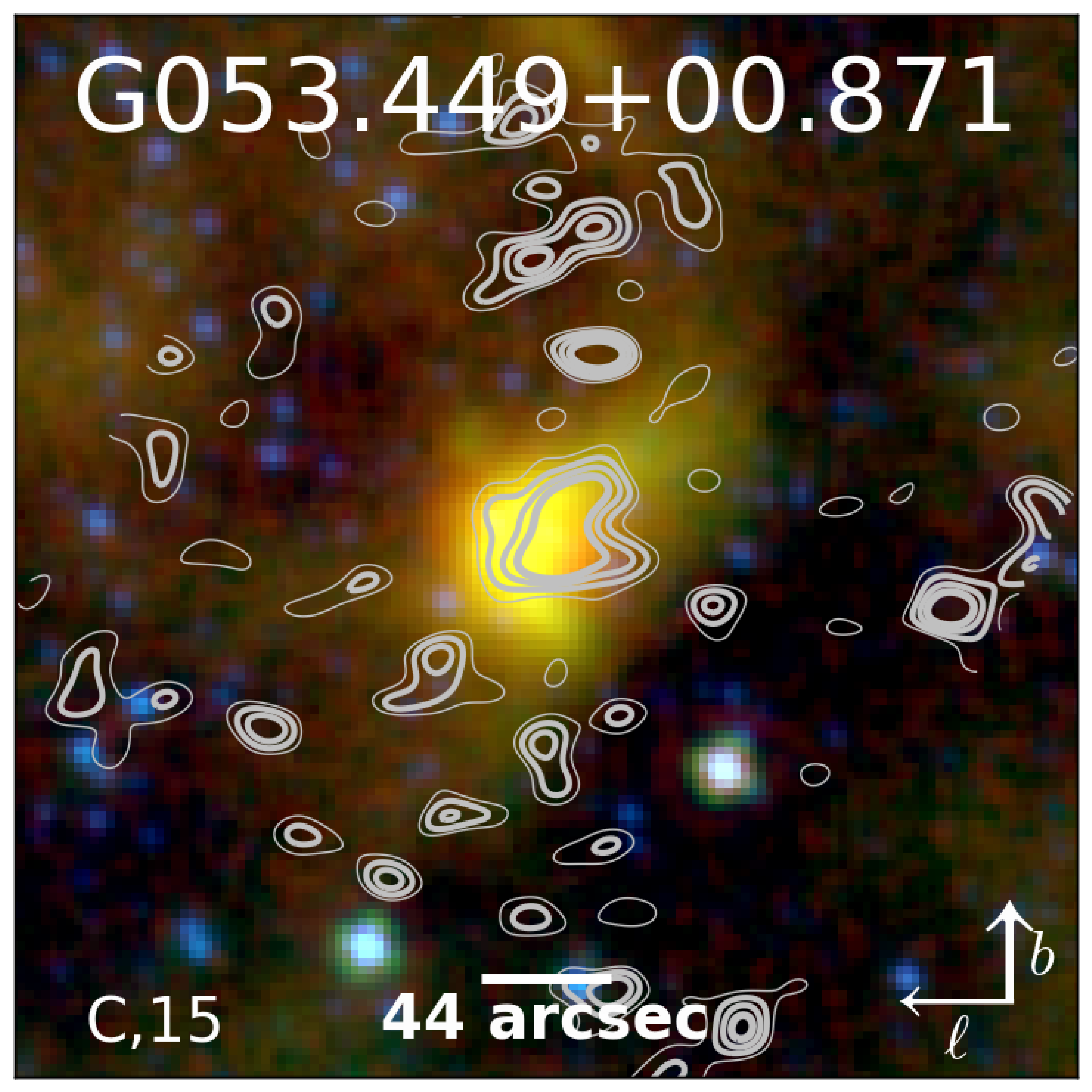}
\includegraphics[width=\figSize]{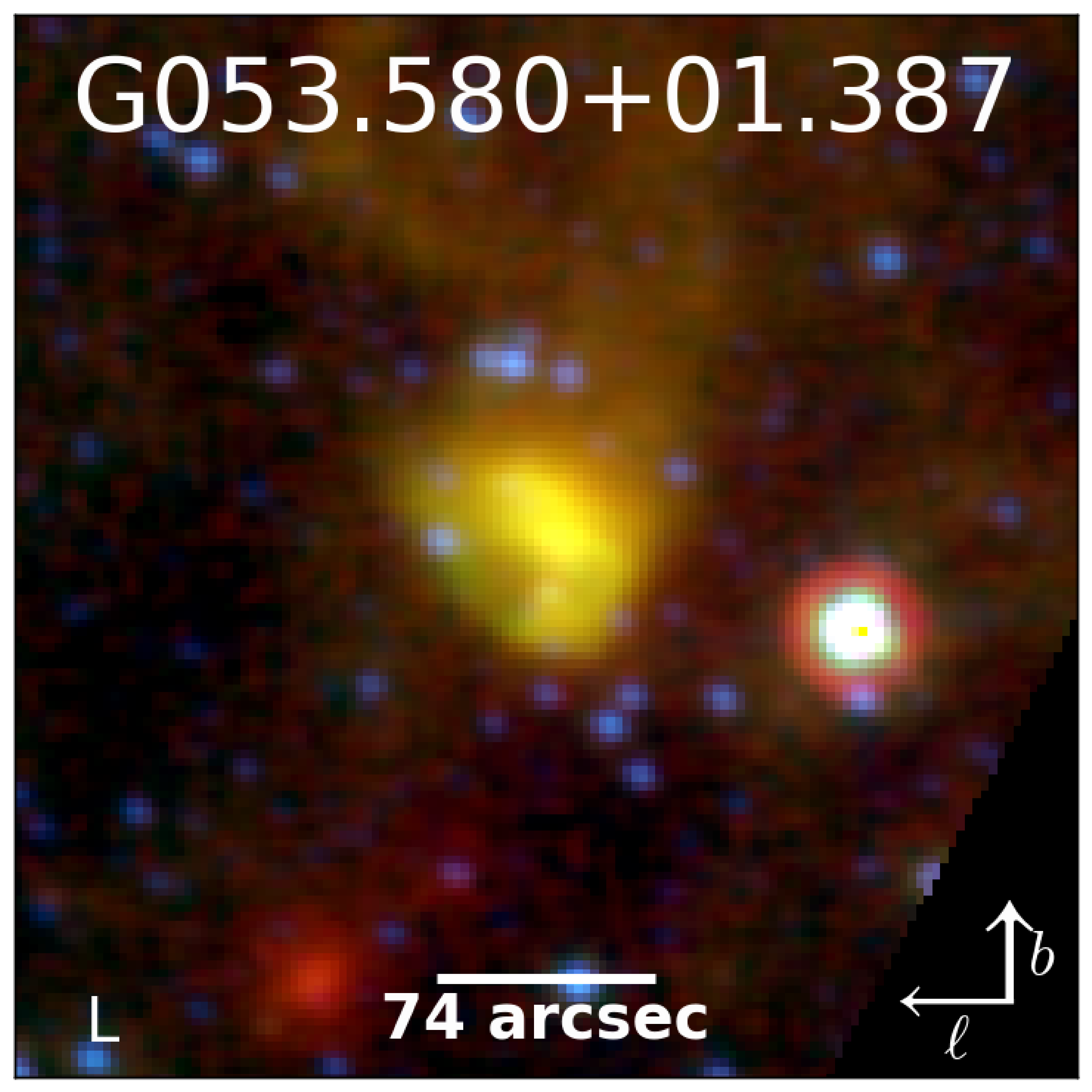}\\
\includegraphics[width=\figSize]{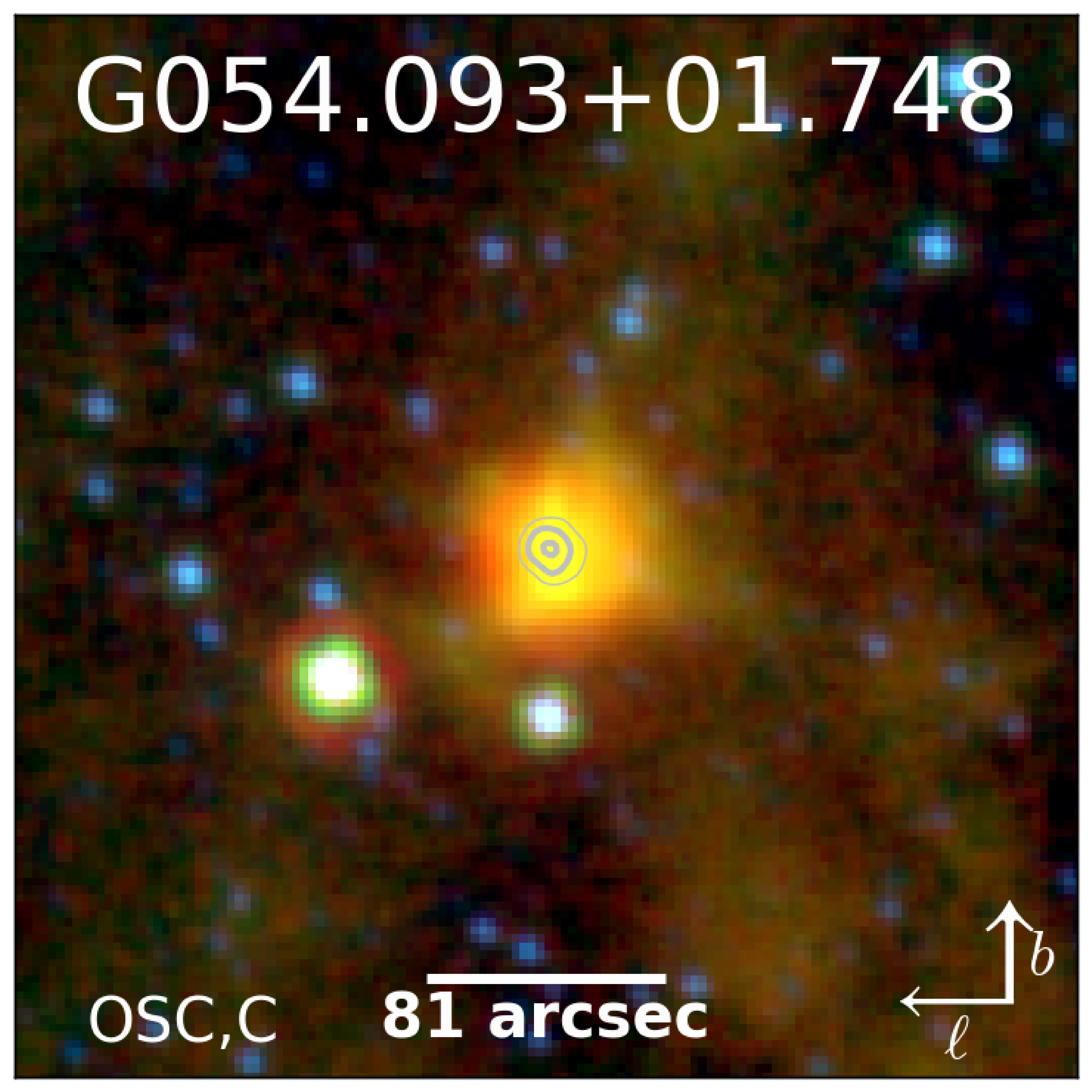}
\includegraphics[width=\figSize]{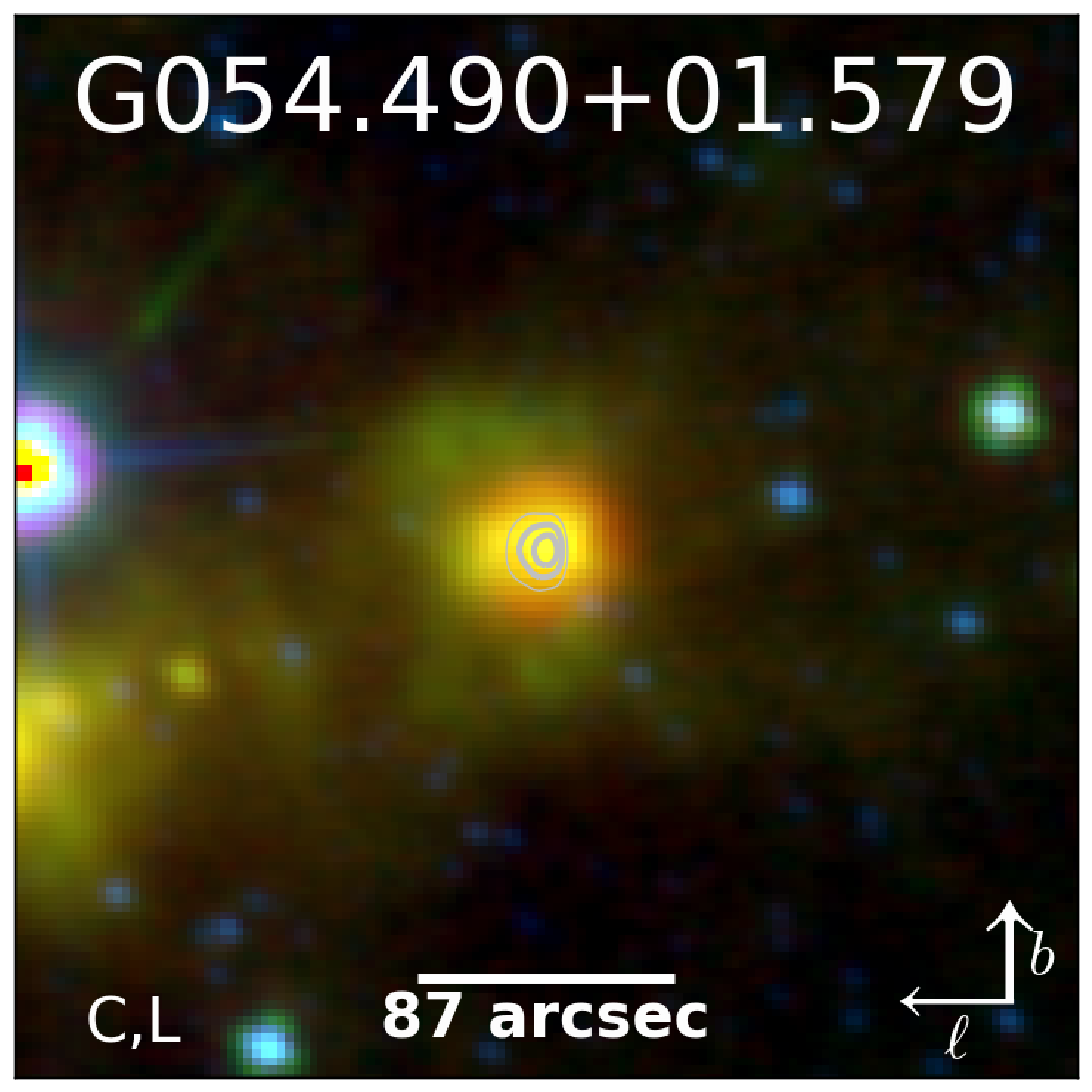}
\includegraphics[width=\figSize]{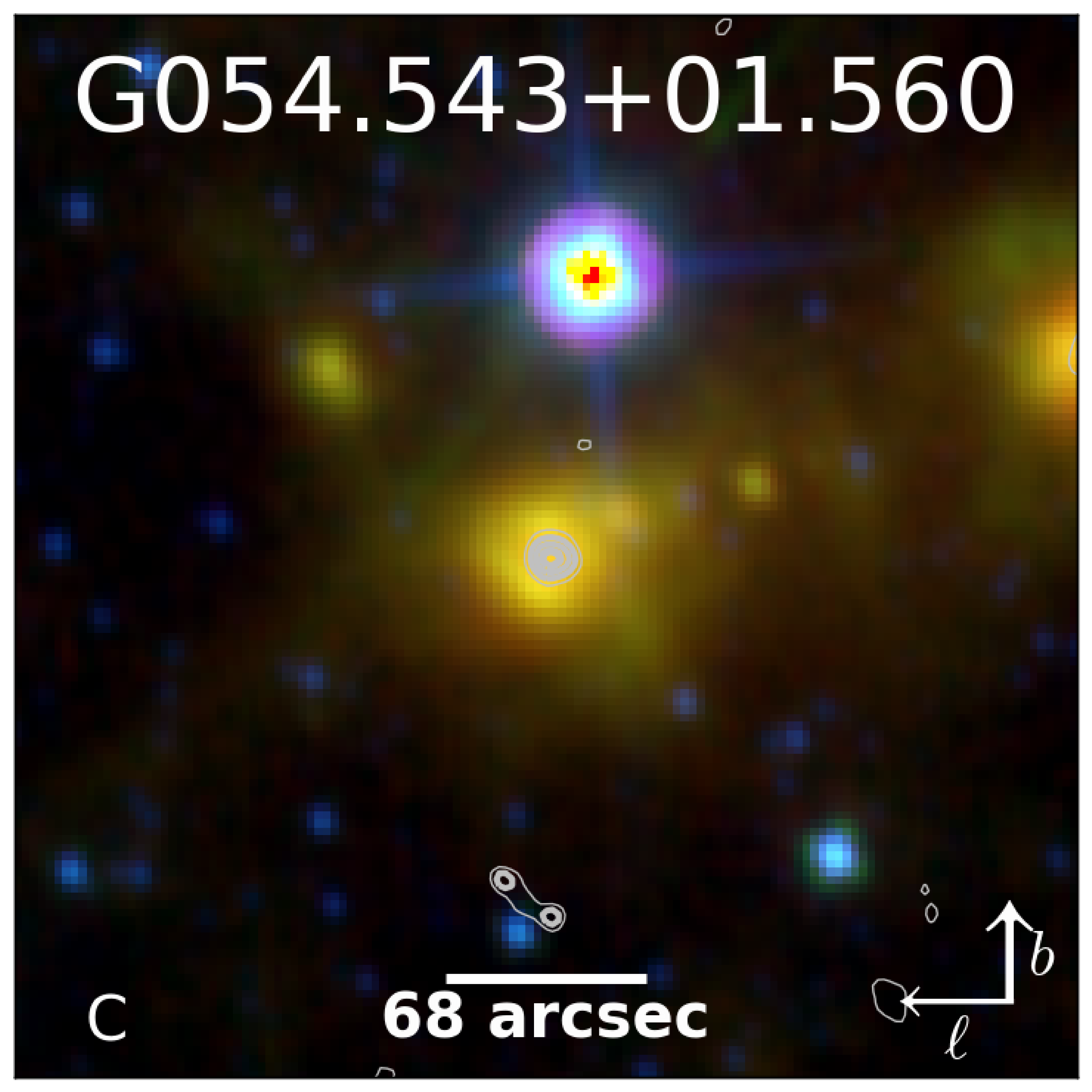}
\end{figure*}
\begin{figure*}[!htb]
\includegraphics[width=\figSize]{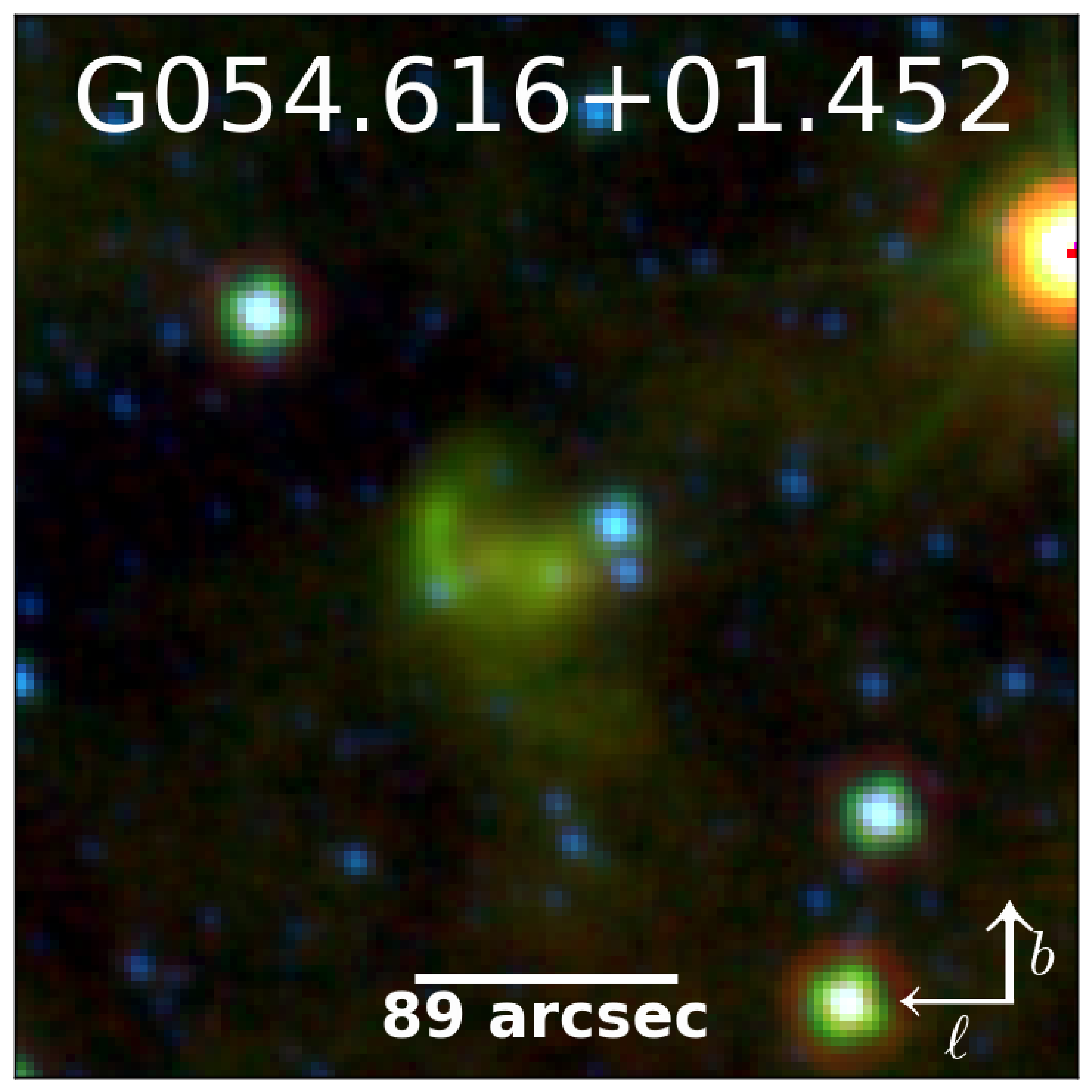}
\includegraphics[width=\figSize]{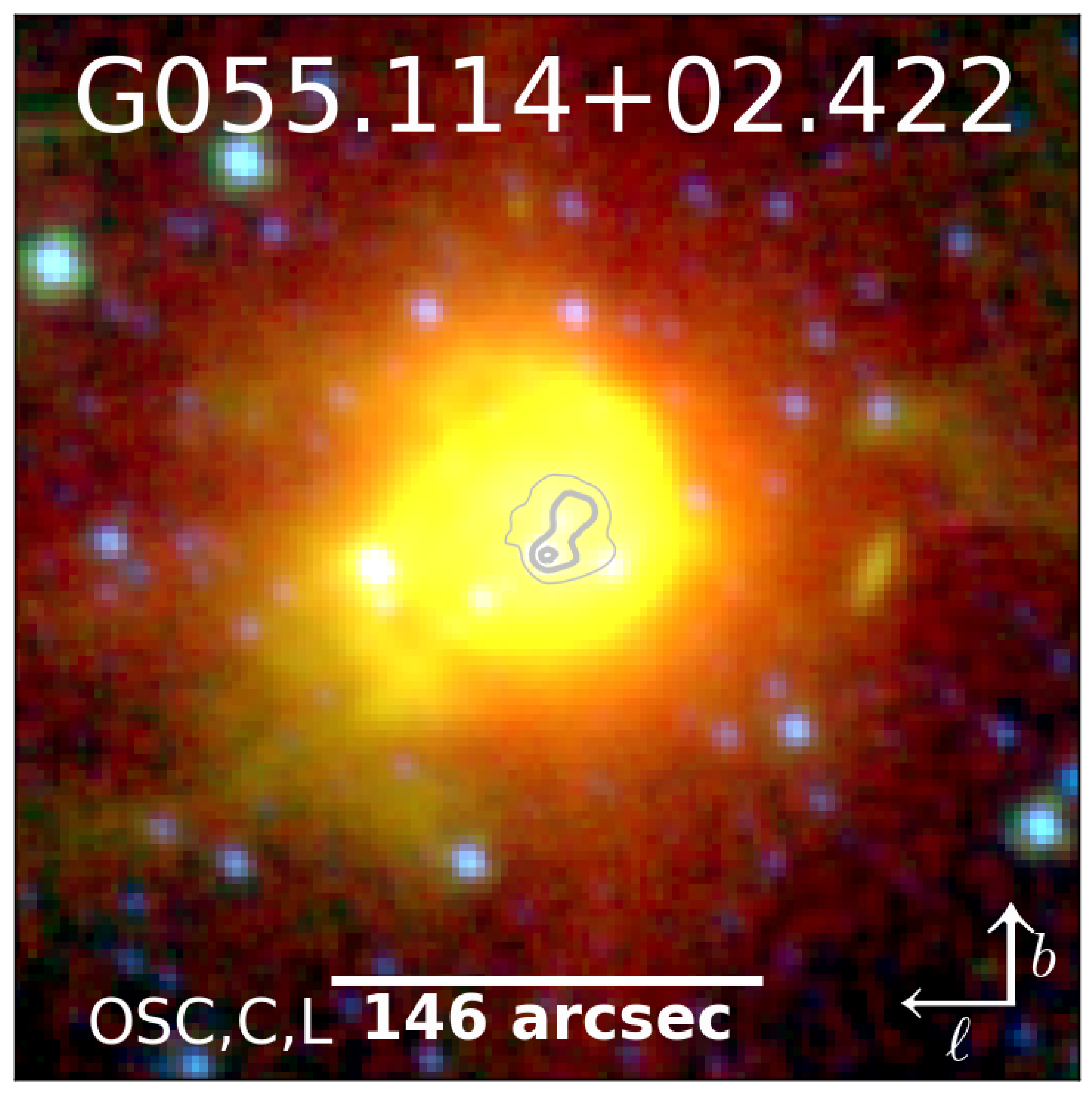}
\includegraphics[width=\figSize]{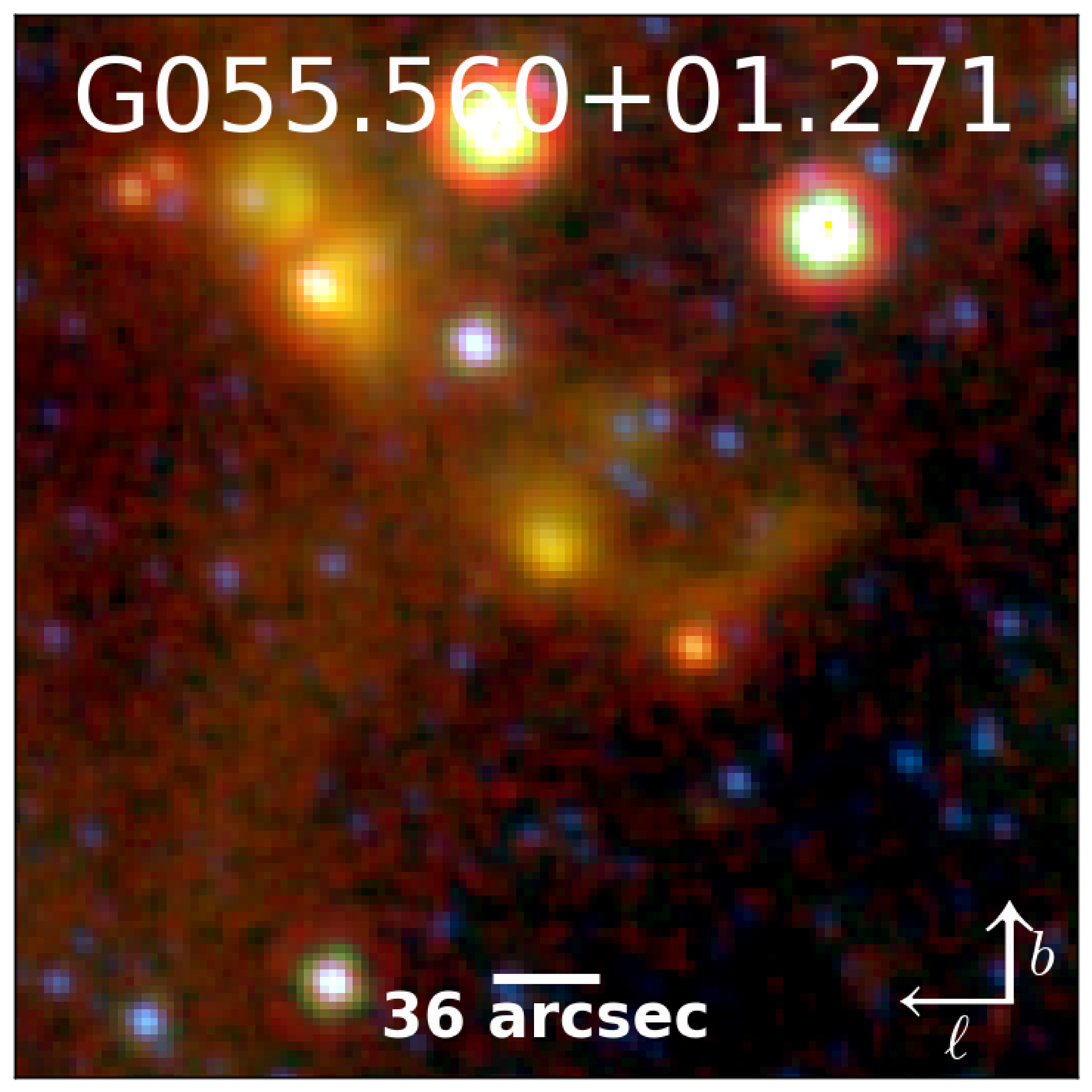}\\
\includegraphics[width=\figSize]{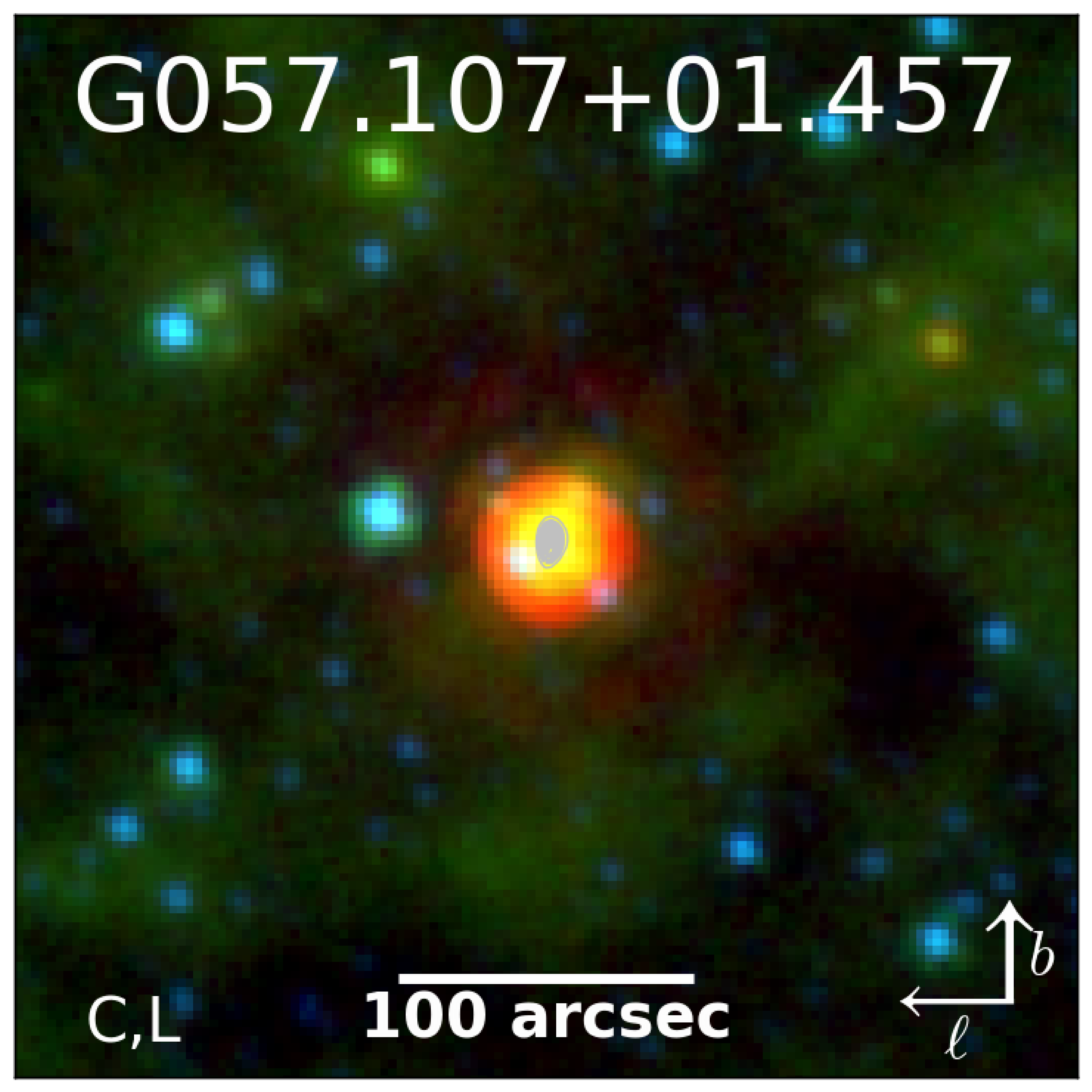}
\includegraphics[width=\figSize]{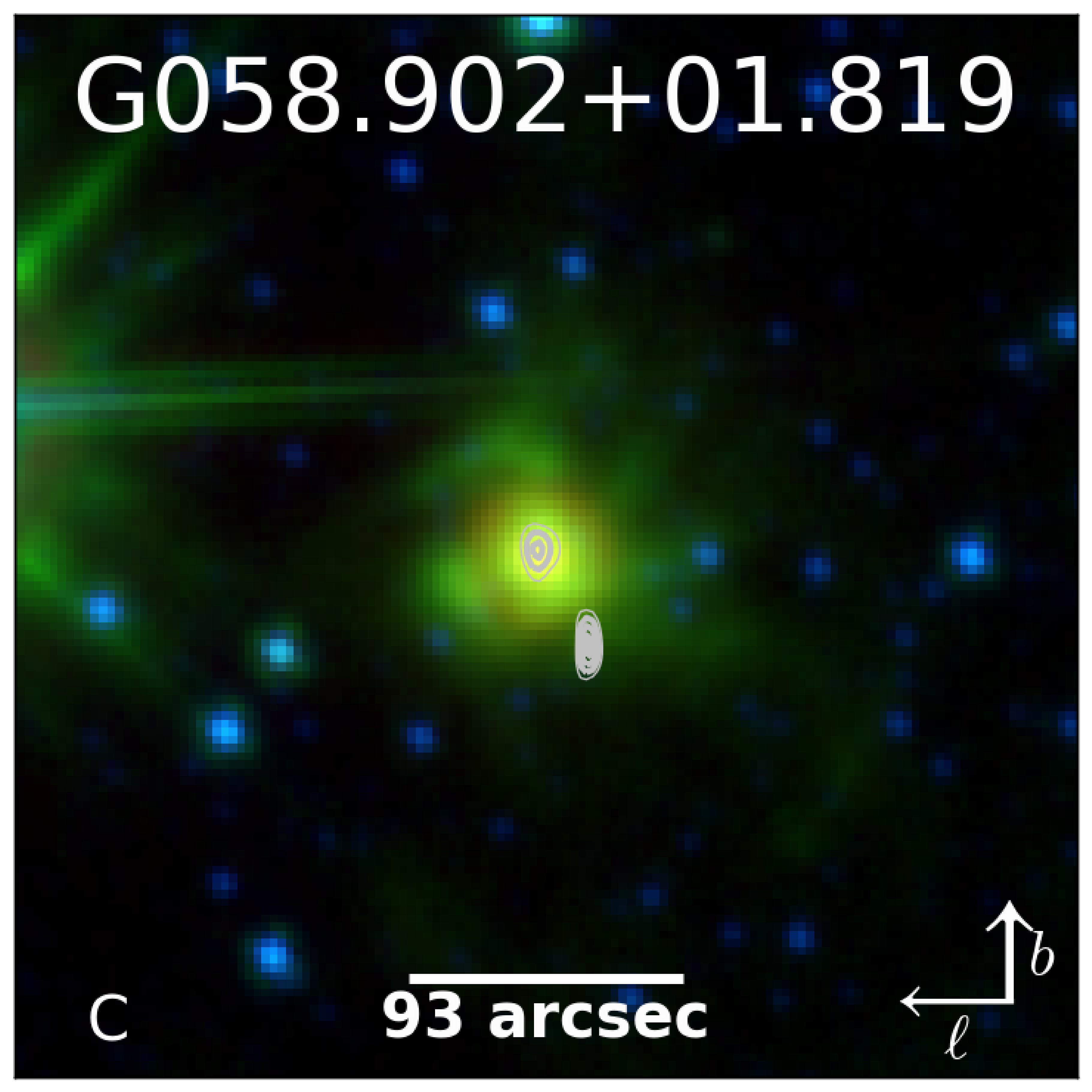}
\includegraphics[width=\figSize]{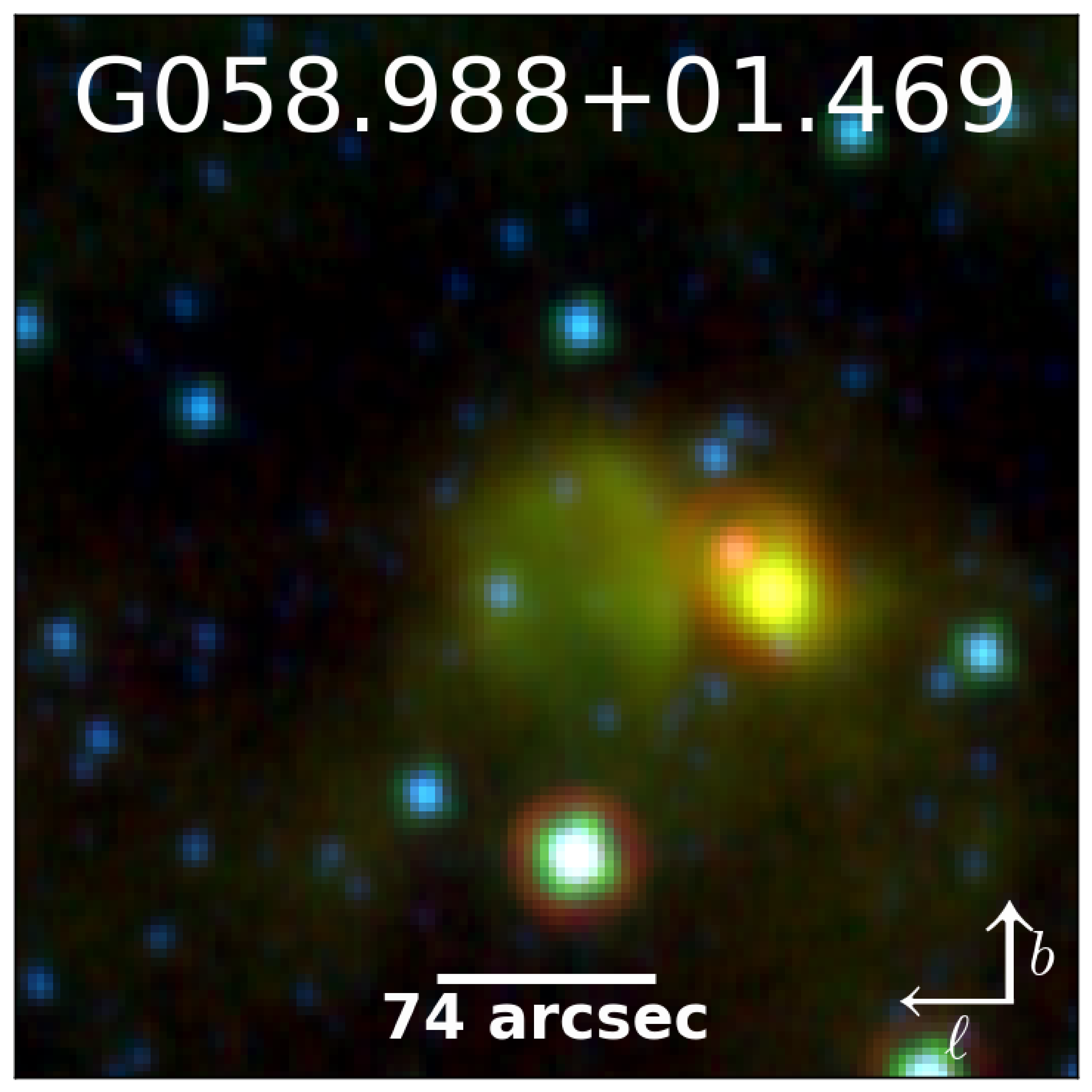}\\
\includegraphics[width=\figSize]{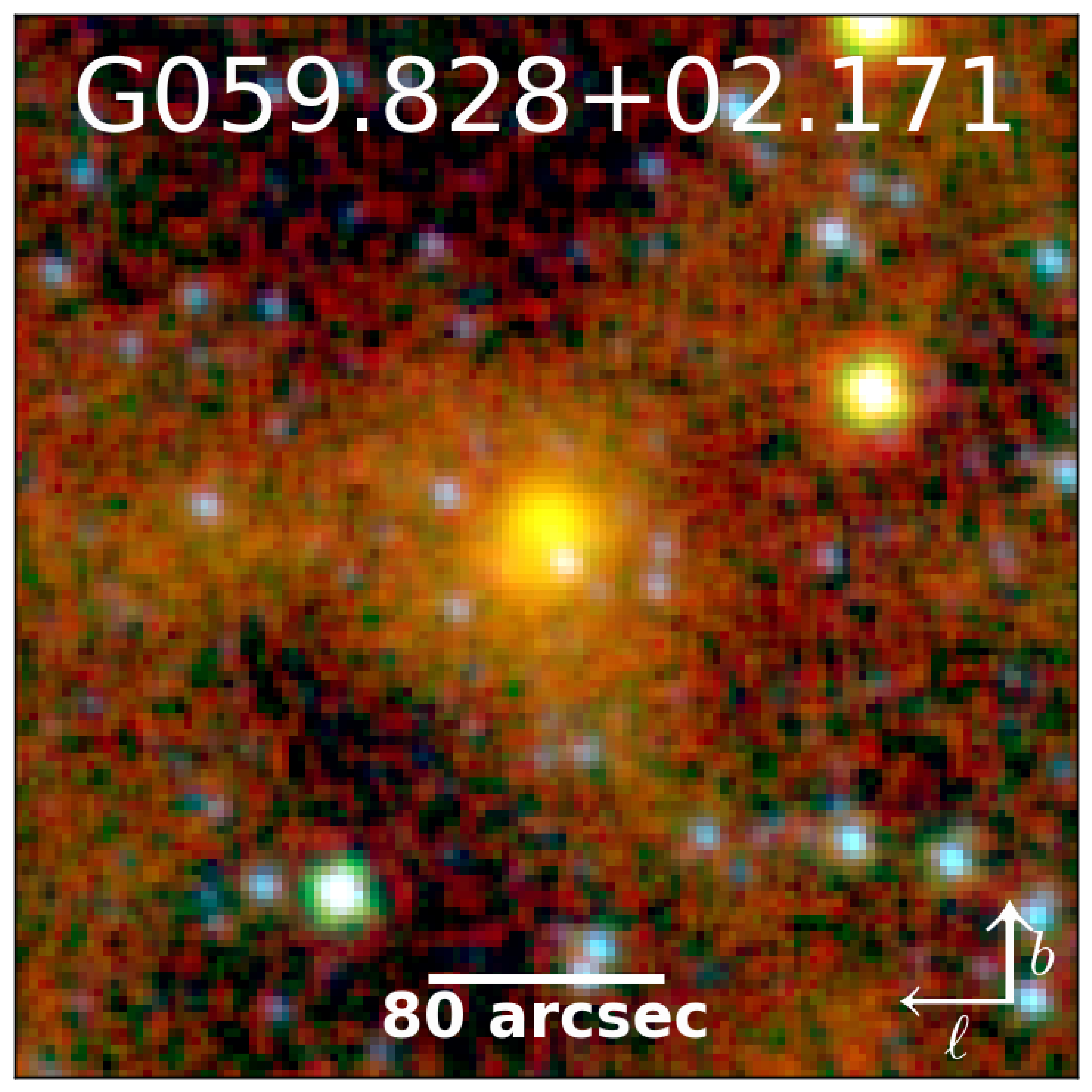}
\includegraphics[width=\figSize]{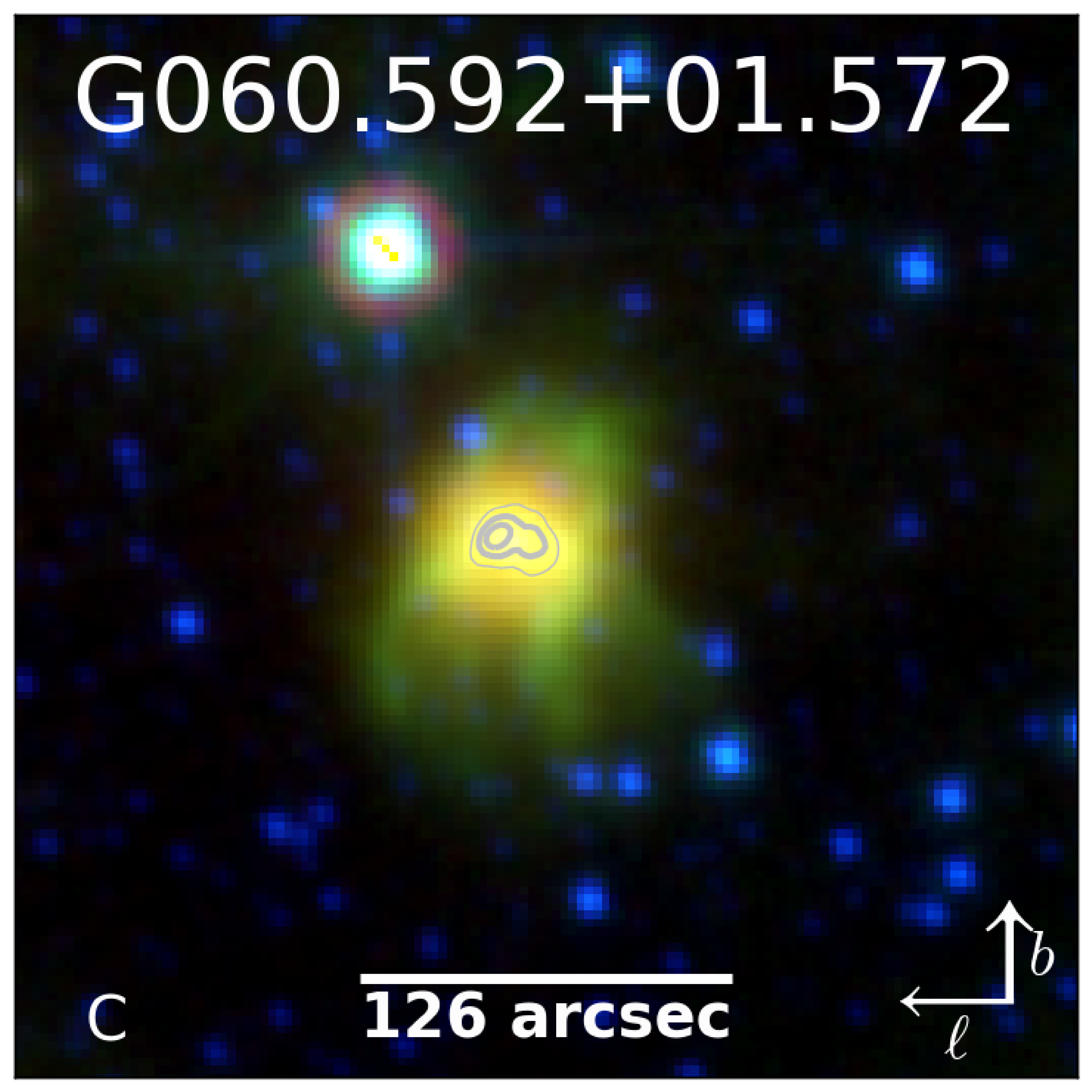}
\includegraphics[width=\figSize]{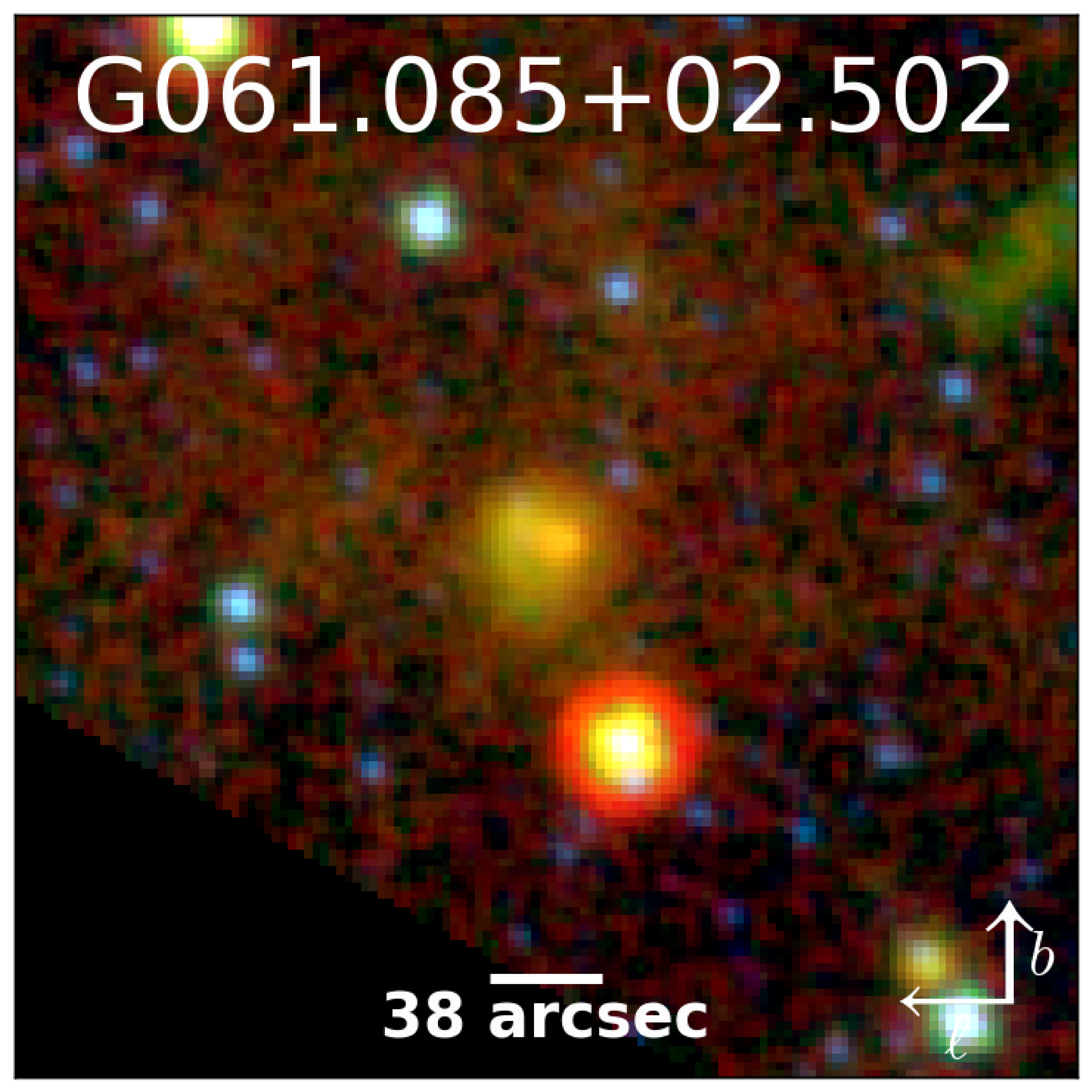}\\
\includegraphics[width=\figSize]{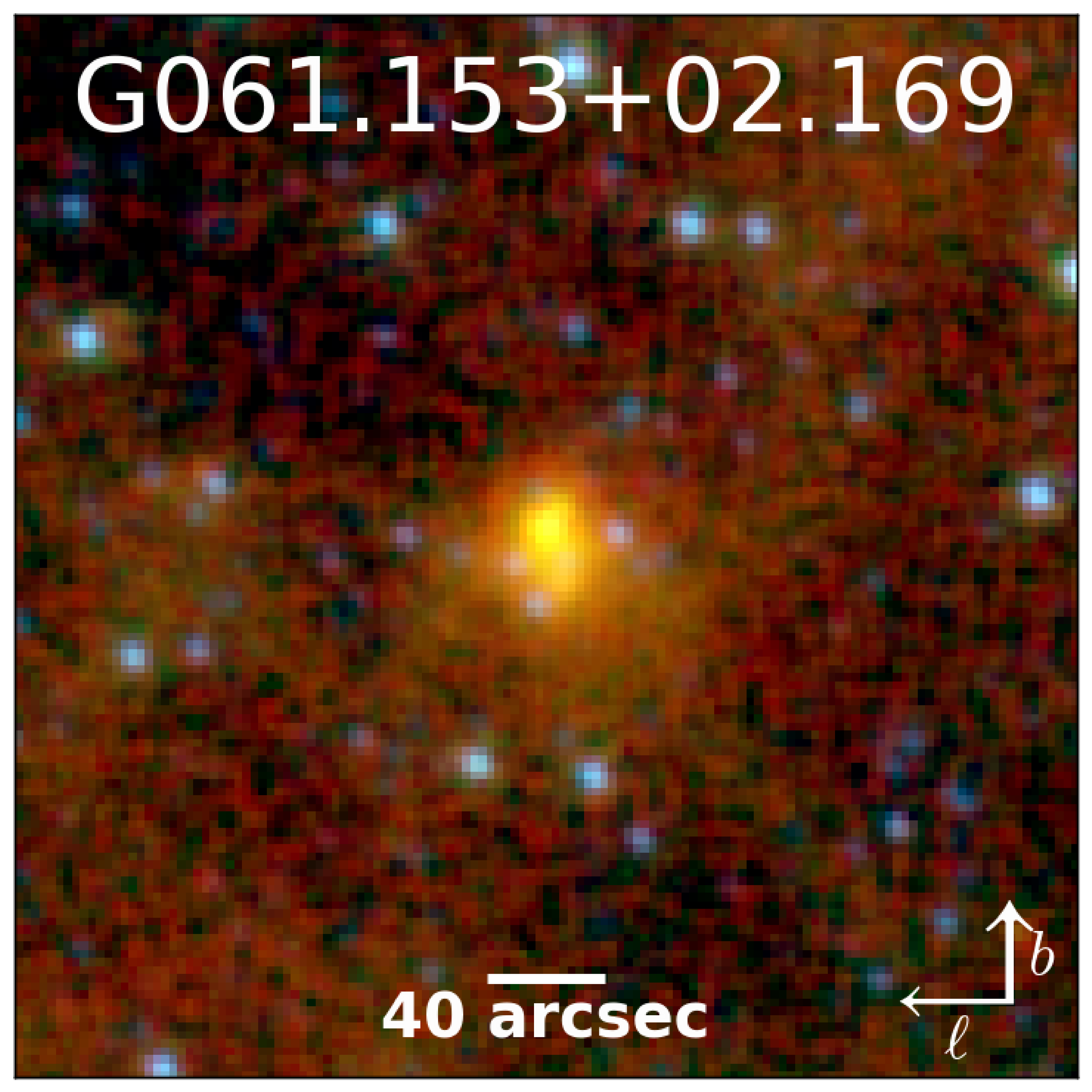}
\includegraphics[width=\figSize]{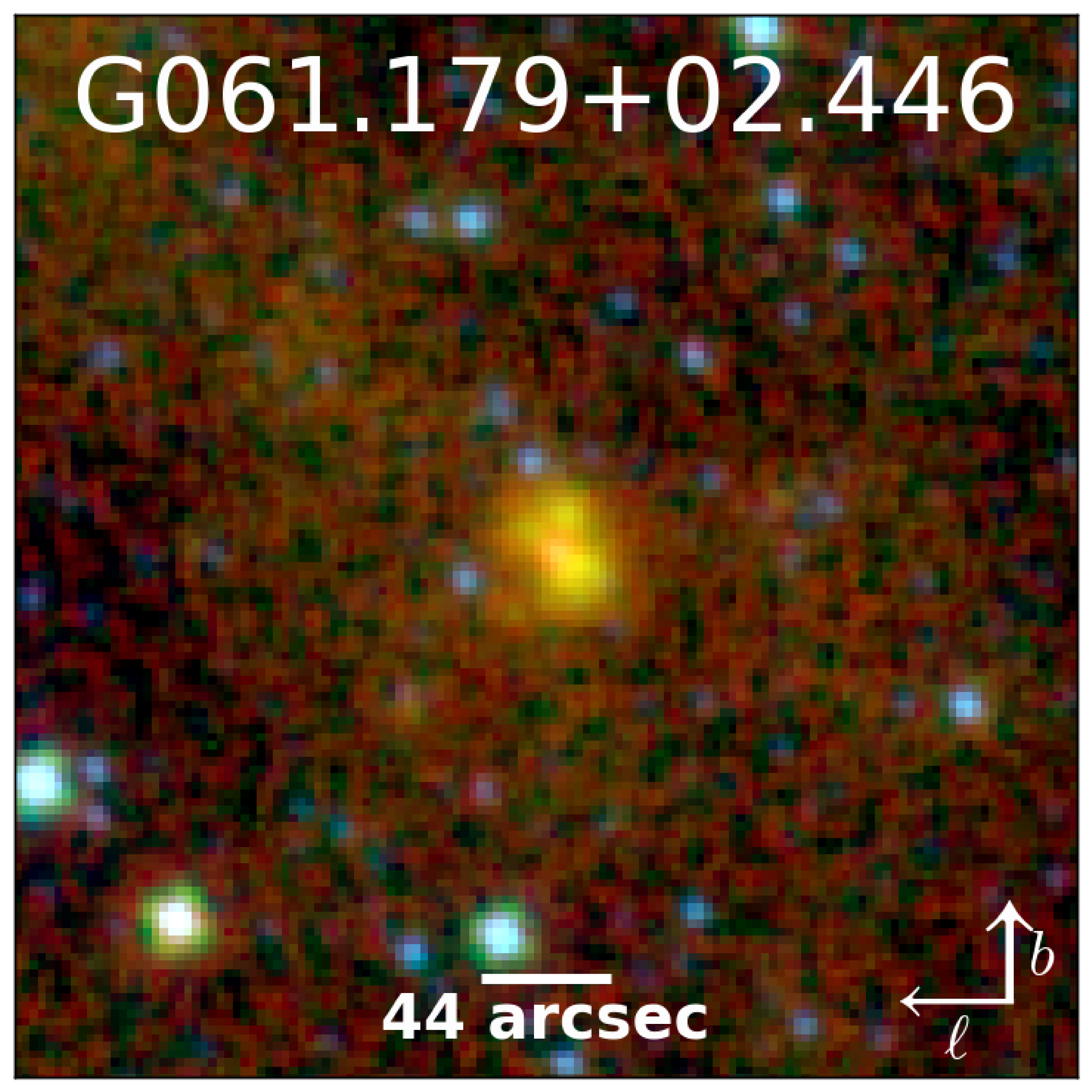}
\includegraphics[width=\figSize]{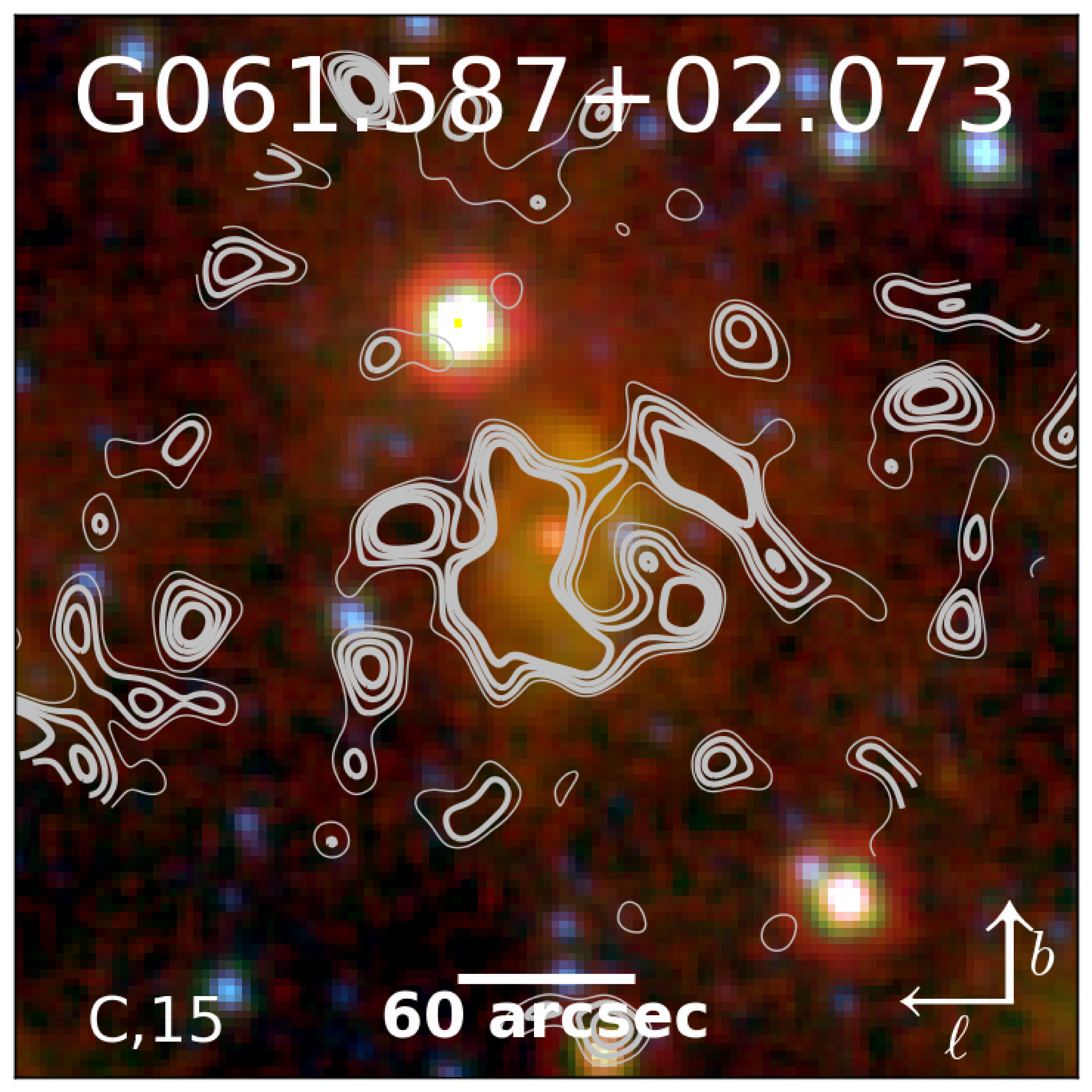}
\end{figure*}
\begin{figure*}[!htb]
\includegraphics[width=\figSize]{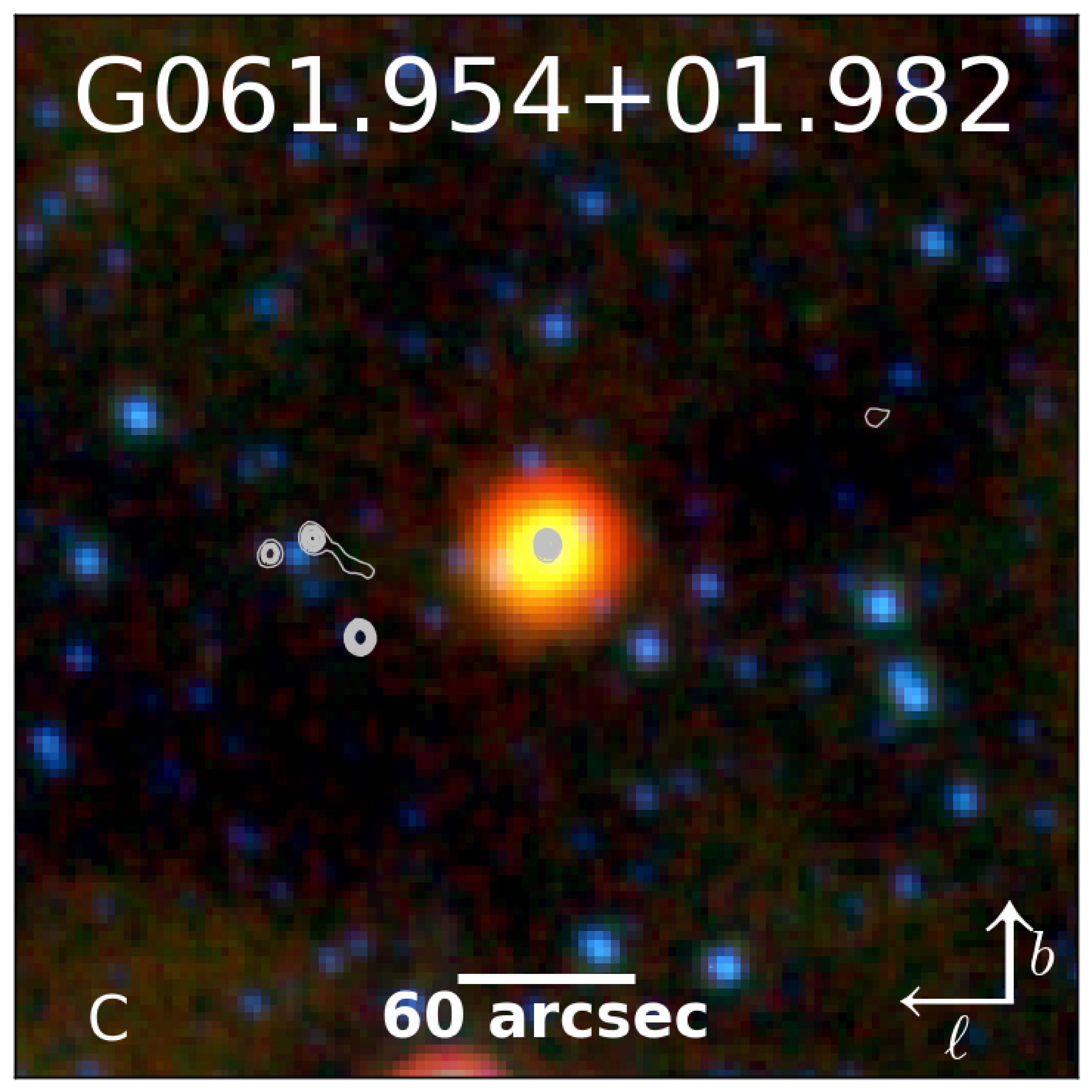}
\includegraphics[width=\figSize]{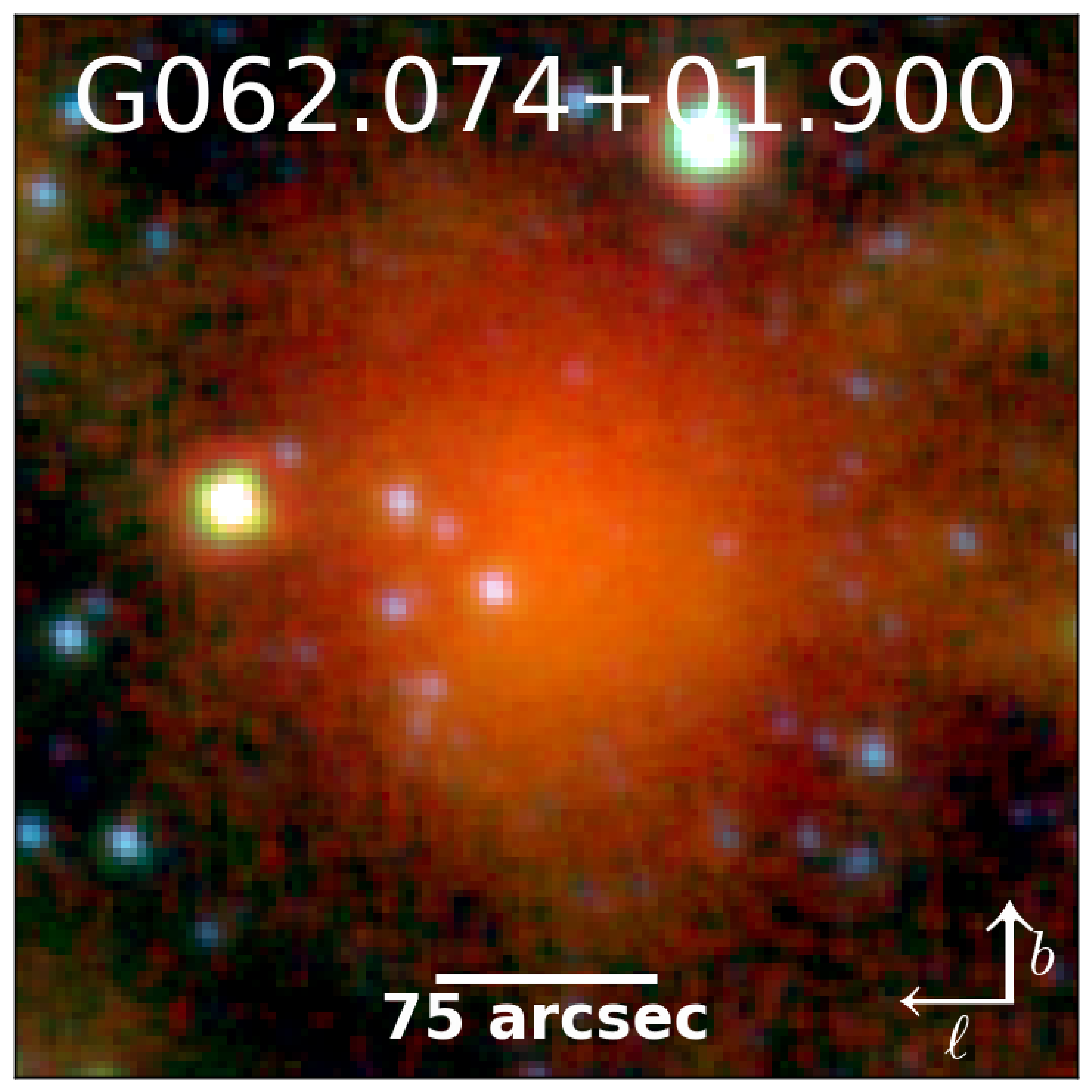}
\includegraphics[width=\figSize]{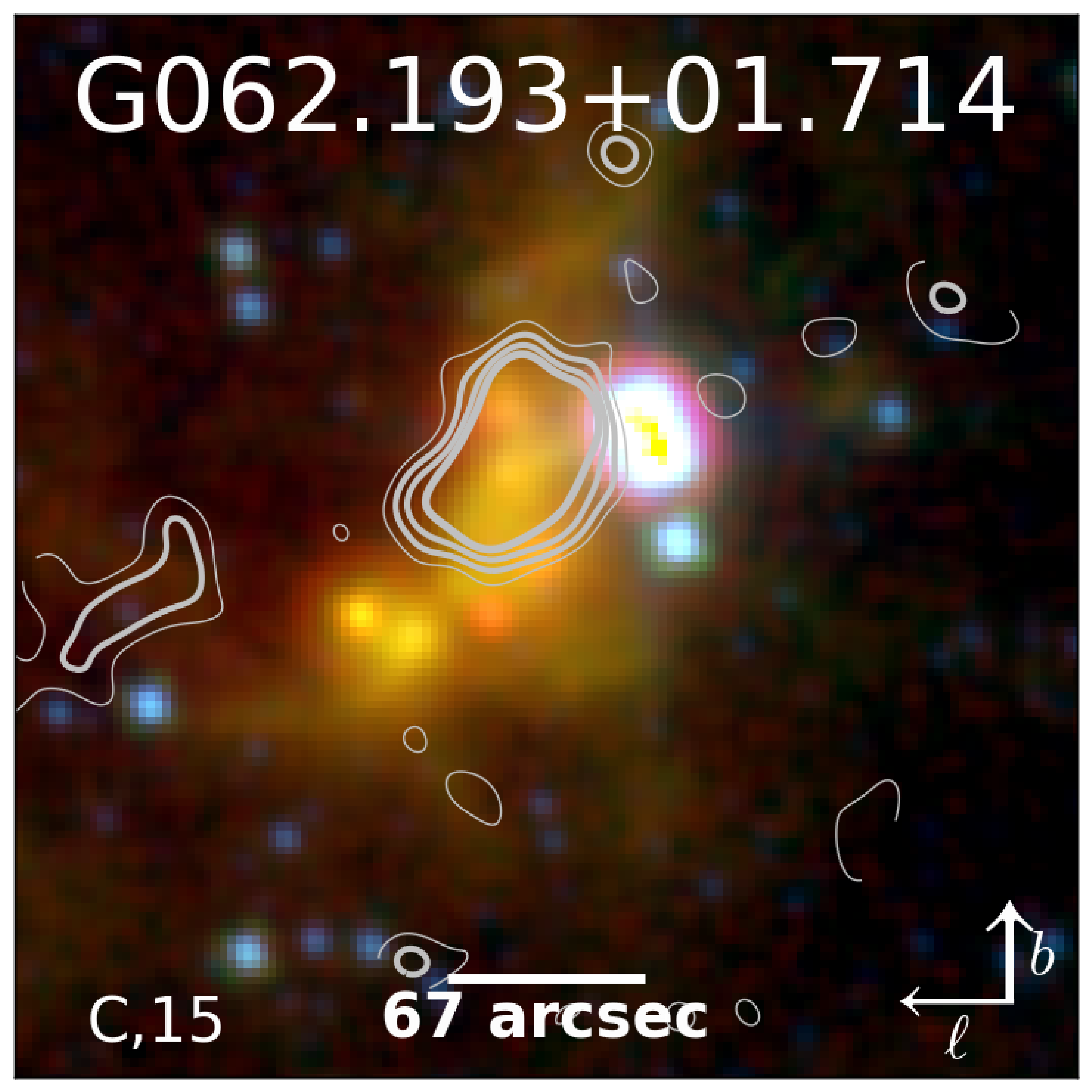}\\
\includegraphics[width=\figSize]{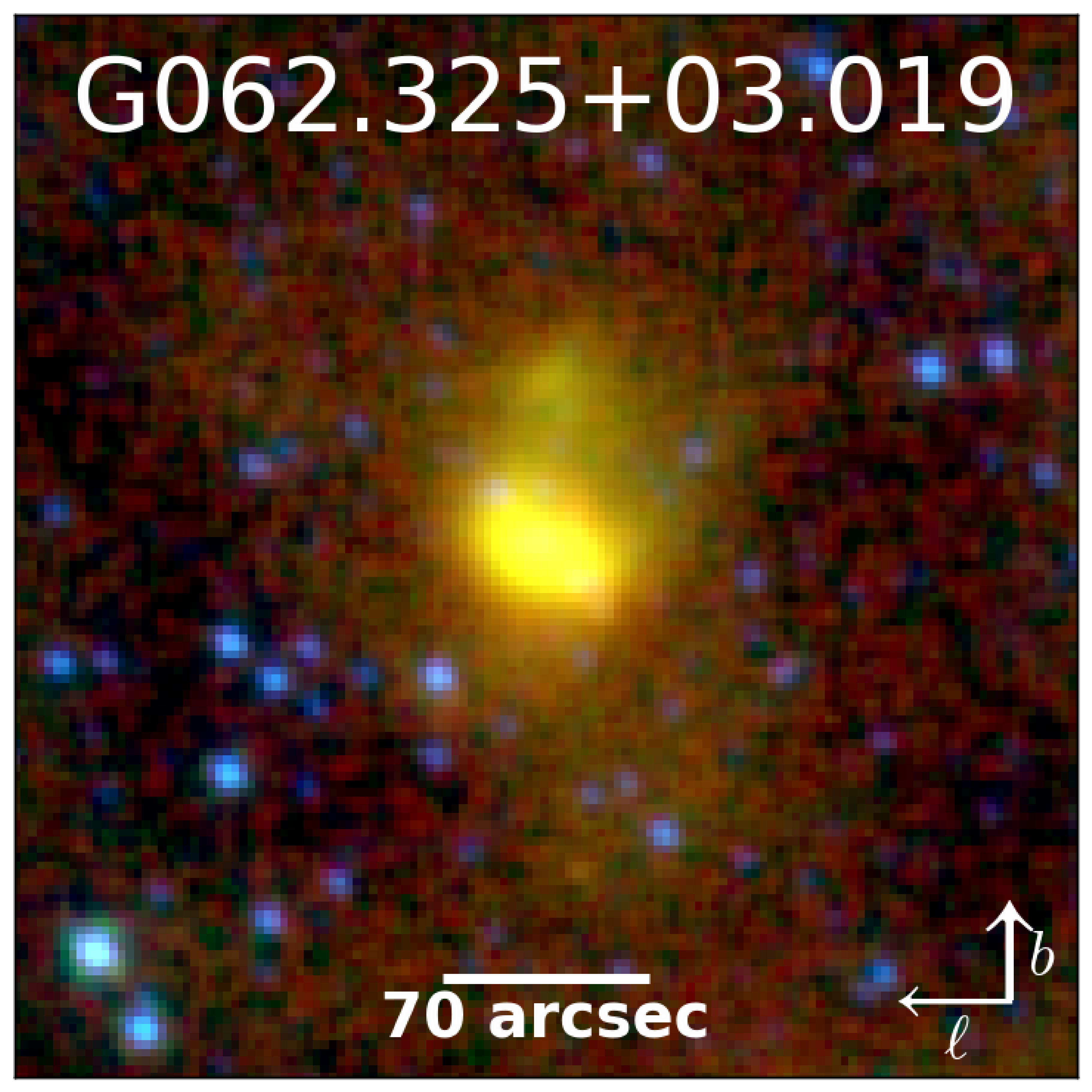}
\includegraphics[width=\figSize]{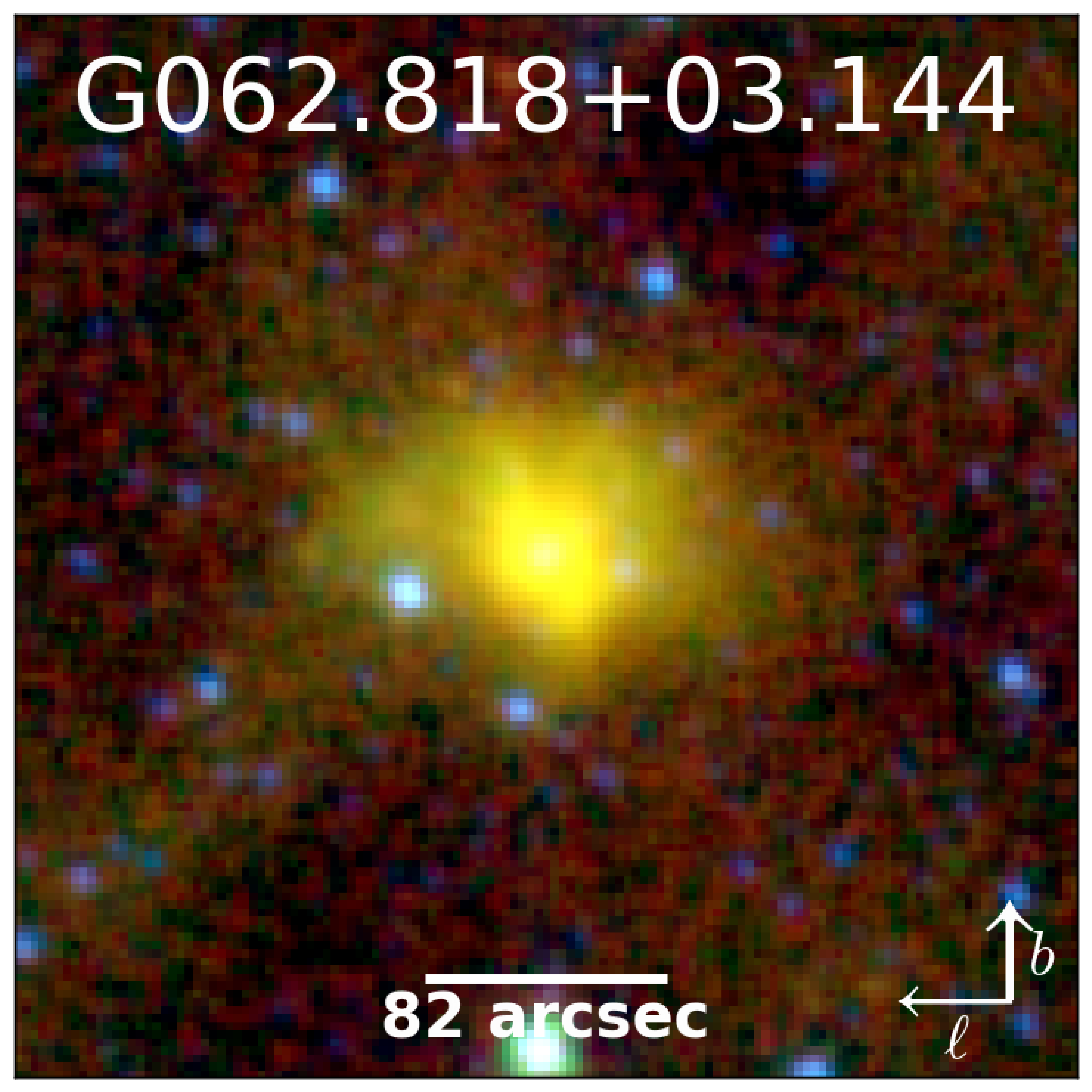}
\includegraphics[width=\figSize]{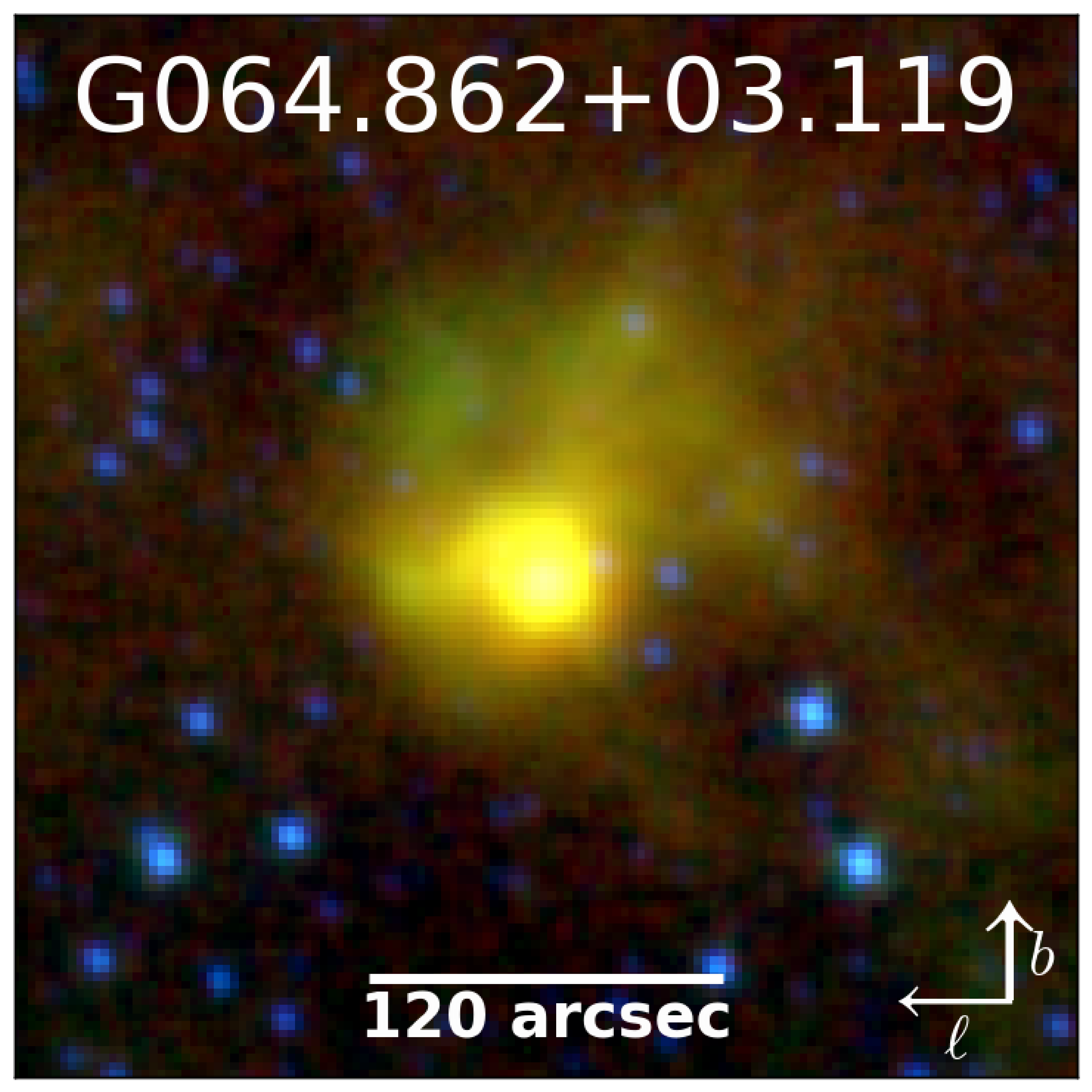}\\
\includegraphics[width=\figSize]{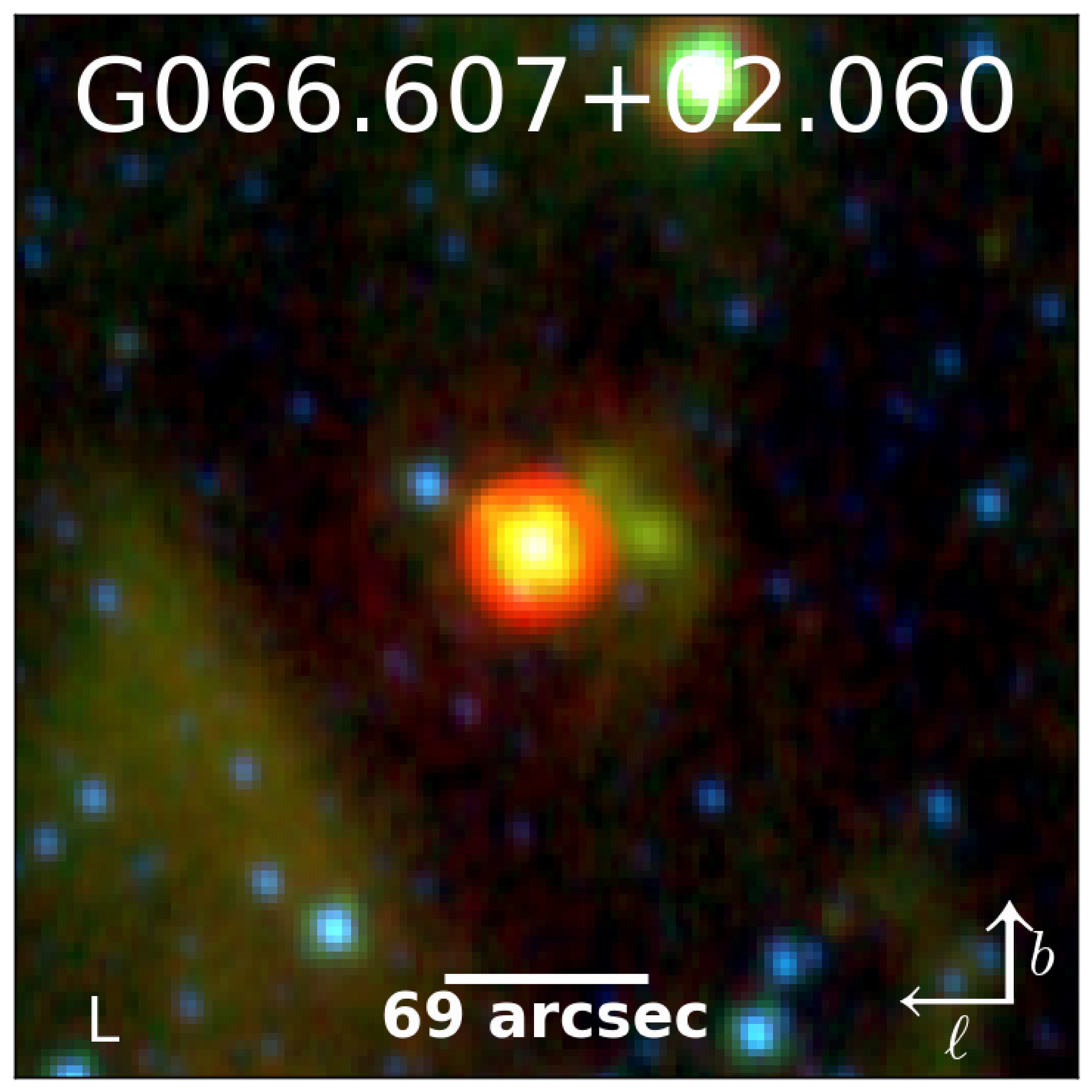}
\includegraphics[width=\figSize]{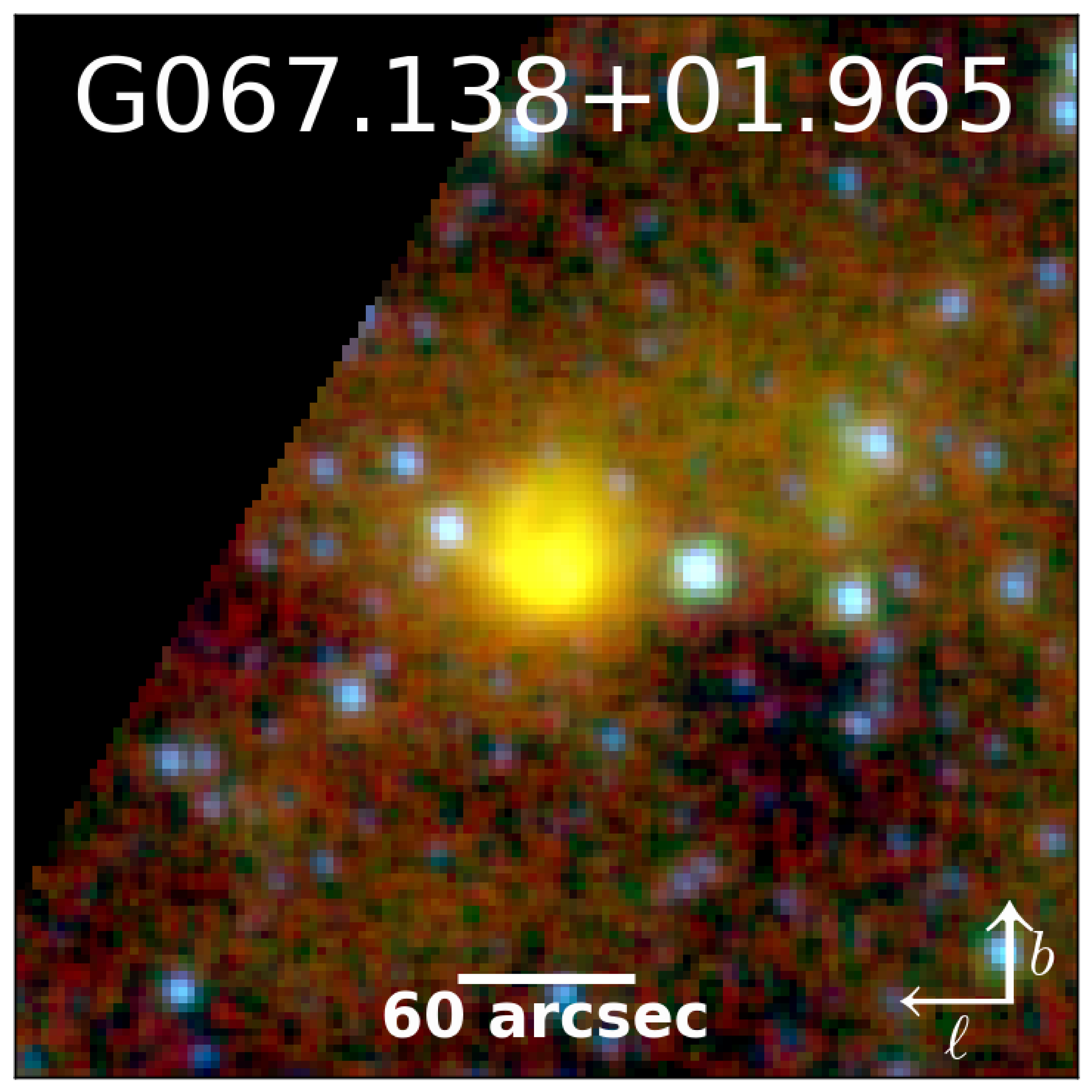}
\end{figure*}

\newcommand{\figTwoSize}{0.5\textwidth}
\pagebreak
\begin{figure*}[!ht]
\caption{\hii\ region molecular spectra. Shown here are four-panel images of spectra for all sources detected in either water maser (H$_2$O 6(1,6)$\rightarrow$5(2,3)) or ammonia (NH$_{3}$ (J,K)=(1,1), (2,2), (3,3)) emission. All intensities are given as the antenna temperature corrected for atmospheric absorption, $T'_A$. Water maser detections (top panel) have red dashed lines indicating the upper and lower ranges of emission with a red arrow showing the velocity of peak maser emission. Ammonia detections (bottom three panels) are overlaid with red Gaussian fits to central and hyperfine emission lines. Water maser emission is smoothed to 0.3 \kms\ wheras ammonia emission is smoothed to 0.75 \kms\/. \label{fig:NH3spectra}}
\hfill\break
\includegraphics[width=\figTwoSize]{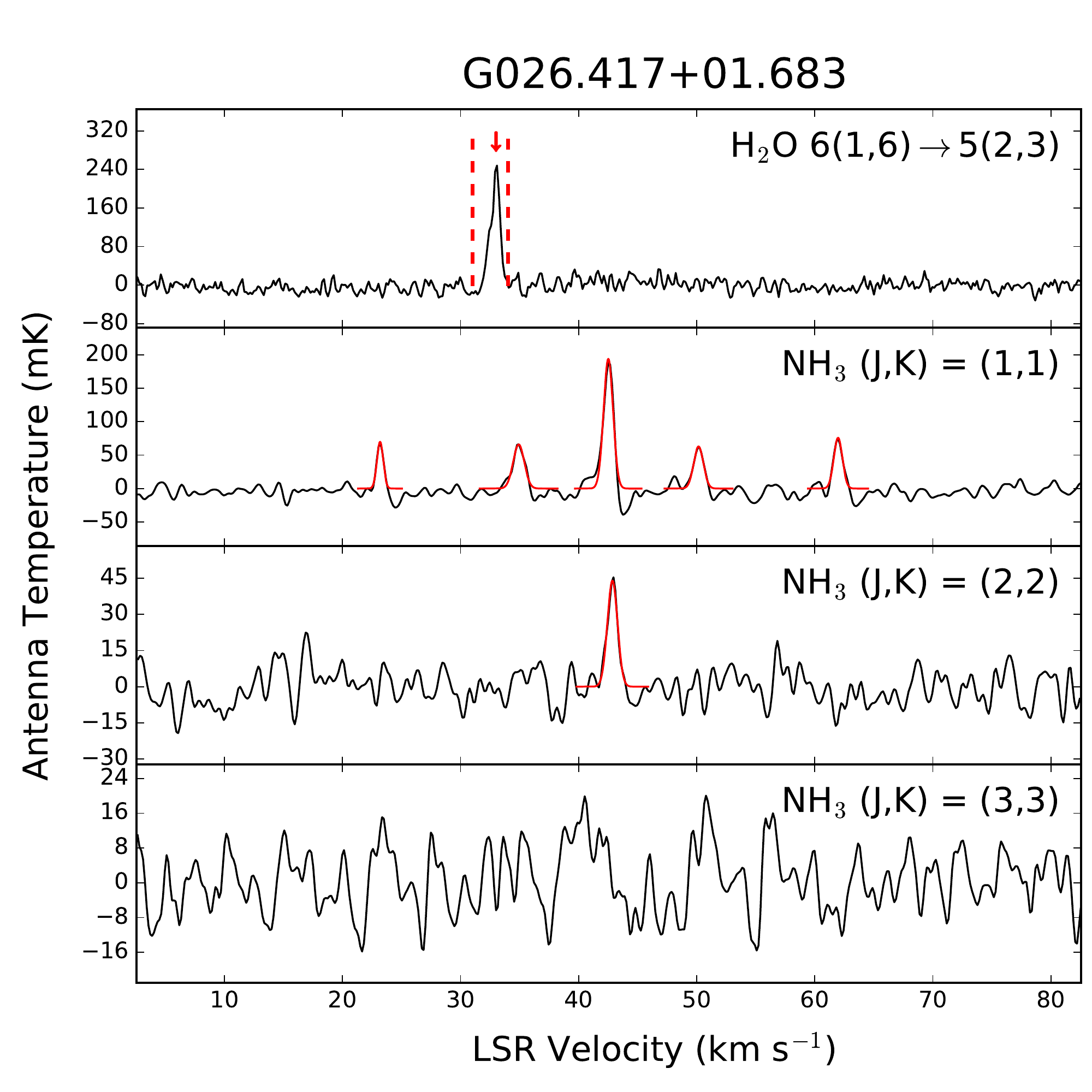}
\includegraphics[width=\figTwoSize]{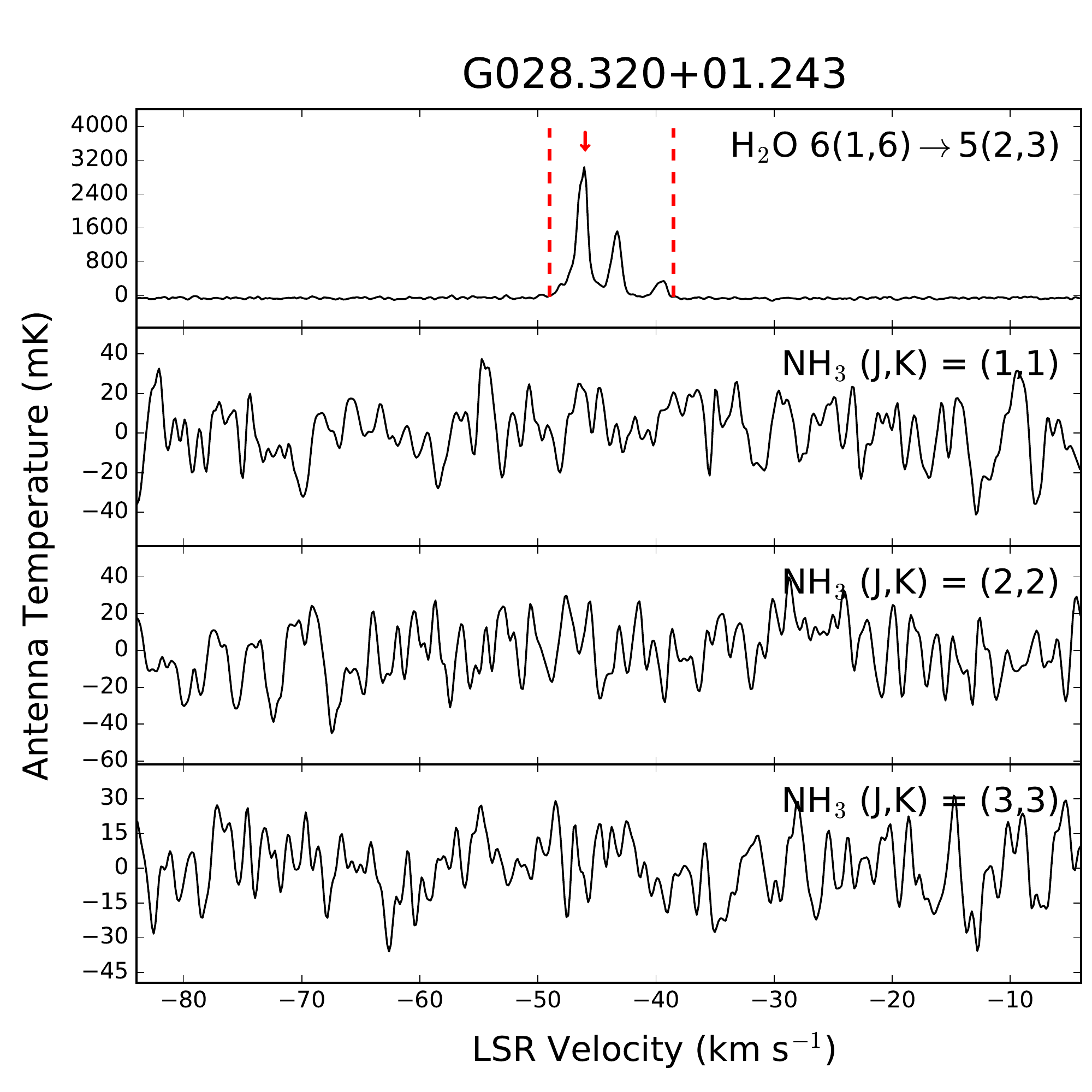}\\
\includegraphics[width=\figTwoSize]{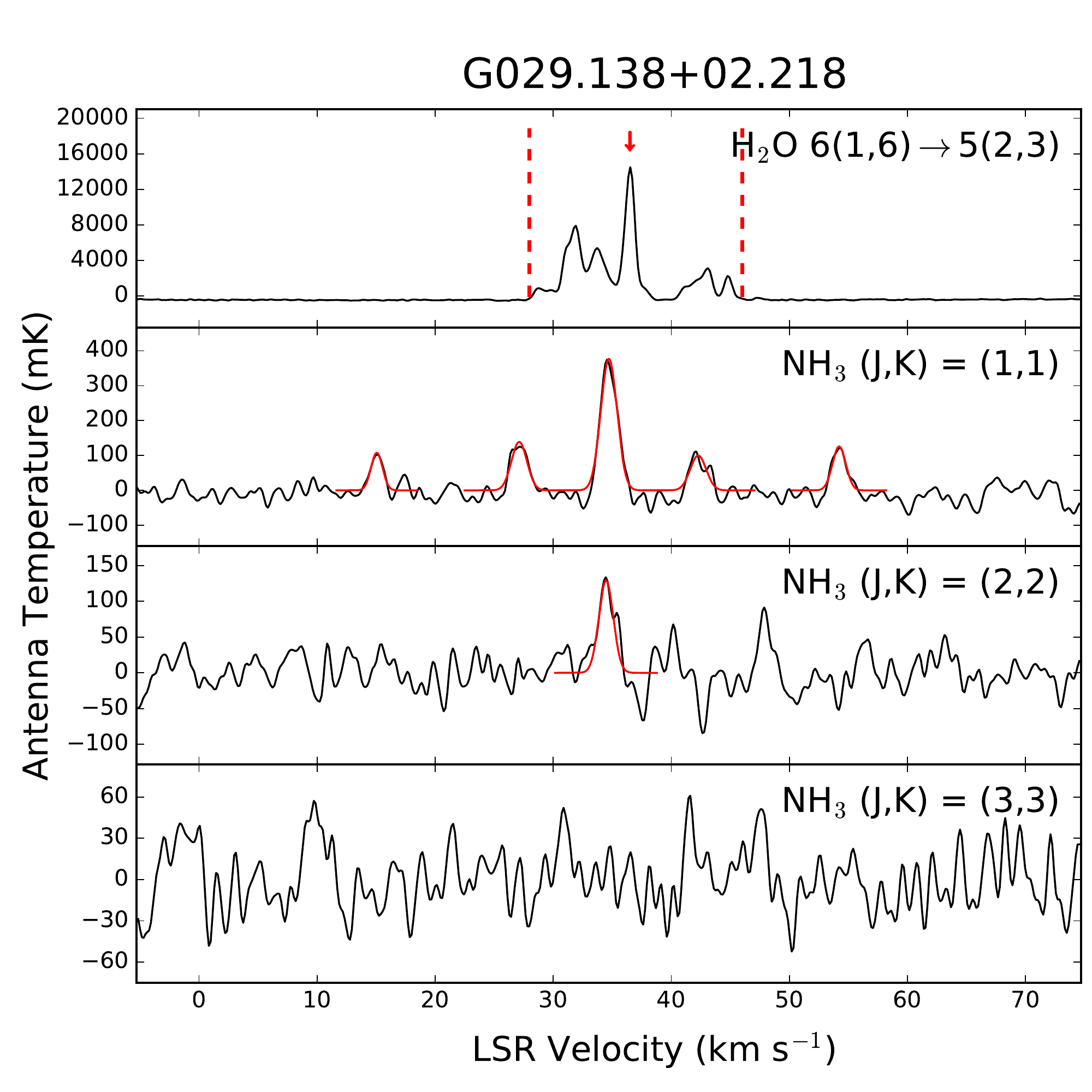}
\includegraphics[width=\figTwoSize]{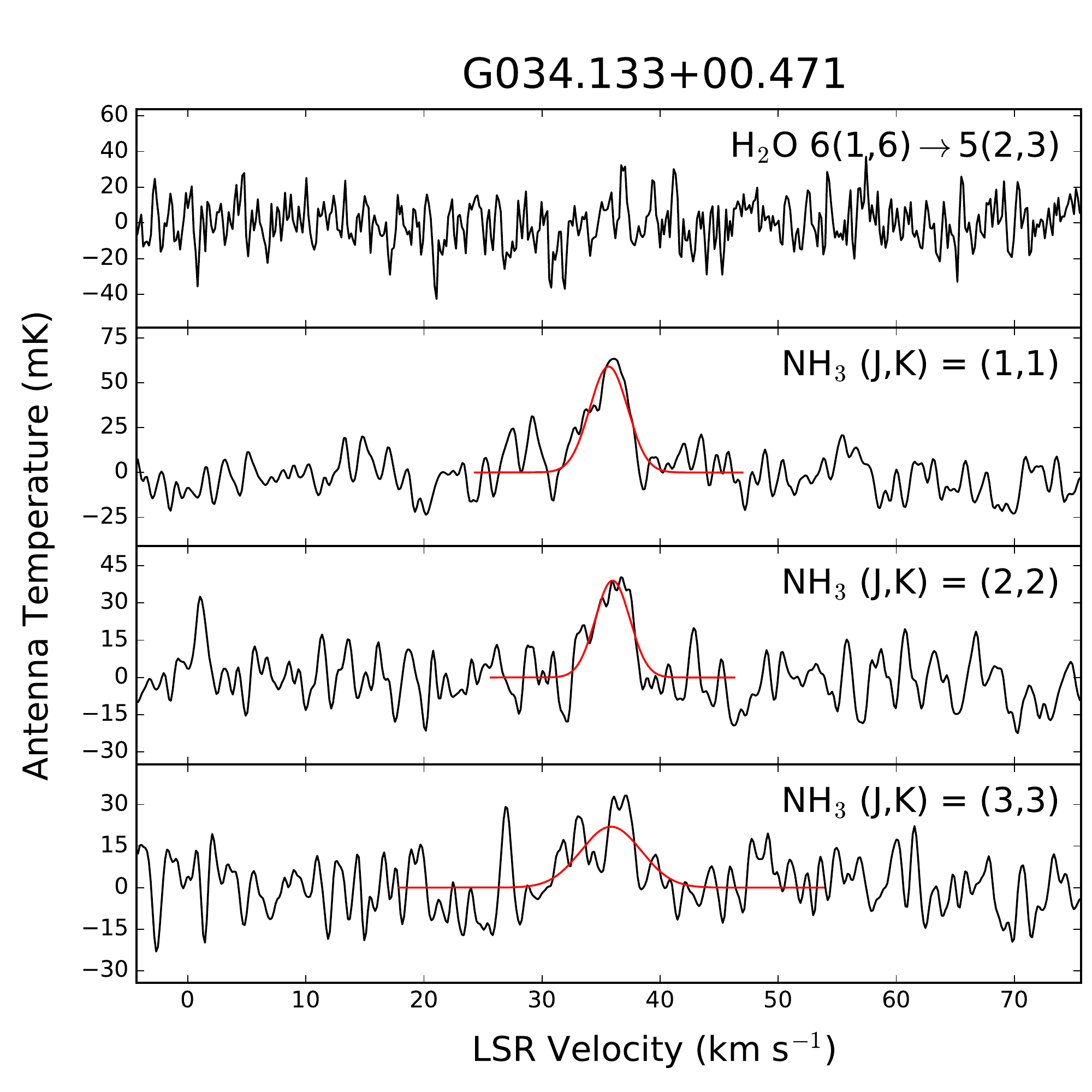}
\end{figure*}
\begin{figure*}[!htb]
\includegraphics[width=\figTwoSize]{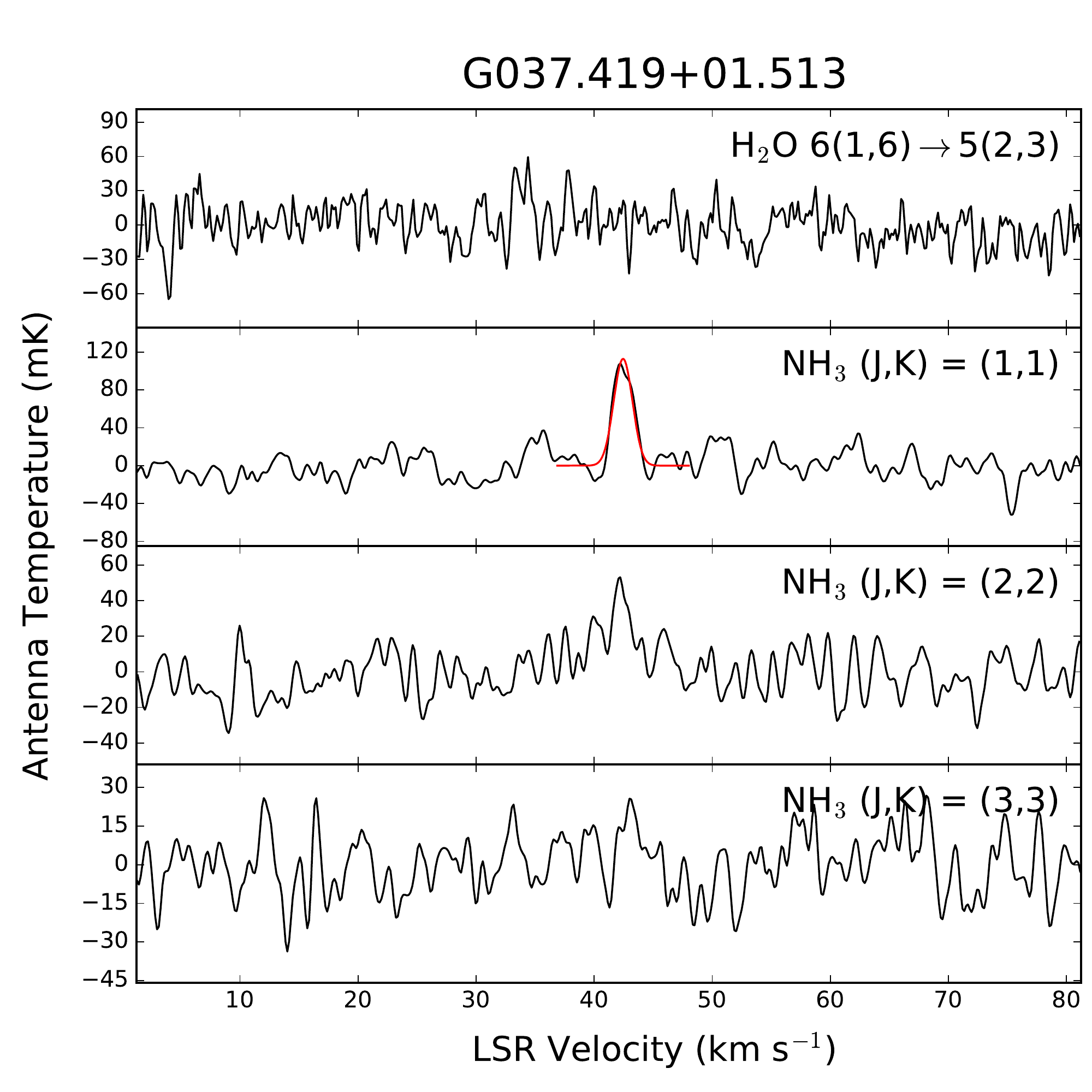}
\includegraphics[width=\figTwoSize]{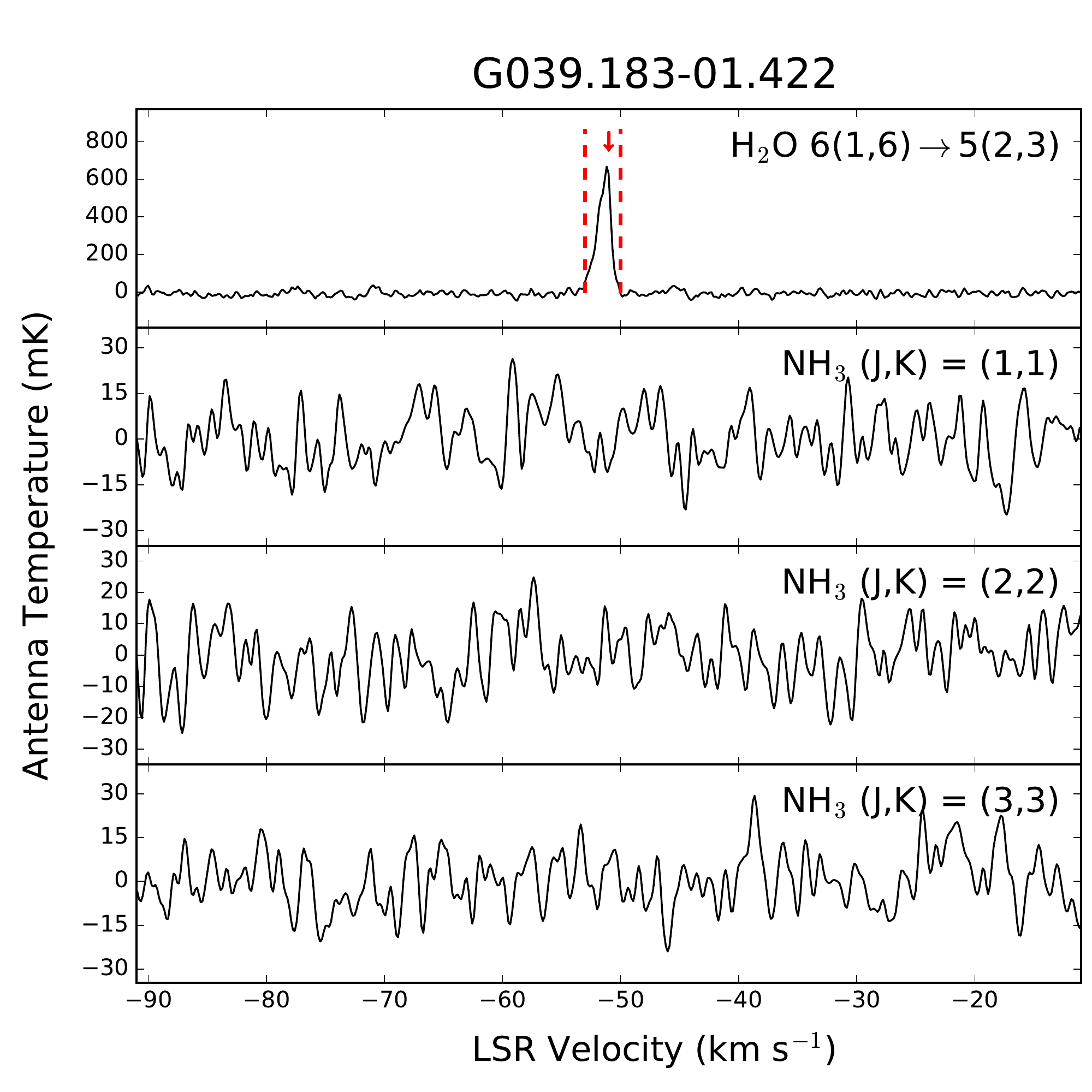}\\
\includegraphics[width=\figTwoSize]{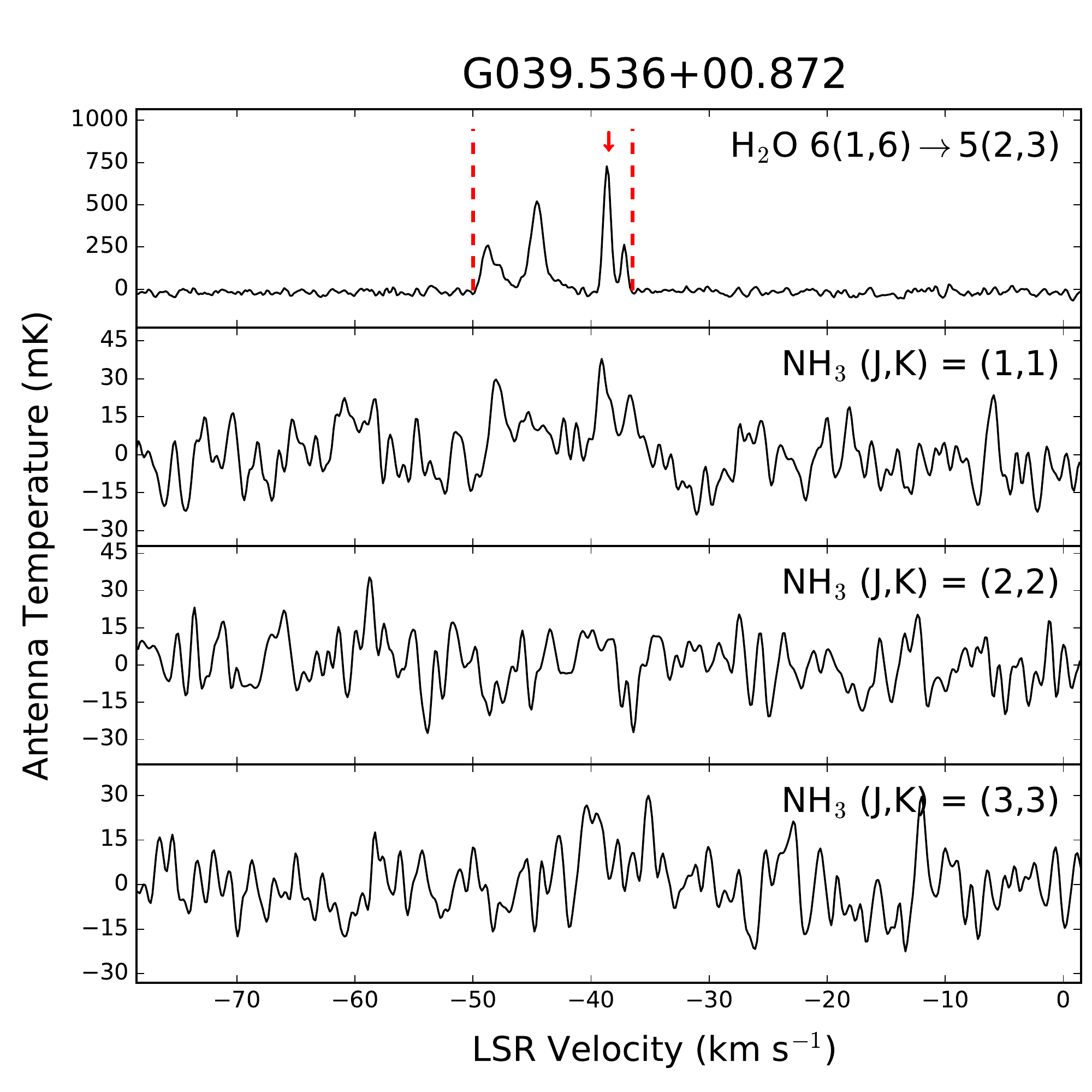}
\includegraphics[width=\figTwoSize]{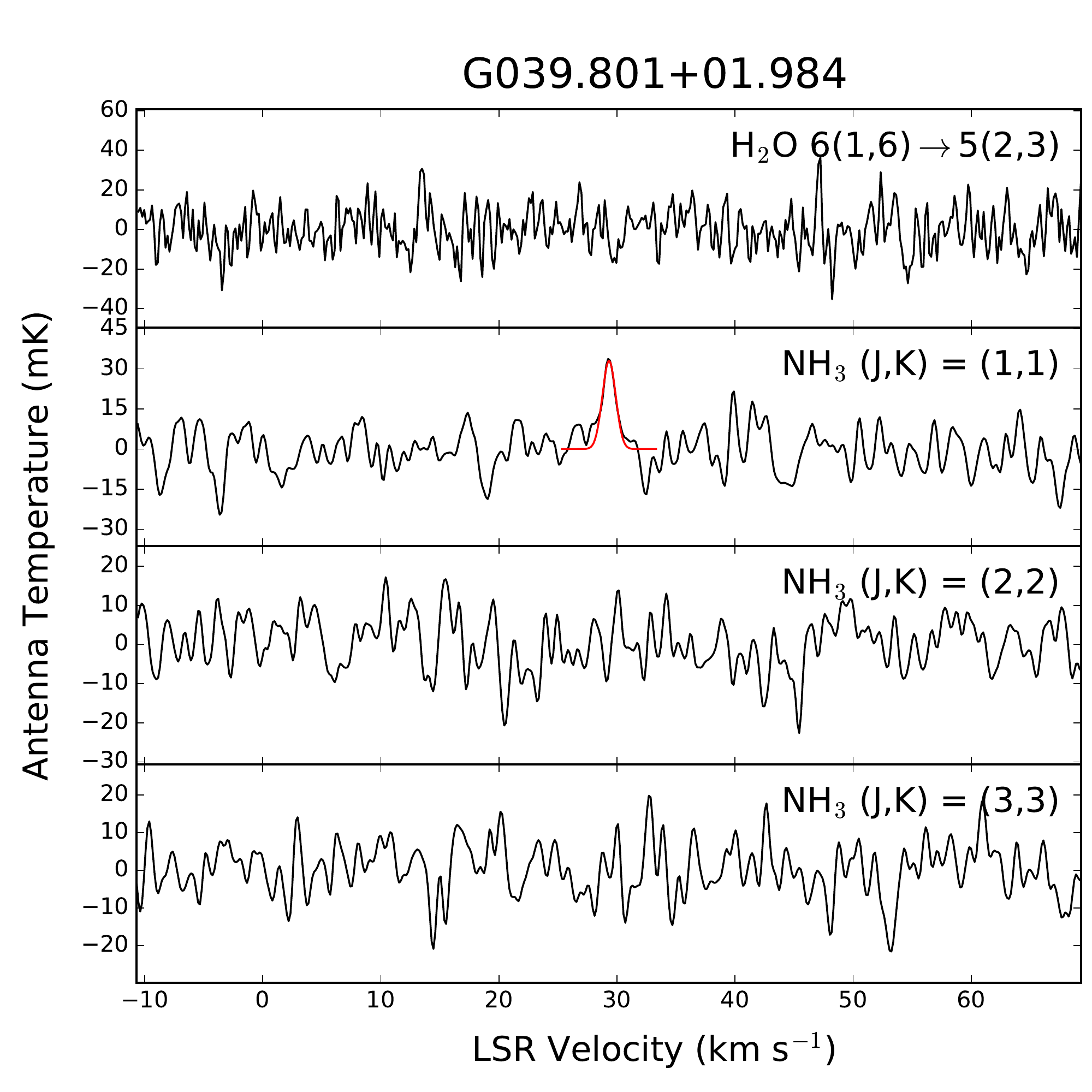}
\end{figure*}
\begin{figure*}[!htb]
\includegraphics[width=\figTwoSize]{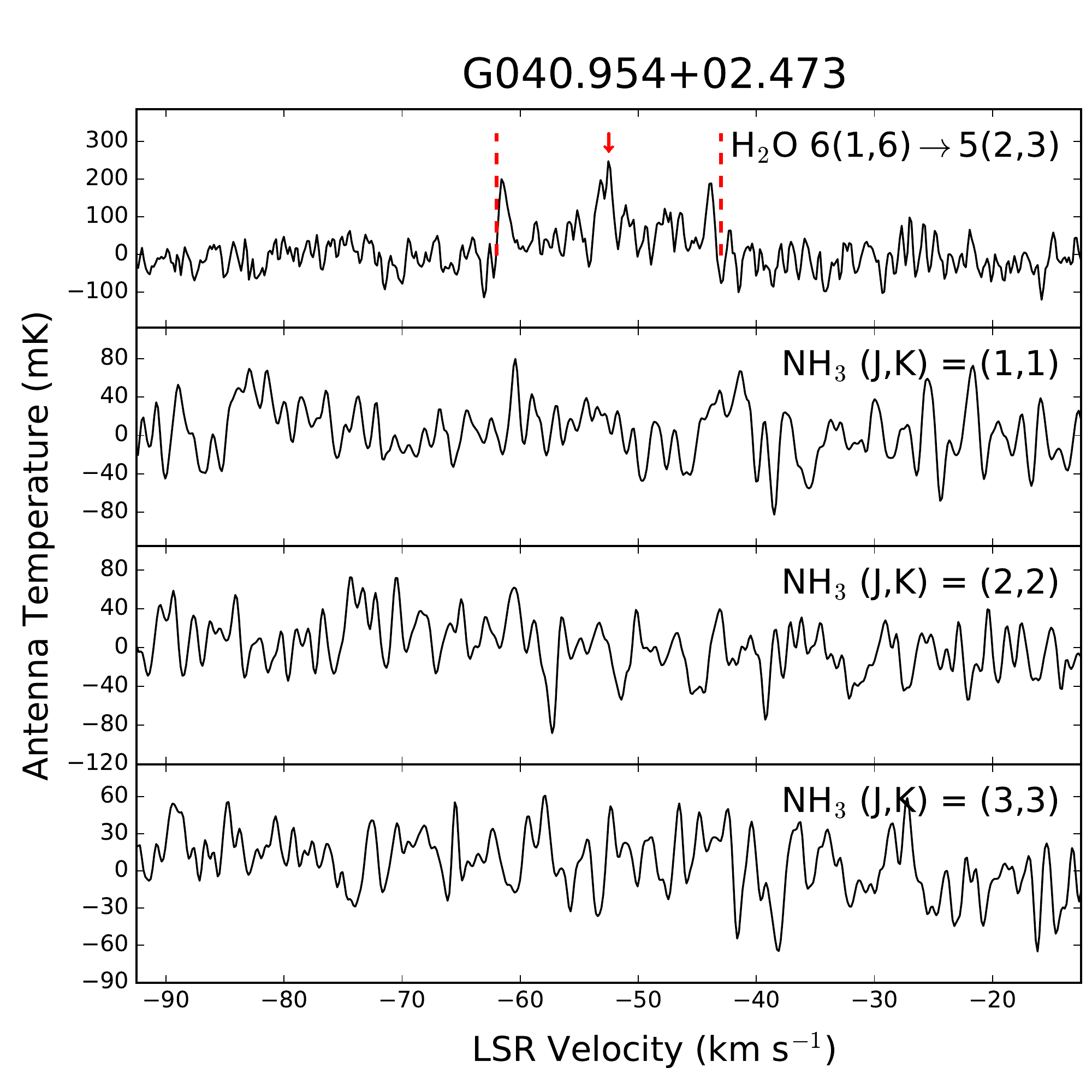}
\includegraphics[width=\figTwoSize]{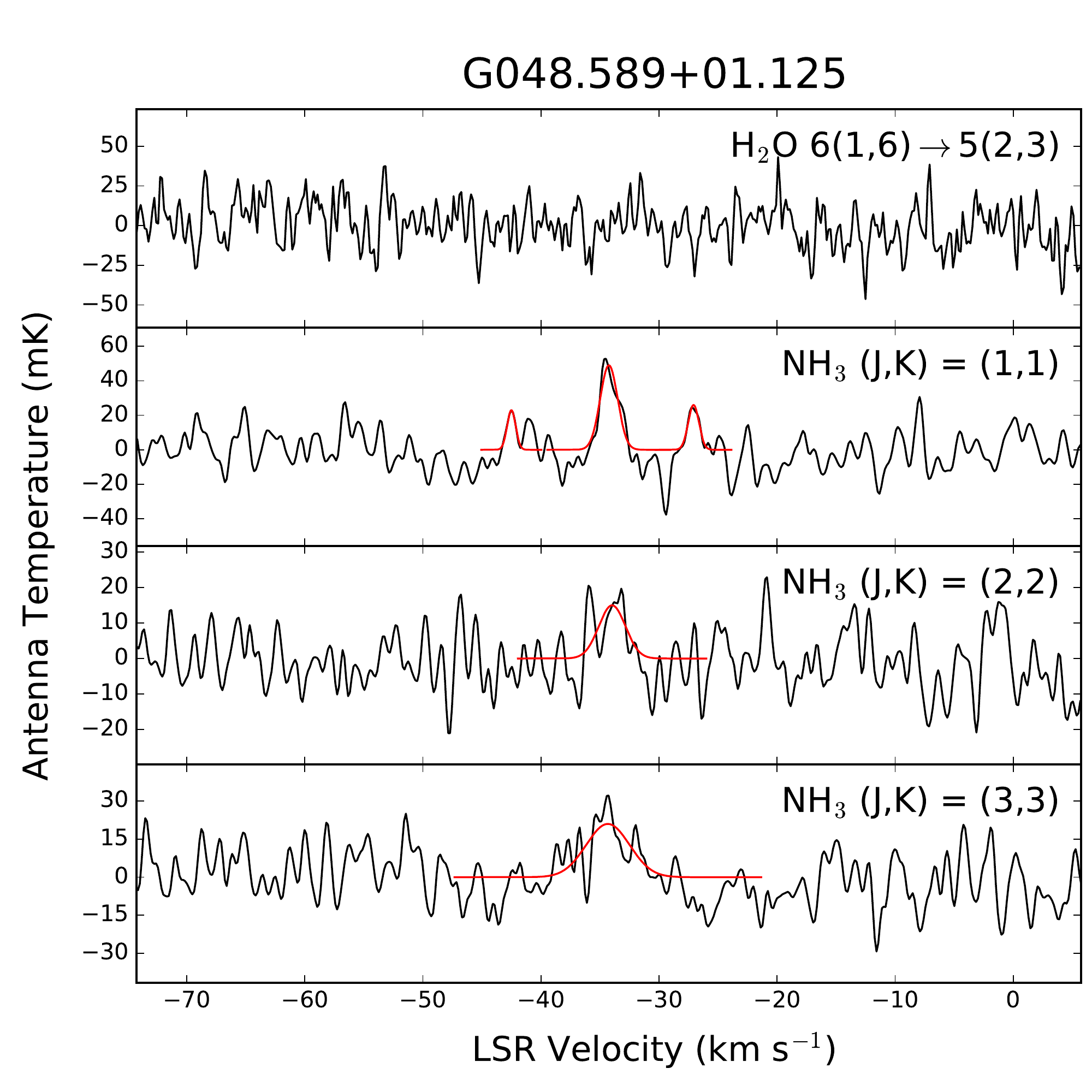}\\
\includegraphics[width=\figTwoSize]{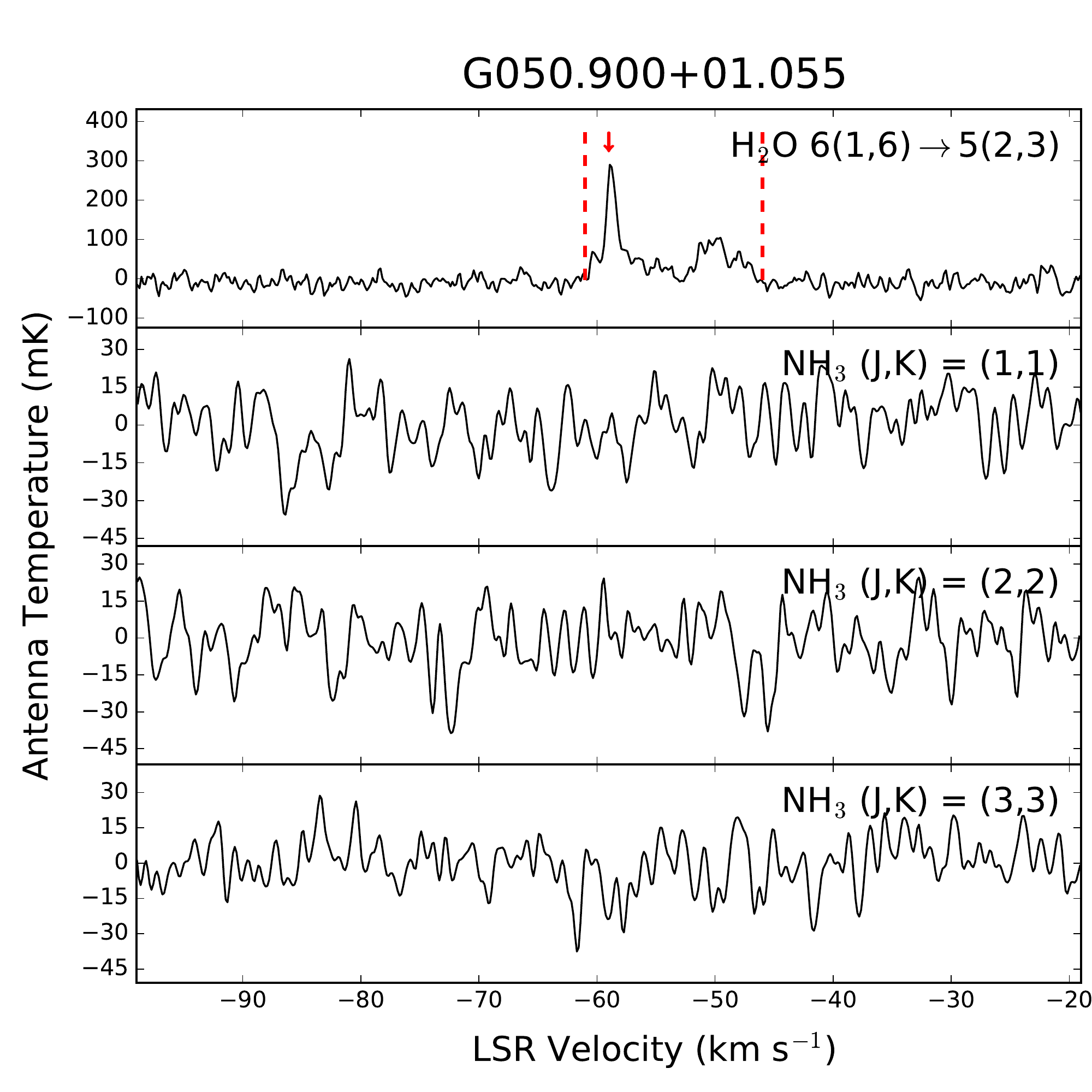}
\includegraphics[width=\figTwoSize]{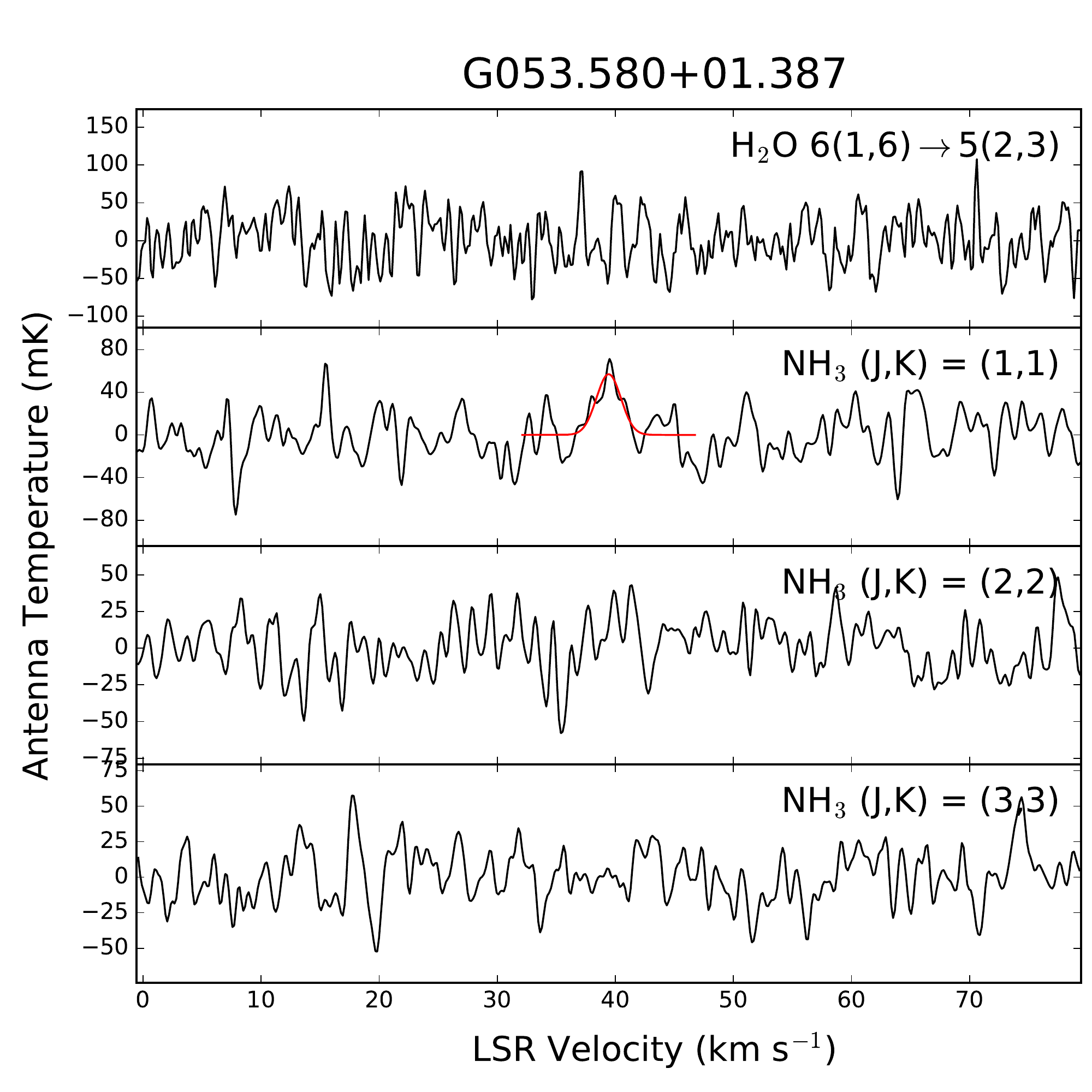}
\end{figure*}
\begin{figure*}[!htb]
\includegraphics[width=\figTwoSize]{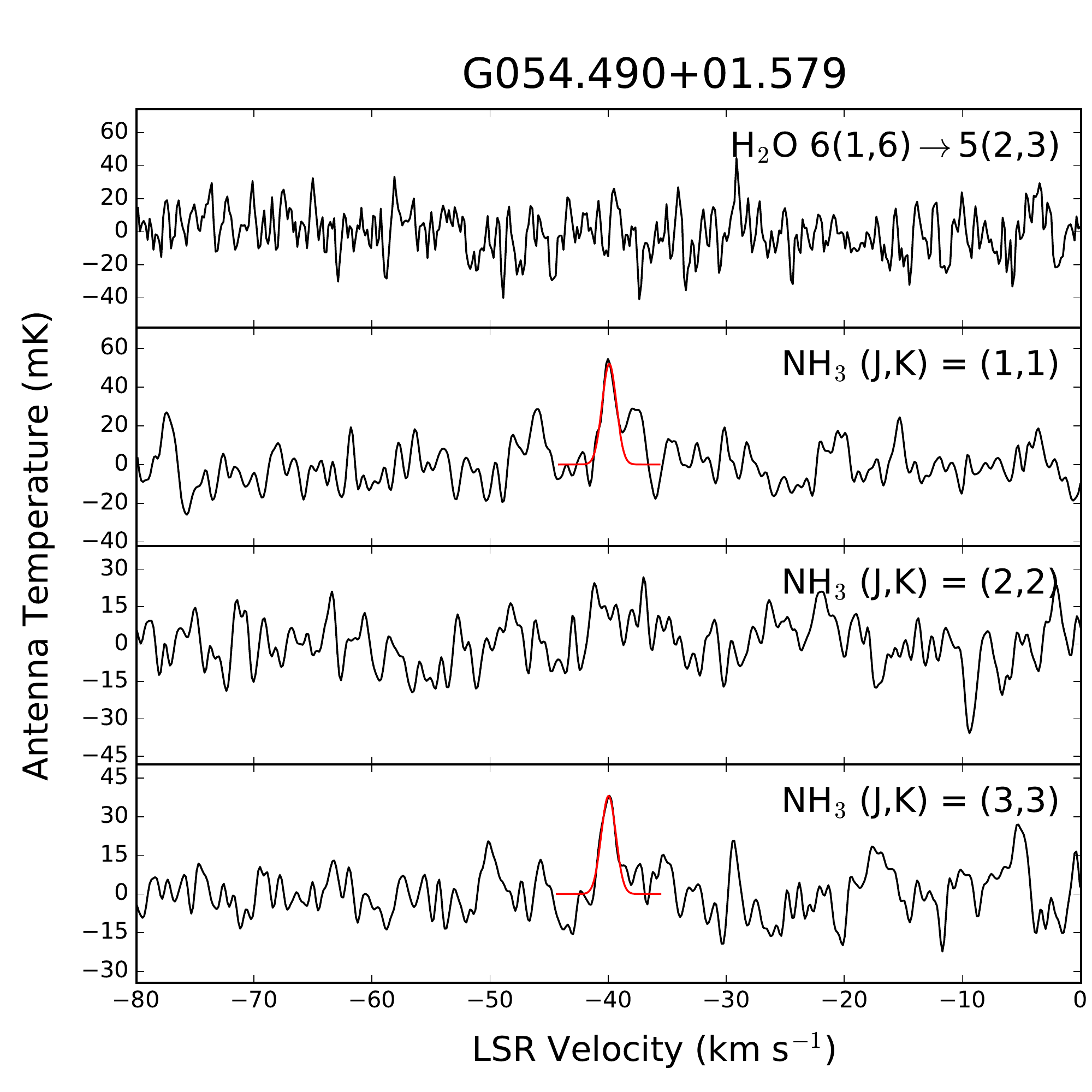}
\includegraphics[width=\figTwoSize]{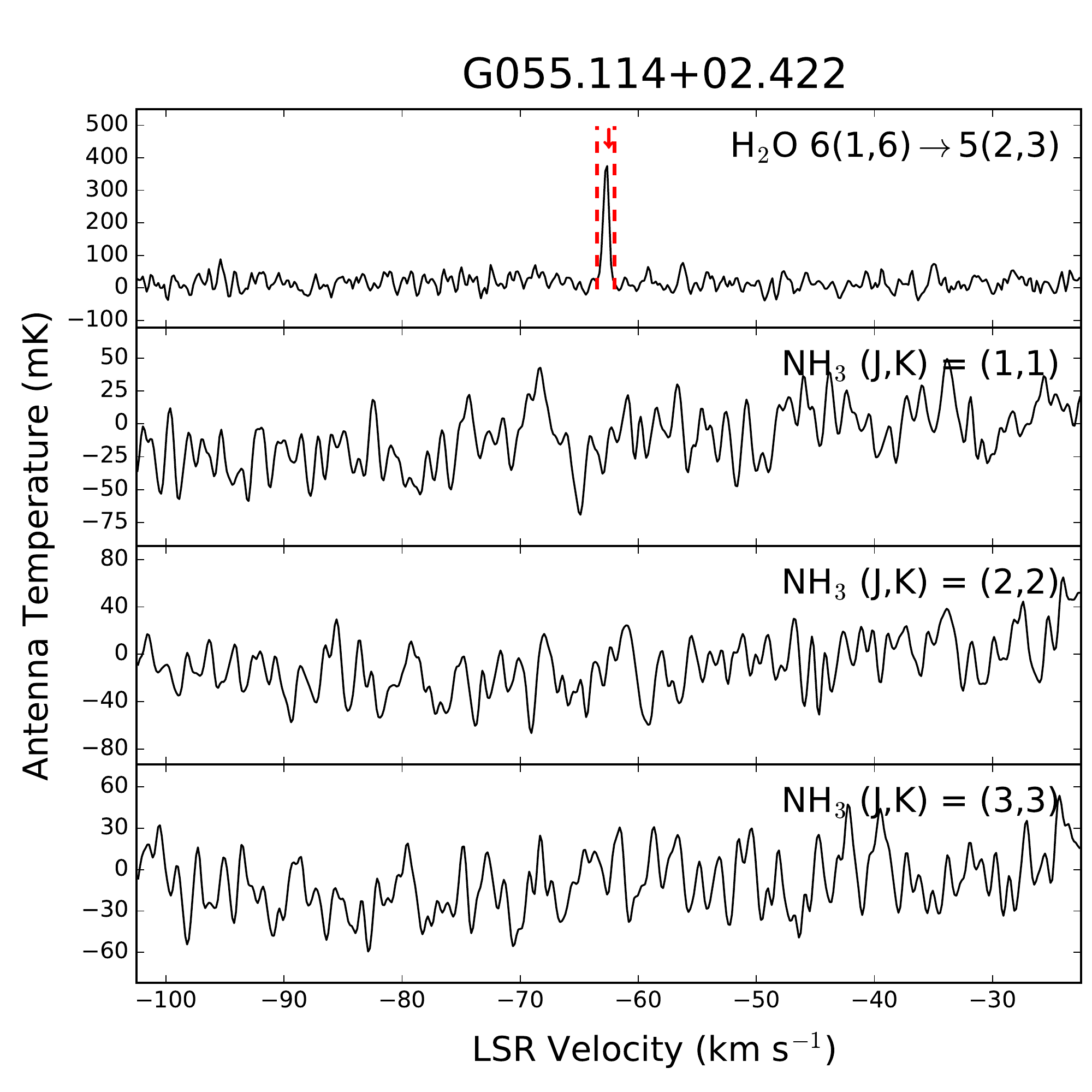}\\
\includegraphics[width=\figTwoSize]{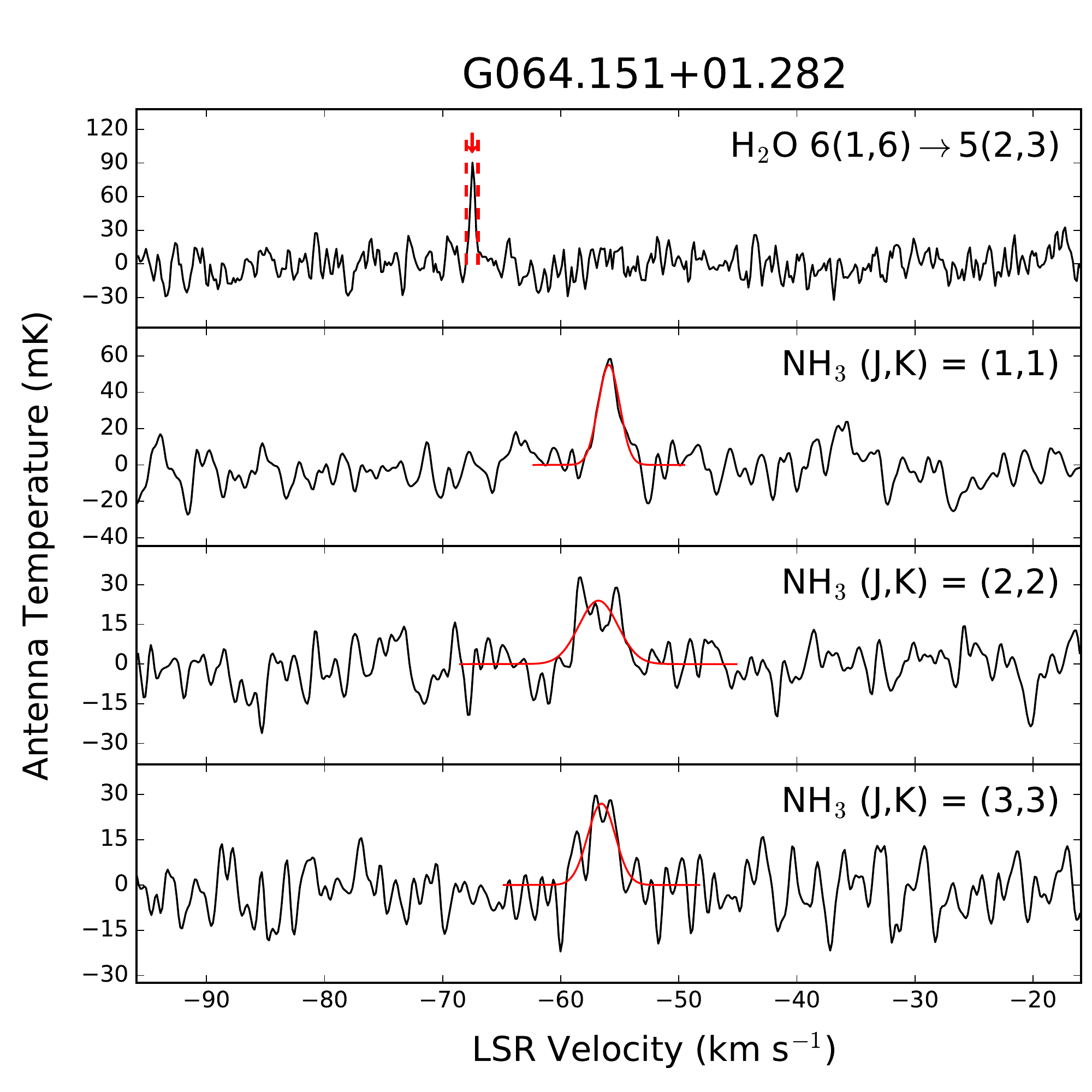}
\includegraphics[width=\figTwoSize]{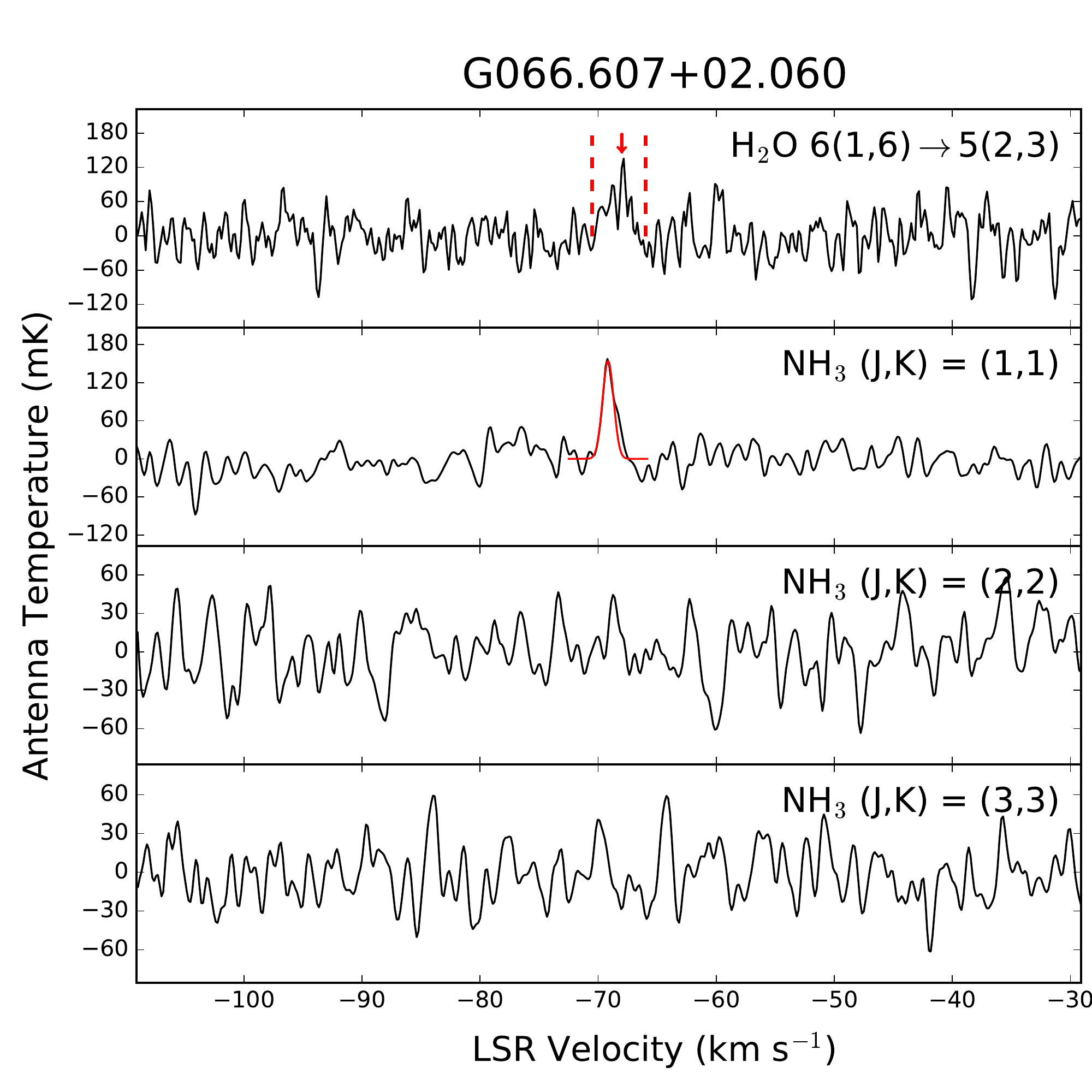}
\end{figure*}

\end{document}